%% file: jhep_paper.tex
\documentclass[a4paper,11pt]{article}
\pdfoutput=1

\usepackage{jheppub}

\usepackage[T1]{fontenc}

\hypersetup{
    colorlinks,
    linkcolor={red!50!black},
    citecolor={blue!50!black},
    urlcolor={blue!80!black}
}

\input{custom_instructions}

\usepackage{longtable}
\usepackage{booktabs}
\usepackage{pdflscape}
\usepackage{listings}
\usepackage{makecell}

\preprint{\begin{tabular}[t]{r}IRMP-CP3-26-28\\COMETA-2026-33\\MLQC4FC-2026-0003\end{tabular}}

\title{\boldmath \newms: automated spin-entangled decays of heavy resonances via the spin-density matrix formalism
}

\author[a]{Spyros Argyropoulos,}
\author[b]{Valentin Durupt,}
\author[b]{Olivier Mattelaer}

\affiliation[a]{Department of Physics, Aristotle University of Thessaloniki, 54124 Thessaloniki, Greece}
\affiliation[b]{Centre for Cosmology, Particle Physics and Phenomenology (CP3), Universit\'e Catholique de
Louvain, Chemin du Cyclotron, Louvain-la-Neuve, B-1348, Belgium}

\emailAdd{spyros.argyropoulos@cern.ch}
\emailAdd{valentin.durupt@uclouvain.be}
\emailAdd{olivier.mattelaer@uclouvain.be}

\abstract{The simulation of processes that contain multiple particles from the decay of unstable resonances becomes computationally challenging both at leading order and beyond. This paper presents \newms, a new version of \ms, a dedicated tool tackling this problem, primarily for events generated with the \mgamc\ event generator. While still based on the Frixione--Laenen--Motylinski--Webber formalism, the new version leverages
spin-density matrices to lift most of the limitations of the previous
implementation. It supports loop-induced processes, $N$-body decays and
samples defined by amplitude-level interference between distinct
hard-process amplitudes. It also introduces the multiple-pole approximation
and a resonance-helicity decomposition, including off-diagonal
helicity-interference contributions. These developments substantially broaden the phenomenological scope of \ms, enabling applications that were previously inaccessible within the framework, several of which are demonstrated in this work.}

\begin{document} 
\maketitle
\flushbottom

\include{intro}

\include{notations}

\include{v1}

\include{v2}

\include{validation}

\include{applications}

\include{performance}

\include{conclusion}

\appendix
\include{appendix}

\acknowledgments

The authors would like to thank Fabio Maltoni and Rikkert Frederix for useful comments on the implementation, Pierre Artoisenet, Stefano Frixione and Eric Laenen for their support, and Quentin Heirebaudt, who implemented and validated the density approach during his master thesis \cite{Heirebaudt:2023msc}. SA acknowledges support from the MaxLHC convention for participation in the \mgamc\ developers' workshop, where part of this work was presented. OM, VD and the \texttt{MadGraph} collaboration are supported by FRS-FNRS (Belgian National Scientific Research
Fund) IISN projects 4.4503.16 (MaxLHC). Computational resources have been provided by the
Consortium des Équipements de Calcul Intensif (CÉCI), funded by the Fonds de la Recherche
Scientifique de Belgique (F.R.S.-FNRS) under Grant No. 2.5020.11 and by the Walloon Region and DR-Weave grant FNRS-DFG No. T019324F (40020485).
This article/publication is based upon work from COST Actions MLQC4FC (CA24146) and COMETA (CA22130), supported by COST (European Cooperation in Science and Technology).

\bibliographystyle{JHEP}
\bibliography{bibliography}

\end{document}

%% file: custom_instructions.tex
\usepackage{listings}
\usepackage{xcolor}
\usepackage{amsmath}
\usepackage{booktabs}
\usepackage{array}
\usepackage{amssymb}
\usepackage{multirow}
\usepackage{cleveref}
\usepackage{pdflscape}
\usepackage[bitstream-charter]{mathdesign}
\usepackage{subcaption}
\usepackage{cleveref}
\crefname{figure}{Fig.}{Figs.} 
\Crefname{figure}{Fig.}{Figs.}
\crefname{equation}{Eq.}{Eqs.} 
\Crefname{equation}{Eq.}{Eqs.}
\crefname{section}{Sec.}{Secs.} 
\Crefname{section}{Sec.}{Sec.}
\crefname{table}{Table}{Tables} 
\Crefname{table}{Table}{Tables}
\usepackage{enumerate}
\crefformat{subfigure}{(#2#1#3)}
\crefrangeformat{subfigure}{(#3#1#4)--(#5#2#6)}

\DeclareSymbolFont{usualmathcal}{OMS}{cmsy}{m}{n}
\DeclareSymbolFontAlphabet{\mathcal}{usualmathcal}

\definecolor{spbg}{RGB}{242,242,242}
\definecolor{spblue}{RGB}{44,72,255}
\definecolor{spgreen}{RGB}{40,150,40}
\definecolor{spteal}{RGB}{60,130,130}
\definecolor{spblack}{RGB}{20,20,20}

\lstdefinelanguage{SciPostPython}{
  language=Python,
  morekeywords={from,import,as,for,while,if,else,elif,return,class,def},
  keywordstyle=\color{spblue}\bfseries,
  commentstyle=\color{spteal}\itshape,
  stringstyle=\color{spgreen},
  basicstyle=\ttfamily\small\color{spblack},
  showstringspaces=false,
  breaklines=true,
  columns=fullflexible
}

\lstdefinestyle{scipostcode}{
  language=SciPostPython,
  backgroundcolor=\color{spbg},
  frame=none,
  rulecolor=\color{black},
  xleftmargin=0.8em,
  xrightmargin=0.8em,
  aboveskip=0.6\baselineskip,
  belowskip=0.3\baselineskip,
  captionpos=b,
  keepspaces=true
}

\newcommand{\mgamc}{{\textsc{MG5aMC}}}
\newcommand{\mg}{{\textsc{MG5aMC}}}
\newcommand{\ms}{{\textsc{MadSpin}}}
\newcommand{\newms}{{\textsc{MadSpin2}}}
\newcommand{\madloop}{{\textsc{MadLoop}}}
\newcommand{\mcal}{\ensuremath{\mathcal{M}}}
\newcommand{\mfull}{\ensuremath{\mathcal{M}_{\rm full}}}
\newcommand{\mprod}{\ensuremath{\mathcal{M}_{\rm prod}}}
\newcommand{\mdec}{\ensuremath{\mathcal{M}_{\rm dec}}}

\newcommand{\opt}[1]{\texttt{#1}}
\newcommand{\code}[1]{\texttt{#1}}

\newcommand{\Tr}{\operatorname{Tr}}

\newcommand{\bk}{\discretionary{}{}{}}
\newcommand{\tabgroup}[1]{%
  \addlinespace[3pt]\midrule
  \multicolumn{4}{@{}l}{\emph{#1}}\\*
  \midrule}


%% file: intro.tex
\section{Introduction}
\label{sec:intro}

The analysis of high-energy physics experiments relies on comparing measured data with theoretical predictions embedded within simulations provided by Monte Carlo event generators \cite{Buckley:2011ms}. These include perturbative calculations of the matrix elements (ME) corresponding to the hard scattering process, stochastic resummation of QCD and QED radiation implemented via parton showers (PS), as well as phenomenological models describing the underlying event and non-perturbative hadronisation.

The increasing amount of data delivered by the Large Hadron Collider (LHC), leading to measurements of ever-improving precision, necessitates in most cases at least next-to-leading order (NLO) calculations of QCD matrix elements in order to reach a theoretical precision comparable to experimental uncertainties.
While state-of-the-art calculations include next-to-next-to-leading order (NNLO) QCD matrix elements matched to PS \cite{Re:2014uua,Campbell:2021svd,Monni:2019whf}, these are so far limited to selected processes. On the other hand, the standard reference for LHC experiments is still provided by NLO QCD calculations matched to PS,
which have been automated in frameworks such as \textsc{Herwig} \cite{Bellm:2015jjp}, \textsc{MadGraph5\_aMC@NLO} (\mgamc) \cite{Alwall:2014hca,Frederix:2018nkq}, \textsc{Powheg} \cite{Nason:2004rx,Frixione:2007vw,Oleari:2010nx,Alioli:2010xd,Banfi:2023mhz}, and \textsc{Sherpa} \cite{Hoeche:2011fd,Sherpa:2019gpd,Sherpa:2024mfk}.
Although in principle any process can be simulated at NLO accuracy in QCD in these frameworks, in practice the problem quickly becomes computationally unfeasible for high-multiplicity final states.

In the presence of resonances, a common strategy to reduce the computational complexity relies on the fact that the particle width is typically small compared to its pole mass. In the strict narrow-width approximation (NWA) \cite{Uhlemann:2008pm,Berdine:2007uv}, the decaying particle is forced to be exactly on-shell and non-resonant diagrams are neglected. This effectively factorises production and decay, significantly simplifying the computation, especially when NLO effects are included only in the production stage. However, in many cases it is crucial to retain off-shell effects (and thus go beyond the strict narrow-width approximation) while preserving full spin correlations.

Several approaches have been developed to address this problem. These include the double-pole approximation, widely used for di-boson production and vector-boson scattering \cite{Stuart:1991xk,Aeppli:1993rs,Denner:2000bj,Denner:2024xul}, where matrix elements are evaluated with on-shell momenta, and off-shell effects are incorporated through a Breit–Wigner smearing; truncated resonance methods \cite{Alwall:2011uj}; the \textsc{BRIDGE} method \cite{Meade:2007js}; the Frixione--Laenen--Motylinski--Webber (FLMW) method \cite{Frixione:2007zp}, as well as density-matrix-based approaches \cite{Richardson:2001df}.
Dedicated implementations of spin-correlated resonance decays are also available in other tools: the internal decay handlers of \textsc{Herwig} and \textsc{Sherpa} \cite{Sherpa:2019gpd}, the factorised-process support of \textsc{Whizard} \cite{Kilian:2007gr,Moretti:2001zz}, the \textsc{EvtGen} package \cite{Lange:2001uf} widely used for heavy-flavour decays, and the \textsc{Tauola}/\textsc{TauSpinner} tools \cite{Jadach:1990mz,Jadach:1993hs,Davidson:2010rw,Czyczula:2012ny} for polarised $\tau$-lepton decays.

In this paper, we present \newms, a new version of the \ms\ code \cite{Artoisenet:2012st} that was originally introduced more than a decade ago and is based on the FLMW method. The new implementation retains the FLMW framework but replaces the computation of the full matrix element with a density-matrix approach. This significantly extends its applicability compared to the previous version, allowing, for example, the treatment of $N$-body decays, loop-induced processes, and samples defined by amplitude-level interference. It also gives access to a decomposition of the resonance-helicity space, including the off-diagonal helicity-interference contributions. It also provides several options for handling off-shell effects, including configurations equivalent to the double-pole approximation and beyond.

The paper is organised as follows. In Sec.~\ref{sec:notation}, we introduce our notation and review the approximations used to describe unstable resonances, clarifying several definitions whose usage is not always consistent in the literature. In Sec.~\ref{sec:v1}, we review the status and limitations of the previous version of the tool. In Sec.~\ref{sec:madspin2}, we describe the new algorithm and its implementation. Sec.~\ref{sec:validation} presents a detailed physical validation of the code, while Sec.~\ref{sec:applications} illustrates the capabilities of \newms\ by presenting several phenomenological applications which were not possible with the previous version. Sec.~\ref{sec:performance} discusses its computational performance. Finally, conclusions are given in Sec.~\ref{sec:conclusions}. We also include three appendices: App.~\ref{seq:commands} corresponds to a short manual, App.~\ref{app:sequential} describes some technical choices of the FLMW method in the presence of multiple particles to decay, and App.~\ref{app:profiling} details the breakdown of the computational performance.

%% file: notations.tex
\section{Notations and approximations}
\label{sec:notation}

For a process (or set of processes) $ab\to p_1\ldots p_n\, d_1d_2$, which proceeds via the exchange of a resonance $X$ in the $s$-channel, the amplitude can be written as (here, without loss of generality, for the case of a spin-1 resonance):
\begin{equation}
    \mathcal{M}(ab\to p_1\ldots p_n\,d_1d_2) =\sum_{\lambda}\mathcal{M}^{\mu}_{\rm prod}\frac{\epsilon_{\mu}^{\lambda}\epsilon_{\nu}^{*\lambda}}{p^2-M^2+iM\Gamma}\mathcal{M}^{\nu}_{\rm dec},
    \label{eq:fact_resonance}
\end{equation}
where $\mathcal{M}^\mu_{\rm prod}\equiv\mathcal{M^\mu}(ab\to p_1\ldots p_n\,X)$, $\mathcal{M}^\mu_{\rm dec}\equiv\mathcal{M^\mu}(X\to d_1d_2)$ and $\lambda$ denotes the polarisation of the intermediate resonance.\footnote{For massive off-shell vector bosons this sum should technically run from one to four, including a polarisation that vanishes when the vector boson is exactly on-shell \cite{Basu:2025zds}. This fourth component is highly suppressed and therefore the code performs the sum only over the transverse and longitudinal polarisations.} The squared amplitude then becomes
\begin{equation}\label{eq:densityMaster}
    |\mathcal{M}(ab\to p_1\ldots p_n\,d_1d_2)|^2 = \frac{1}{(p^2-M^2)^2+M^2\Gamma^2}\sum_{\lambda,\lambda'}\rho^{\lambda\lambda'}_{\rm prod}\rho^{\lambda\lambda'}_{\rm dec},
\end{equation}
with 
\begin{equation}
\rho^{\lambda\lambda'}_{\rm prod/dec}  \equiv  
\mathcal{M}_{\rm prod/dec}^{\mu}\mathcal{M}_{\rm prod/dec}^{*\nu}   \epsilon^{\lambda}_{\mu} \epsilon_{\nu}^{*\lambda'}
\end{equation}
the spin-density matrix for the production/decay process. In general, the amplitude will have a non-trivial dependence on $(p_a\cdot p_{d_i}),(p_b\cdot p_{d_i})$, known as \textit{production spin correlations}, and on $(p_{d_i}\cdot p_{d_j})$, known as \textit{decay spin correlations}. These spin correlations are encoded in the off-diagonal elements of the spin-density matrix.  

As explained in the introduction, the calculation of the full amplitude, especially at higher orders in perturbation theory, is usually performed with certain approximations. These are explained in the following subsections.

\subsection{Narrow-width and decay-chain approximations}

For resonances that are sufficiently narrow ($\Gamma\ll M$), the first approximation is to neglect non-resonant diagrams, such that all considered amplitudes contain the resonances that appear in Eq.~\ref{eq:fact_resonance}. This is called the decay-chain approximation (DCA) and is used extensively throughout this work.

One can further employ the narrow-width approximation (NWA) to integrate out the propagator in Eq.~\ref{eq:densityMaster}, yielding
\begin{equation}\label{eq:NWA_old}
    \frac{f(p^2)}{(p^2-M^2)^2+M^2\Gamma^2}\xrightarrow[\Gamma\to0]{}\frac{f(M^2)\delta(p^2-M^2)}{2\Gamma M},
\end{equation}
where $f$ is a generic smooth function.

A strict application of the NWA implies that the invariant-mass distribution of the decaying particle will be a delta function at the pole mass, which may be inadequate for observables sensitive to the resonance line shape or to kinematic boundaries.

When using this limit in Eq.~\ref{eq:densityMaster}, one should note that the production and decay do not fully factorise due to the off-diagonal elements of the spin-density matrices.
However, if one integrates over the full phase space of the decay, the integrated decay density matrix becomes proportional to the identity and production spin correlations cancel in the inclusive cross-section, yielding the branching ratio formula:
\begin{equation}\label{eq:NWA}
    \sigma_{\rm full} \equiv\int_{-\infty}^{+\infty}\frac{dp^2}{2\pi}\frac{\sigma(p^2)}{(p^2-M^2)^2+M^2\Gamma^2}\xrightarrow[\Gamma\to0]{}\frac{\sigma(M^2)}{2\Gamma M}=\sigma_{\rm prod}\times {\rm BR}.
\end{equation}
It should be noted that although in the NWA production spin correlations vanish in the inclusive cross-section, they survive in differential distributions and after imposing phase-space cuts.

In contrast, one can choose to neglect the spin correlations locally, i.e. before the integration over the full phase space. We dub this approximation ``No Spin Approximation'' (NSA) which, at the amplitude-squared level, is equivalent to the following:
\begin{equation}\label{eq:NSA2}
    |\mathcal{M}(ab\to p_1\ldots p_n\,d_1d_2)|^2 \approx \frac{1}{M^2\Gamma^2}\frac{1}{2s_X+1}\Tr\left[ \rho_{\rm prod}\right]\cdot\Tr\left[ \rho_{\rm dec}\right],
\end{equation}
where $s_X$ is the spin of the resonance $X$.

Multiple approaches have been developed to improve upon the strict NWA, preserving both spin-correlation and off-shell effects (see \cite{Carrivale:2025mjy} for an in-depth comparison):
\begin{itemize}
    \item the first one is the DCA introduced above, as implemented in \mgamc. In this approach a full phase-space integration (on the $n+2$ final-state particle phase space) of the production plus decay amplitude squared (Eq.~\ref{eq:densityMaster}) is performed without further approximation, i.e.\ with off-shell momenta. The drawback of using off-shell momenta lies in the fact that the amplitudes are not gauge invariant, which can lead to an ambiguity in the tail prediction depending on the gauge choice. Consequently, this method typically implements a cut that restricts the invariant mass of the decaying resonance to be close to the pole\footnote{In \textsc{MadSpin} this cut is implemented with the parameter \texttt{bwcutoff}.}.
    
    \item The second method is based on a pole expansion, usually referred to as the Multiple Pole Approximation (MPA) \cite{Stuart:1991xk,Aeppli:1993rs}.
    In this approach, the full phase-space integration is performed as well but, before using Eq.~\ref{eq:fact_resonance}, one changes the kinematics of the phase-space points so that the momenta of the decaying particles are exactly on-shell. The production/decay amplitudes and polarisation vectors are evaluated with these on-shell momenta, while the denominator is kept off-shell to ensure the Breit-Wigner shape.
    
     This construction preserves gauge invariance at the level of the pole expansion. However, the result depends on the choice of reshuffling/projection map used to relate the off-shell and on-shell kinematics, which also makes the prediction not uniquely defined in the off-shell tail, even though the invariant-mass distribution exactly follows the Breit-Wigner one by construction.

    We will also call MPA an equivalent method (up to reshuffling effects), where the phase-space integration is done independently for the production and the decay, with on-shell momenta, and evaluating Eq.~\ref{eq:fact_resonance}. Then, in a second step, the four-momenta of the full event are reshuffled so that the invariant mass of the decaying particle follows a Breit-Wigner distribution. The two methods would be strictly identical if the reshuffling were bijective, which is not possible. This second method --implemented in \newms-- is subject to a potential dead zone, as will be demonstrated later on (see Sec.~\ref{subsec:offshell}).

     \item For completeness, we will also mention here the approximation employed by the ancestor of \textsc{MadSpin}: \textsc{BRIDGE} \cite{Meade:2007js}, where partial spin-correlation effects were kept by replacing each of the density matrices with its diagonal elements. However, we will not compare to that method in this paper.
     
\end{itemize}

 While the first method is widely used at leading-order (LO) accuracy, beyond LO the MPA method is generally preferred \cite{Carrivale:2025mjy}. The original \textsc{MadSpin} code \cite{Artoisenet:2012st} employed a variation of the DCA, known as the FLMW approximation (discussed in the next section), while the new version of the code also contains an implementation of the MPA.

\subsection{The FLMW procedure}\label{subsec:FLMW}

Motivated by the requirement to model spin-correlation effects to NLO accuracy, Frixione, Laenen, Motylinski and Webber have proposed a procedure to extend the DCA approach \cite{Frixione:2007zp}. 

This procedure is based on an extra unweighting step that uses the tree-level amplitudes associated with the $\mathbb{H}/\mathbb{S}$ NLO events \cite{Frixione:2002ik}. One should note that this method is not fully NLO accurate for several reasons, namely it is based on LO decay matrix elements, it neglects loop effects between production and decay and it uses the underlying Born matrix elements in the unweighting of $\mathbb{S}$ events.

In detail, the FLMW method takes a production event (which can be either weighted or unweighted) and generates an associated weighted decay event. Then it merges the production and decay events, reshuffling the four-momenta so that the propagator virtuality follows a Breit-Wigner distribution.
A weight is then associated with the events:
\begin{equation}\label{eq:ms_weight}
\frac{|\mathcal{M}_{\rm full}|^2}{|\mathcal{M}_{\rm prod}|^2 }w_{\rm shuffle},
\end{equation}
where $w_{\rm shuffle}$ is the Jacobian associated with the reshuffling procedure and with the generation of the phase-space of the decay event (see App.~\ref{app:sequential} for more details). Since this weight is bounded \cite{Frixione:2007zp}, one can use a standard accept/reject mechanism to shape the decay kinematics according to the full density matrix encoded in the numerator. In this procedure, only the decay events are accepted/rejected, allowing one to keep the same number of events as in the original sample; if that sample was unweighted, it will remain unweighted.

As in the MPA, the reshuffling procedure in the FLMW approximation is not unique either, and differences in the reshuffling method can lead to differences in the final prediction. However, this effect is expected to be milder in the case of FLMW since the accept/reject will wash out some of those differences.
On the other hand, since the FLMW method evaluates the matrix element using off-shell momenta for the propagator, it is sensitive to gauge fixing, like the DCA.

An overview of the various approximations described above is given in Table~\ref{tab:madspin_modes_combined}.

%% file: v1.tex
\section{Running modes in \ms\ (version 1) and their limitations}\label{sec:v1}

The previous version of \ms\ has three different running modes, selected via the \opt{spinmode} option:
\texttt{full} (also selectable as \texttt{madspin}), \texttt{none} and \texttt{onshell}.
These were developed starting from different code bases and correspond to different approximations built on the NWA, as explained below.
\subsection{The \texttt{full}/\texttt{madspin} mode}\label{subsec:full_v1}

The original implementation of \ms\ uses the FLMW procedure. The main differences between the FLMW paper and its implementation in \ms\ can be summed up in three points:
\begin{itemize}
    \item First, while in the original FLMW procedure the maximum weight was calculated analytically, \ms\ uses a numerical estimate obtained by sampling the weight for a set of production events, each combined with a large number of decay configurations. While a numerical evaluation is slower than an analytical estimate, its advantages are that it can be applied to a larger class of processes and it allows one (through the distribution of weights) to gauge the validity of the NWA (in the sense that phase-space points where the NWA breaks down would lead to very large weights). 

    \item Second, while the reshuffling algorithm was not specified in the FLMW paper, \ms~\cite{Artoisenet:2012st} implements a quite complex one, first selecting a single diagram (based on the single diagram enhancement strategy \cite{Maltoni:2002qb}) and then using the channel mapping associated with that diagram to reshuffle the momenta.

    \item Finally, in \ms\ the production matrix element is evaluated with off-shell momenta (mainly for technical reasons), leading to a small theoretical mismatch in the evaluation of the amplitude. It has been checked in multiple cases that this mismatch is negligible.
\end{itemize}

The limitations of this implementation are the following:
\begin{enumerate}
    \item The phase-space integrator for the decay is limited to sequences of two-body decays (whose phase-space volume is flat). This is sufficient for SM decays but a clear limiting factor for some beyond-the-Standard-Model (BSM) scenarios (e.g. modelling decays in models with compressed spectra).
    \item The method requires the computation of $\mfull$ via \mgamc.
     While \mgamc\ is quite generic and can handle a large class of models/processes, it has limitations, in particular when handling the decay-chain approximation. The two main limiting cases are processes with loop-induced production and/or decay, and the case where the squared production matrix element is not guaranteed to be positive, as for the interference-only contributions arising in the Standard Model Effective Field Theory (SMEFT) (cf.\ Sec.~\ref{subsec:valid_interf}).\footnote{Support for polarised decays was also a limitation for a long time, which was lifted at LO in \mgamc~\cite{BuarqueFranzosi:2019boy}.}

    \item Since the reshuffling is performed at fixed $\sqrt{\hat{s}}$, the algorithm cannot populate some parts of the phase space (in particular the sub-threshold region), leading to either a dead zone or an under-populated region.
    This is also a strong limitation for a $2\to 1$ process, like $pp\to Z, Z\to \ell^+\ell^-$, where the mass of the decaying resonance is fixed by the partonic centre-of-mass energy.

\end{enumerate}
The first two of these limitations are lifted in \newms, while the last one remains and is discussed in more detail in Sec.~\ref{subsec:offshell}.

\subsection{The \texttt{none} mode}

This mode was introduced in 2014, when \mgamc\ started to support loop-induced processes \cite{Hirschi:2015iia}: since the \texttt{full} mode cannot handle such processes (limitation 2 above), the idea was to provide a reasonable way to handle the decay of the Higgs boson, whose dominant production channel is loop-induced.
It is the most basic mode, containing neither finite-width effects nor spin correlations, and corresponds to the NSA described above.

In practice, this mode generates unweighted decay events at LO accuracy, leveraging the phase-space integrator of \mgamc.
Then it attaches one of the generated decay events to a production event without any type of reshuffling or acceptance/rejection factor.

While this mode is only relevant for the decay of particles of zero spin (like the Higgs boson), it lifts both issues 1 and 2 described above, i.e. it can handle 3- and higher-body decays as well as loop-induced processes.

\subsection{The \texttt{onshell} mode}\label{subsec:onshell}

The \texttt{onshell} mode was a first attempt to allow 3-body decays with spin correlations, albeit with the compromise of keeping the decaying resonances exactly on-shell \cite{Mattelaer:2016ynf}. Besides that, this mode cannot handle the computation of loop-induced processes, and therefore its usefulness is restricted to niche cases, such as the decay of narrow resonances that have 3-body decays, which appear for instance in SUSY models with compressed spectra.

As in the \texttt{none} mode, the \texttt{onshell} mode uses \mgamc\ to generate unweighted decay events (allowing arbitrarily long decay chains). However, in this mode each decay event is associated, for the accept/reject phase with the weight
\begin{equation}\label{eq:onshell_reweighting}
    \frac{|\mfull|^2}{|\mprod|^2|\mdec|^2}.
\end{equation}
Notice that unlike Eq.~\ref{eq:ms_weight}, Eq.~\ref{eq:onshell_reweighting} does not contain the Jacobian factor since the decay events are already unweighted and no reshuffling is used. 

Note that this mode has a technical issue in the proper handling of resonances that decay into particles that are also present in the production event: the \texttt{onshell} mode fails to correctly distinguish such identical particles and averages the full matrix element over all the combinations that are compatible with the given decay. This limitation is lifted in \newms, and we will come back to it in more detail in Sec.~\ref{subsec:valid_iden} when validating the new code for that particular case.

While this mode is the least frequently used one in \ms, it constitutes the foundation on which we have based \newms, which alleviates most of the aforementioned limitations.

%% file: v2.tex
\section{The new \textsc{MadSpin2} code}
\label{sec:madspin2}

The updated version of \ms, presented here, is included in \mgamc\ v3.8.0 and later and will also be part of the \textsc{MadGraph7} release. It contains two major updates in the calculation of matrix elements and in the simulation of off-shell effects, which allow one to overcome most of the limitations presented above.

All legacy run modes described in Sec.~\ref{sec:v1} are retained in the new code. To distinguish them from the new spin-density-matrix
implementations, the legacy \texttt{madspin} and \texttt{onshell} modes
have been renamed to \texttt{madspin\_v1} and \texttt{onshell\_v1}, respectively, while
the \texttt{none} mode remains unchanged. The new implementations are
described below.

\subsection{Calculation of matrix elements via spin-density matrices}

While the previous \ms\ version required the calculation of the full matrix element, given by Eq.~\ref{eq:fact_resonance}, the new implementation relies only on the evaluation of the spin-density matrices $\rho^{\lambda\lambda'}_{\rm prod}$ and $\rho^{\lambda\lambda'}_{\rm dec}$ from Eq.~\ref{eq:densityMaster}, which can be separately evaluated by \mgamc\ for every process. This is a new feature of \mgamc\ which also allows the study of quantum information observables at colliders \cite{Durupt:2025wuk}.

For the FLMW method, the density matrix needs to be evaluated with off-shell momenta, causing a small technical complication. \mgamc\ uses \textsc{HELAS}~\cite{Murayama:1992gi,deAquino:2011ub} to evaluate the spinors and polarisation vectors. Such an implementation assumes on-shell particles since, instead of computing the invariant mass from the four-momenta, such routines directly use the pole mass -- for numerical reasons. To lift this limitation, we introduced a new syntax that instructs \mgamc\ (and \textsc{HELAS}) to compute the invariant mass entering these routines directly from the four-momenta.\footnote{The new syntax corresponds to adding a star after the particle name (e.g.\ \texttt{p p > h, h > z z*}). Such syntax is only supported in standalone mode, for the evaluation of matrix elements or density matrices at fixed phase-space points. It is not available in the standard event-generation or cross-section computation modes.}

Finally, the spin-density matrix implementation in \newms\ improves the treatment of processes that contain identical particles that appear in both production and decay, such as $pp\to Z(d\bar{d})d$. A detailed explanation and a validation of such processes is presented in Sec.~\ref{subsec:valid_iden}.

\subsection{Off-shell effects}\label{subsec:offshell}

In \ms, the reshuffling was based on the single diagram enhancement method \cite{Maltoni:2002qb}: a diagram is first selected, and the associated reshuffling keeps the invariant mass of every propagator of that Feynman diagram fixed. \newms\ instead uses a slightly simpler method, inspired by parton-shower algorithms and based on the information written in the Les Houches Event (LHE) file \cite{Alwall:2006yp}: the reshuffling leaves the invariant mass of every intermediate particle recorded in the file untouched.\footnote{In practice, the decision whether or not to write such a particle into the LHE output is actually based on the single diagram enhancement method and on a check of whether the particle is sufficiently on-shell, making the two methods quite similar in spirit.}
To achieve this, we split the input event into a sequence of production event plus decay events (following the information written in the LHE file) and then apply the \textsc{RAMBO} algorithm \cite{KLEISS1986359} on each of those parts sequentially.

\begin{figure}[htbp]
\begin{tabular}{cc}
  \centering
    \includegraphics[width=0.47\textwidth]{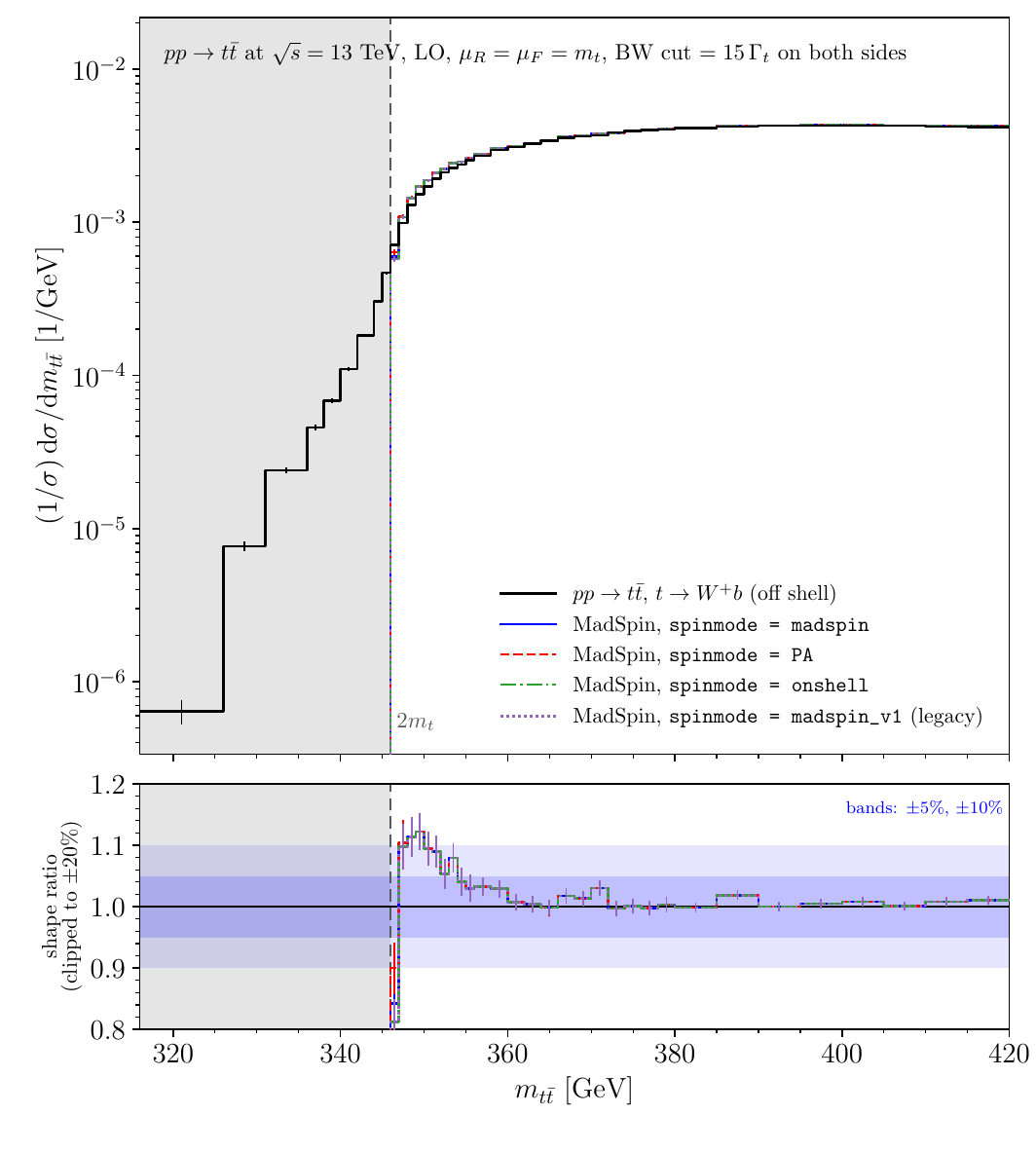} &
    \includegraphics[width=0.47\textwidth]{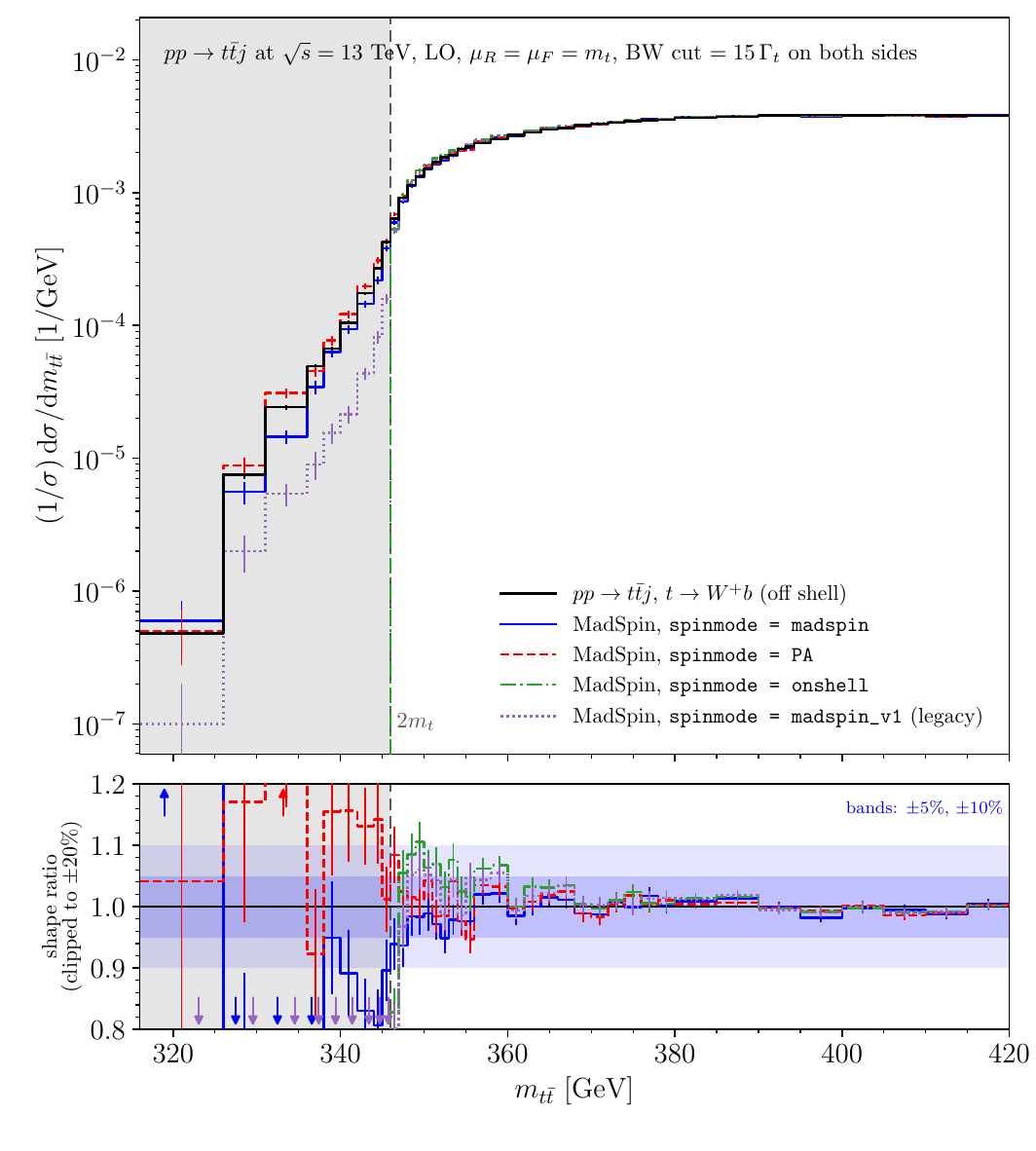}
    \end{tabular}
  \caption{The invariant-mass distribution $m(t\bar{t})$, for the $pp\to t\bar{t}$ (left) and $pp\to t\bar{t}j$ process (right). In both panes, the solid black curve corresponds to the off-shell reference computed with \mgamc, while the coloured curves correspond to \newms\ with \texttt{spinmode=madspin} (solid blue), \texttt{PA} (dashed red), \texttt{onshell} (dash-dotted green) and the legacy \texttt{madspin\_v1} (dotted purple). In the region close to the top-pair production threshold shown here, the NWA is not valid and reshuffling effects become important.}
  \label{fig:reshuffling}
\end{figure}

A common limitation of both \ms\ and \newms\ is that the reshuffling procedure keeps the original $\sqrt{\hat{s}}$ fixed: configurations whose resampled masses would require a larger partonic energy are not populated, leading to a ``dead zone'' close to the production threshold.
To illustrate this issue, Fig.~\ref{fig:reshuffling} compares the various \newms\  predictions (and \code{madspin\_v1}) against the pure \mgamc\ prediction for the invariant-mass distribution $m(t\bar{t})$ in $pp\to t\bar{t}$ events (left) and $pp\to t\bar{t}j$ events (right). Both plots are done at fixed scale $\mu_R=\mu_F=m_t=173$~GeV,  the $t\bar{t}j$ sample is using the following cuts: $p_T(j)>20$~GeV, $|\eta(j)|<5$. Each sample has at least one million unweighted events.

In the left plot, the displayed variable is by construction equal to $\sqrt{\hat{s}}$; this explains both why all the \ms\ curves sit exactly on top of each other and why none of them contains a single event below the threshold ($m(t\bar{t})<2m_t$).\footnote{One can remark that close to the threshold the curves are not fully identical; this is due to a few events carrying an overweight, which can be due to the breaking of the NWA in those regions.}

On the other hand, the extra jet in the $pp\to t\bar{t}j$ process can absorb the recoil from the momentum reshuffling,  allowing this region to be populated (with the obvious exception of the \texttt{onshell} mode). In this region, where the NWA is not valid, neither \code{madspin} nor \code{PA} reproduces the reference convincingly, although both perform considerably better than \code{madspin\_v1}, which strongly underestimates this region. This bias is due to the diagram selection, which keeps $m_{t\bar{t}}$ invariant for more than fifty percent of the sample. Note that while the new reshuffling works better for this specific case (and in particular with \code{PA}), one should not generalise the conclusion to other processes and/or observables.

\subsection{Summary of different run modes}

\newms\ supports six modes: the three legacy modes discussed above (\texttt{none}, \texttt{madspin\_v1} and \texttt{onshell\_v1}) and the three new density-matrix modes described below.
The new \texttt{onshell} mode in \newms\ has the same physics content (NWA) as the \texttt{onshell\_v1} mode but it uses the density matrix for the evaluation of the full matrix element. This also allows one to unambiguously calculate the matrix element for processes with identical particles in production and decay.

The new \texttt{madspin} mode follows the FLMW method using  \textsc{MadEvent} for the generation of the decay events, the  spin-density matrix to evaluate the full matrix element and the new reshuffling algorithm.  In this case the weight associated with the events is the same as Eq.~\ref{eq:onshell_reweighting} but includes the Jacobian associated with the reshuffling
\begin{equation}
\frac{|\mfull|^2}{|\mprod|^2|\mdec|^2}w_{\rm shuffle}.
\end{equation}

Additionally, a new mode \texttt{PA} (pole approximation) is introduced, which corresponds to the MPA. This mode is technically the \texttt{onshell} mode followed by the same reshuffling as the one done in the new \texttt{madspin} mode.

The three new modes also provide a new feature, where additional event weights can be added to the output LHE file when decaying an \emph{unpolarised} sample; each of these weights corresponds to the weight the event would carry assuming a given polarisation state for the decaying resonances. Note that this differs from generating a polarised sample (via the \mgamc\ syntax of \cite{BuarqueFranzosi:2019boy}): a single unpolarised sample provides the partial weight of every polarisation state, reducing the number of samples needed to study polarisation effects (see App.~\ref{subsec:polarised_syntax} for technical details). This feature is denoted by ``Helicity decomp.'' in Table~\ref{tab:madspin_modes_combined}.

A summary of the available run modes in \newms\ and the associated features and approximations is given in Table~\ref{tab:madspin_modes_combined} and the full list of options associated with each implementation is reported in Appendix~\ref{seq:commands}.

\begin{landscape} \begin{table}[p] \centering \footnotesize \setlength{\tabcolsep}{3pt} \renewcommand{\arraystretch}{1.15} \caption{Comparison of unstable-particle approximations and corresponding run modes in \ms\ and \newms.} \label{tab:madspin_modes_combined} \begin{tabular}{lccccc cccccc} \toprule & \multicolumn{5}{c}{\textbf{Theory approximation}} & \multicolumn{6}{c}{\textbf{\ms\ implementation / capabilities}} \\ \cmidrule(lr){2-6} \cmidrule(lr){7-12} \textbf{Approximation} & \textbf{Non-res.} & \textbf{Gauge} & \textbf{Spin} & \textbf{Off-shell} & \textbf{No reshuf.} & \textbf{\ms} & \textbf{\newms} & \textbf{Ident.} & \textbf{3-body} & \textbf{Loop-ind.} & \textbf{Helicity} \\ & \textbf{diagrams} & \textbf{inv.} & \textbf{corr.} & \textbf{effects} & \textbf{dependence} & \textbf{spinmode} & \textbf{spinmode} & \textbf{particles} & \textbf{decays} & \textbf{/interf.} & \textbf{decomp.} \\ \midrule Full ME (CMS) & \(\checkmark\) & \(\checkmark\) & \(\checkmark\) & \(\checkmark\) & \(\checkmark\) & \multicolumn{2}{c}{only possible in \mgamc} & exact & \(\checkmark\) & \(\checkmark\) & LO only \\ \midrule \multicolumn{12}{c}{\ms\ (v1)}\\ \midrule NSA (NWA) & \(\times\) & \(\checkmark\) & \(\times\) & \(\times\) & \(\checkmark\) & \texttt{none} & \texttt{none} & exact & \(\checkmark\) & \(\checkmark\) & \(\times\) \\ DCA (NWA) & \(\times\) & \(\times\) & \(\checkmark\) & \(\times\) & \(\checkmark\) & \texttt{onshell} & \texttt{onshell\_v1} & average & \(\checkmark\) & \(\times\) & \(\times\) \\ FLMW & \(\times\) & \(\times\) & \(\checkmark\) & \(\checkmark\) (ME rew.) & \(\checkmark/\times\) & \texttt{madspin} & \texttt{madspin\_v1} & exact & \(\times\) & \(\times\) & \(\times\) \\ \midrule \multicolumn{12}{c}{\newms}\\ \midrule DCA (NWA density) & \(\times\) & \(\times\) & \(\checkmark\) & \(\times\) & \(\checkmark\) & -- & \texttt{onshell} & exact & \(\checkmark\) & \(\checkmark\) & \(\checkmark\) \\ FLMW (density) & \(\times\) & \(\times\) & \(\checkmark\) & \(\checkmark\) (ME rew.) & \(\checkmark/\times\) & -- & \texttt{madspin} & exact & \(\checkmark\) & \(\checkmark\) & \(\checkmark\) \\ MPA & \(\times\) & \(\checkmark\) & \(\checkmark\) & \(\checkmark\) (PA + BW smear.) & \(\times\) & -- & \texttt{PA} & exact & \(\checkmark\) & \(\checkmark\) & \(\checkmark\) \\ \bottomrule \end{tabular} \vspace{1mm} 
\begin{minipage}{0.96\linewidth} \textbf{Notes:} 
``Non-res.'' denotes non-resonant diagrams. ``Gauge inv.'' indicates whether the approximation is gauge invariant. ``No reshuf. depend.'' indicates the absence of a dependence on the momentum reshuffling or pole-projection prescription. ``Ident. particles'' refers to the treatment of identical particles in the production and decay. ``Helicity decomp.'' specifies whether the method supports helicity decomposition at LO/NLO. ``ME rew.'' refers to the reweighting procedure explained in Sec.~\ref{subsec:FLMW}, while ``PA+BW smear.'' refers to the pole approximation followed by a Breit-Wigner smearing as explained in Sec.~\ref{subsec:offshell}. The label ``Full ME (CMS)'' indicates the full matrix element with off-shell currents within the complex-mass-scheme framework \cite{Denner:1999gp,Denner:2005fg}. \end{minipage} \end{table} \end{landscape}

%% file: validation.tex
\section{Validation}
\label{sec:validation}

A validation of the \newms\ run modes is performed for each new feature that is introduced in the new code: the calculation of the matrix elements via the spin-density matrix (Sec.~\ref{subsec:valid_onshell}), the handling of identical particles in production and decay (Sec.~\ref{subsec:valid_iden}), the polarisation-aware decay (Sec.~\ref{subsec:validation-polarised}) and the loop-induced processes (Sec.~\ref{subsec:valid_ms_loop_induced_density}). The validation of pure amplitude interference terms is presented later, together with their phenomenological applications, in Sec.~\ref{subsec:valid_interf}.
Unless otherwise specified, all the samples used in this section are generated for $pp$ collisions at $\sqrt{s}=13\,$TeV using the NNPDF2.3LO PDF set ($\alpha_s(M_Z)=0.13$) \cite{Ball:2012cx,Ball:2013hta}, and the renormalisation and factorisation scales are set event by event via the Catani--Krauss--Kuhn--Webber (CKKW) back-clustering procedure \cite{Catani:2001cc}, the default in \mgamc.

\subsection{On-shell processes}\label{subsec:valid_onshell}

The old \texttt{onshell\_v1} and the new \texttt{onshell} implementations differ only in the way the matrix elements are evaluated: directly in the former, and via the convolution of the production and decay spin-density matrices in the latter. The two implementations are therefore expected to yield identical results.

These two modes were compared for a wide array of processes, covering all of the functionalities that are provided with the \mgamc\ framework. This included processes with LO or NLO matrix elements with up to four decaying resonances of spin 0, 1/2 and 1. Both coloured and colourless resonances were tested, produced in QCD or weak processes, as well as new physics models. Configurations in which all decaying resonances have the same spin were tested, as well as configurations with resonances of different spins, for which the helicity indices associated with the different particles have different dimensions. The validation also included (nested) decay chains (e.g. \texttt{decay t > w+ b , (w+ > e+ ve)}), 3-body decays and processes with multi-particle labels in the production and the decay (e.g. \texttt{decay t > l+ vl b}).

\begin{figure}[htp]
    \centering
    \includegraphics[width=0.5\textwidth]{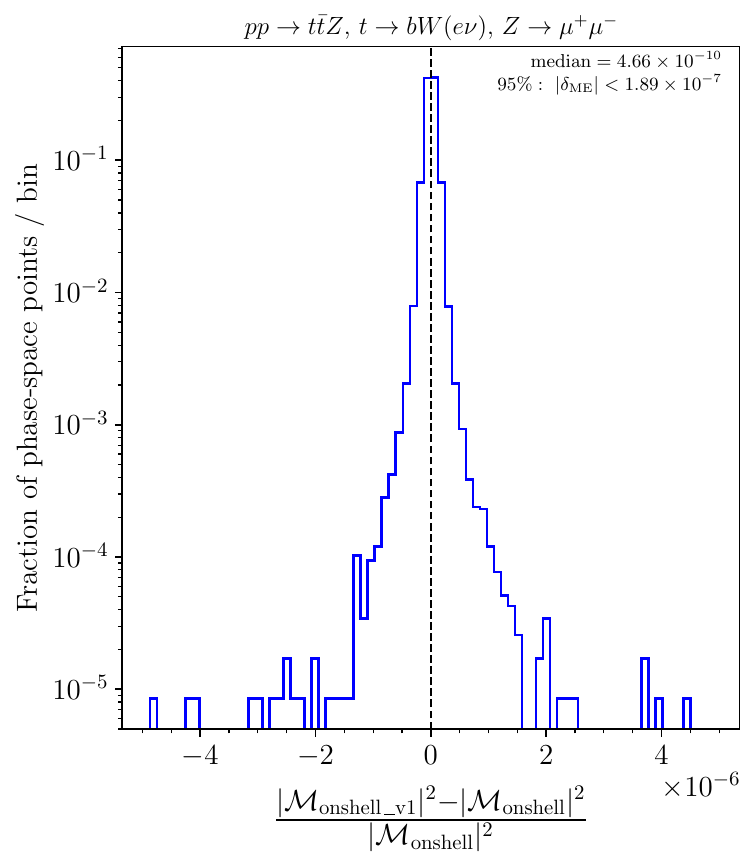}
    \caption{Relative difference between squared amplitudes for the $pp\to t\bar{t}Z$, $t\to W(e\nu)b$, $Z\to \mu^+\mu^-$ process evaluated with the \texttt{onshell} and \texttt{onshell\_v1} modes. The comparison is based on $100\,000$ events.}
    \label{fig:valid_ttz}
\end{figure}

The matrix elements of the two calculations, compared point-by-point in phase space, were found to agree to an accuracy of at least $\mathcal{O}(10^{-6})$. An example plot showing the validation results for $pp\to t\bar{t}Z$ followed by a fully leptonic $t\bar{t}$ decay into electrons and positrons and a $Z$ decay to muons is shown in Fig.~\ref{fig:valid_ttz}. \footnote{ The precision is limited to single precision since we have limited the momenta to such precision.}

The new code provides the functionality to compare the ME calculated via the convolution of the production and decay spin-density matrices to the one that corresponds to the full process, for debugging purposes. This functionality is turned off by default, but can be turned on using the switch \texttt{density\_debug} (See \cref{sub:manual}).

\subsection{Processes with identical particles in production and decay}\label{subsec:valid_iden}

As explained in Sec.~\ref{sec:v1}, in the \texttt{onshell\_v1} mode the full matrix element of the production and decay event has to be evaluated. In this calculation there is no knowledge of which particles come from the decay of a resonance. When identical particles appear both in the production event and in the decay of an unstable resonance, the calculation of the full matrix element therefore becomes ambiguous.

To illustrate this, let us consider the process $pp \to dZ$, where $p$ denotes a quark or a gluon. If the $Z$ boson then decays via $Z\to d\bar{d}$, the full matrix element to be evaluated corresponds to $\mathcal{M}_{\text{full}}(pp\to dd\bar{d})$. Since the $d$ quark appears more than once in the full event, the following permutations are possible
\begin{equation}\label{eq:ME1}
    \mathcal{M}_{\text{full,1}}=\mathcal{M}_{\text{full}}(pp\to d_{\text{prod}} d_{\text{dec}}\bar{d}_{\text{dec}}) 
\end{equation}
and  
\begin{equation}\label{eq:ME2}
    \mathcal{M}_{\text{full,2}}=\mathcal{M}_{\text{full}}(pp\to d_{\text{dec}} d_{\text{prod}}\bar{d}_{\text{dec}}),
\end{equation}
where the last two particles should correspond to the particles assigned to the $Z$ resonance and the subscripts `prod' and `dec' indicate a particle that is present in the production or decay process respectively. Here $\mathcal{M}_{\text{full,1}}$ corresponds to the correct assignment and $\mathcal{M}_{\text{full,2}}$ to the wrong assignment, where the $d_{\rm dec}$ quark from the $Z$ decay is incorrectly assigned to the production event.

Since the \texttt{onshell\_v1} mode is unable to determine which of $\mathcal{M}_{\text{full,1}}$ or $\mathcal{M}_{\text{full,2}}$ should be evaluated, the code provides four options: calculating the matrix element from the first assignment, averaging the squared matrix elements calculated from all possible assignments, taking the maximum of these, or crashing (see App.~\ref{sub:manual}). In the first case, the evaluated ME depends on the ordering of the particles in the generation command. More specifically if one uses
\begin{lstlisting}[style=scipostcode,language=Python,caption={},label={}]
    generate p p > d z
\end{lstlisting}
then $\mathcal{M}_{\text{full}}(pp\to dd\bar{d})=\mathcal{M}_{\text{full,1}}$, yielding the correct result, while if one generates the process with the command
\begin{lstlisting}[style=scipostcode,language=Python,caption={},label={}]
    generate p p > z d
\end{lstlisting}
then $\mathcal{M}_{\text{full}}(pp\to dd\bar{d})=\mathcal{M}_{\text{full,2}}$ yielding a wrong result. In order to avoid yielding a systematically wrong result and to ensure that the evaluated matrix elements are independent of the ordering of the particles in the generation command, the default mode (\texttt{average}) corresponds to taking an average of the squared ME, i.e. $|\mathcal{M}_{\text{full}}(pp\to dd\bar{d})|^2=\frac{|\mathcal{M}_{\text{full,1}}|^2+|\mathcal{M}_{\text{full,2}}|^2}{2}$. The motivation behind this choice is that, in the NWA, the correctly assigned (resonant) amplitude dominates the average, so that the contribution of the wrong assignments is suppressed.\footnote{Note that the factor of two does not matter for the accept/reject method used here.}

In the calculation of the matrix elements via the spin-density matrix, the production and decay ME are evaluated separately, so no assignment ambiguity arises: the ME are always correctly evaluated, independently of the ordering of the particles in the generation command.

\begin{figure}[htp]
    \centering
    \includegraphics[width=0.5\textwidth]{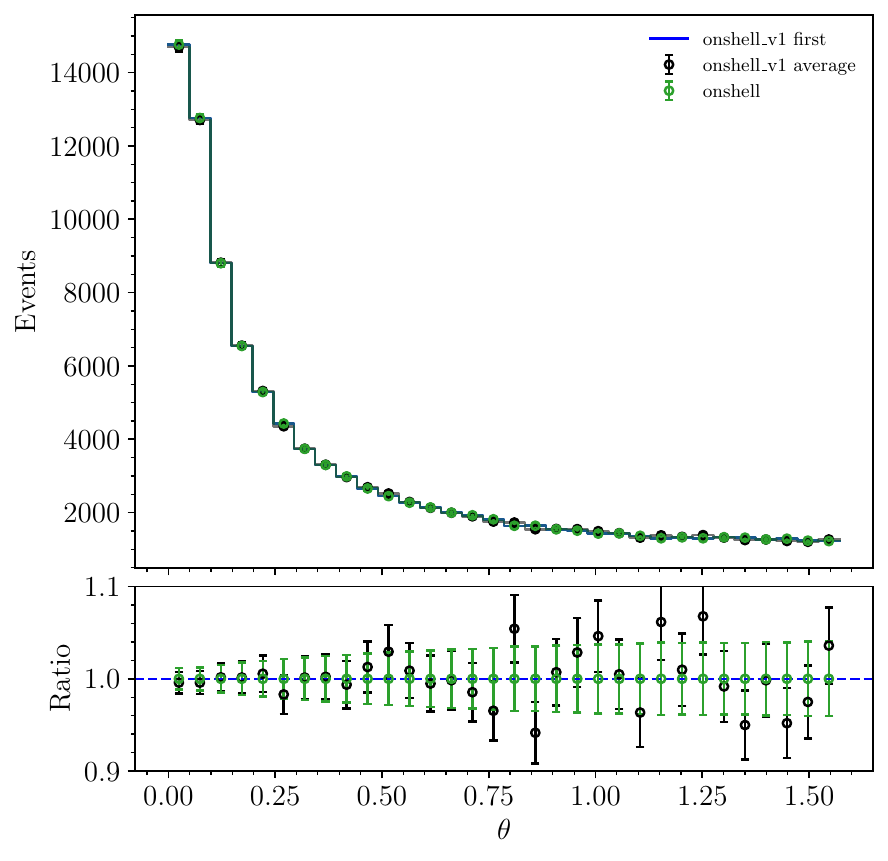}
    \caption{Polar angle of the $d$ quark in the $pp\to dZ$ process, with $Z\to d\bar{d}$. The \texttt{onshell} run mode is compared to the \texttt{onshell\_v1} calculation, with two different run options: either taking the first assignment of the final-state particles to the decaying resonance (which is by design the correct ordering for this process and therefore matches the density-matrix computation) or averaging the ME from all possible assignments for the $d$ quark (the previous fail-safe). The density-matrix result and the \texttt{first} option are in exact agreement, as expected, while the \texttt{average} option agrees with them within the statistical uncertainty.}
    \label{fig:valid_ident}
\end{figure}

A comparison of the aforementioned modes for the $pp\to dZ(d\bar{d})$ process is shown in Fig.~\ref{fig:valid_ident}. 

\subsection{Decay of polarised resonances and helicity interference}
\label{subsec:validation-polarised}

Since the production spin-density matrix is
contracted against the decay density matrices (Eq.~\ref{eq:densityMaster}), \newms\ can be
instructed to restrict that contraction to a chosen subspace of the
helicity indices.  For two spin-$1/2$ particles each index runs over
$\{-,+\}$, so the contraction carries $4\times4=16$ entries.  Thus the spin-density matrix can be decomposed
into nine blocks that are closed under transposition, and therefore
individually real: each particle contributes either one of the two
diagonal projectors $D^{\pm}=\{\pm,\pm\}$, or the off-diagonal pair
$I=\{(+,-),(-,+)\}$.  The four blocks $(D^{\pm},D^{\pm})$ are the
familiar polarised cross sections; the remaining five contain at least
one $I$ index and carry no cross section of their own, integrating to
zero by construction and corresponding to helicity interference \cite{Basu:2026zmz}.  The card syntax used to select an individual
block is given in Appendix~\ref{subsec:polarised_syntax}.

In this subsection, we test that we can calculate each block independently and that their sum reproduces exactly the unpolarised prediction. For this we use $pp\to t\bar t$ at $\sqrt{s}=13\,$TeV with both tops decayed leptonically in the \texttt{madspin} mode. The
charged leptons are maximal spin analysers, which is what makes the
decay distributions sensitive to the production density matrix.
Spin observables are defined in the helicity basis $(\hat k,\hat n,\hat
r)$ in the parent rest frame, with $\hat k$ the top direction in the
$t\bar t$ frame, $\hat n$ the direction normal to the production plane and $\hat
r=\hat n\times\hat k$~\cite{Bernreuther:2015yna}.

The three observables used in this validation are:
\begin{description}
  \item[$\cos\theta^{k}_{\ell^+}$] the cosine of the angle of the $\ell^+$ with respect
    to $\hat k$, measured in the rest frame of its parent top.  This is a
    single-particle polarisation observable: it probes the diagonal of
    the density matrix for one top and is blind to correlations between
    the two tops.
  \item[$\cos\theta^{n}_{\ell^+}\cos\theta^{n}_{\ell^-}$] the product of the
    cosines of the angles of the $\ell^\pm$ with respect to $\hat n$, each measured in the
    rest frame of its parent top.  The \emph{mean} of this distribution gives
    the transverse--transverse spin correlation coefficient,
    $C_{nn}=9\,\langle\cos\theta^{n}_{\ell^+}\cos\theta^{n}_{\ell^-}\rangle$.
    That coefficient is a pure double
    interference term \cite{Bernreuther:2015yna},
    \begin{equation}
        C_{nn} = \mathrm{Tr}[\rho\,\sigma_n\otimes\sigma_n] =
    2\,\mathrm{Re}\,\rho(+-,-+) - 2\,\mathrm{Re}\,\rho(++,--).
    \end{equation}
    Both entries flip the helicity of the top \emph{and} of the antitop, and no diagonal entry of $\rho$
    contributes.
  \item[$\Delta\phi(\ell^+,\ell^-)$] the azimuthal separation of the two
    leptons in the laboratory frame.  Unlike the frame-defined
    coefficients above, this is not a pure spin observable, but it inherits
    a substantial off-diagonal contribution and is the observable most
    commonly used experimentally to expose $t\bar t$ spin correlations.
\end{description}

\begin{figure}[htbp]
  \centering
  \begin{tabular}{cc}
    \includegraphics[width=0.47\textwidth]{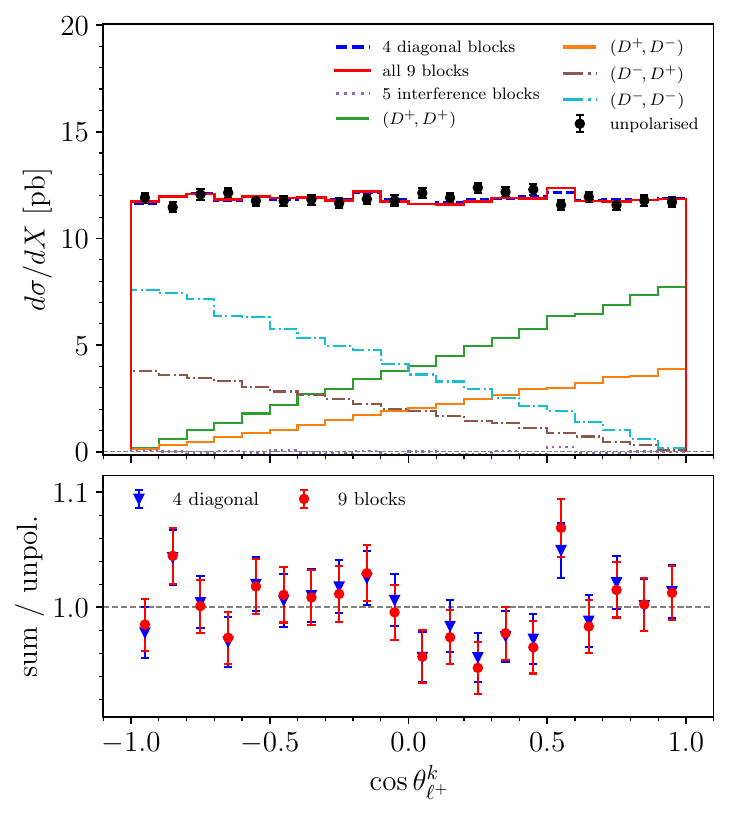} &
    \includegraphics[width=0.47\textwidth]{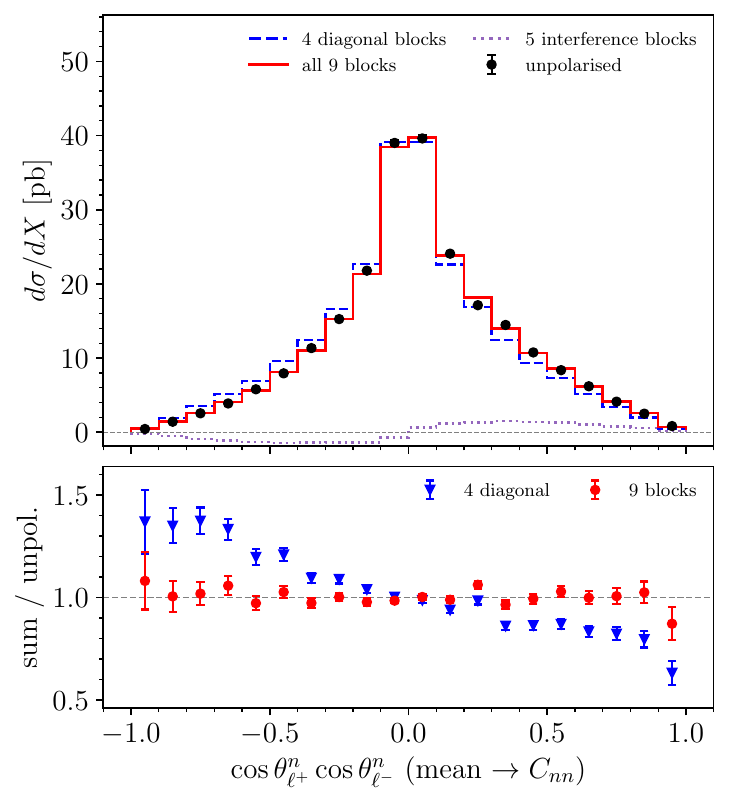} \\[1ex]
    \includegraphics[width=0.47\textwidth]{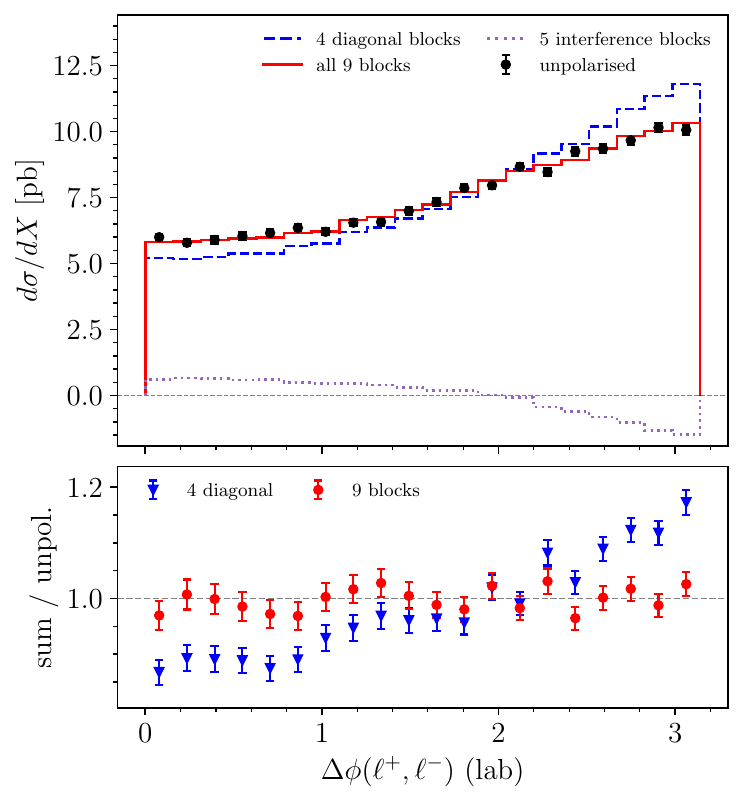} &
    \includegraphics[width=0.47\textwidth]{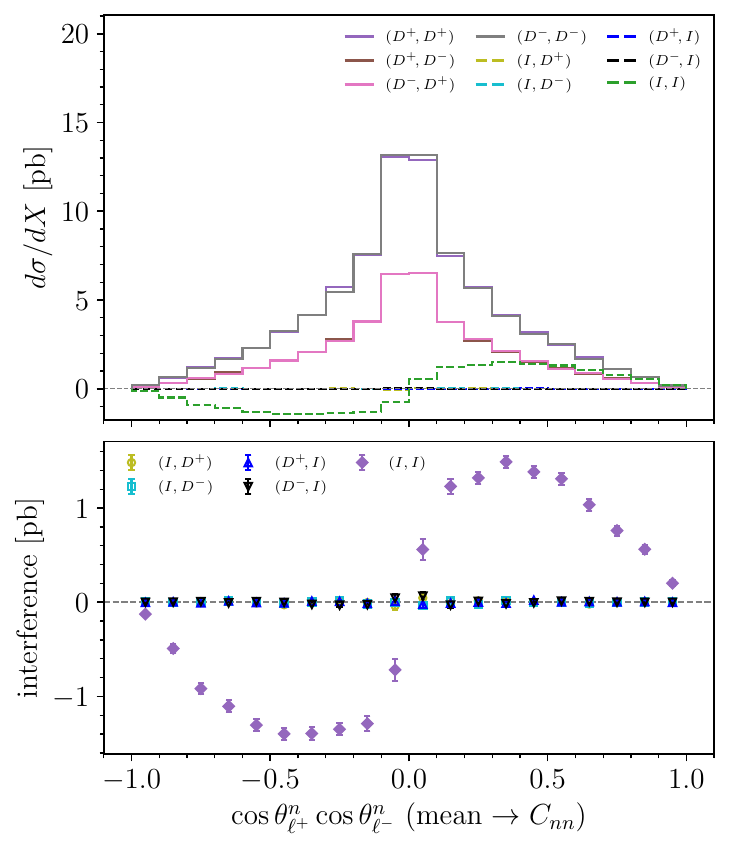} \\
  \end{tabular}
  \caption{Closure of the spin-density block decomposition for
    $pp\to t\bar t$ with both tops decayed leptonically:
    $\cos\theta^{k}_{\ell^+}$ (upper left),  $C_{nn}$ (upper right) and $\Delta\phi(\ell^+,\ell^-)$ (lower left).  In these three panes the lower sub-pane shows
    the ratio to the unpolarised sample for the four-block and the
    nine-block sums.  Lower
    right: the same $C_{nn}$ distribution decomposed into the nine
    individual blocks, showing that the whole effect is carried by the
    $(I,I)$ block while the other interference terms are consistent with zero.}
  \label{fig:density-closure}
\end{figure}

The results are shown in Figure~\ref{fig:density-closure}. 
The first three panes show, for each observable, the
unpolarised reference together with two reconstructions: the sum of the
four diagonal blocks alone, and the sum of all nine.  The lower sub-pane of
each pane gives the ratio to the unpolarised sample.
For
$\cos\theta^{k}_{\ell^+}$ the four diagonal blocks already reproduce the
unpolarised distribution: the interference blocks contribute nothing
that this observable can see, and the four-block and nine-block curves
lie on top of one another.  
For $\cos\theta^{n}_{\ell^+}\cos\theta^{n}_{\ell^-}$ and
$\Delta\phi(\ell^+,\ell^-)$ the four-block sum fails visibly — the
ratio acquires a pronounced slope — and is restored to unity only once
the five helicity interference blocks are added. Control distributions insensitive to spin, such as $p_T(t)$ and
$m(t\bar t)$, close with either reconstruction and are not shown.

The fourth pane of Fig.~\ref{fig:density-closure} decomposes $\cos\theta^{n}_{\ell^+}\cos\theta^{n}_{\ell^-}$
into all nine contributions separately, and it shows where the effect
actually lives.  The four diagonal blocks are the tall curves: they carry
the whole cross section.  What vanishes for them
is not the distribution but its \emph{first moment}, i.e. each of the
curves is symmetric around zero, and the mean is compatible with zero (i.e. $C_{nn}=0$). 
The four mixed blocks with a single $I$ index are zero in the stronger sense,
bin-by-bin.  The entire $C_{nn}$ signal is therefore carried by the
$(I,I)$ block, in which \emph{both} helicity indices are off-diagonal: it
supplies the entire first moment,
$\langle\cos\theta^{n}_{\ell^+}\cos\theta^{n}_{\ell^-}\rangle=+0.03654\pm0.00056$,
corresponding to $C_{nn}=0.329\pm0.005$, in agreement with the NLO SM
prediction of $0.326\pm0.002$~\cite{Bernreuther:2015yna} and with the CMS
measurement of $0.329\pm0.020$~\cite{CMS:2019nrx}.  That is
the expected behaviour — $C_{nn}$ is a genuine two-particle correlation,
so it cannot be built from a block that leaves one particle's index
diagonal.

\subsection{Spin-density matrices for loop-induced processes}
\label{subsec:valid_ms_loop_induced_density}

To support loop-induced processes within \newms, we implemented the evaluation of spin-density matrices from squared one-loop amplitudes, as required for loop-induced leading-order processes. While the technical implementation is quite different from the tree-level case, the underlying idea and the user interface are essentially the same.

We use a special mode of \madloop~\cite{Hirschi:2011pa} (called the optimised output \cite{Hirschi:2015iia}) where the amplitudes are projected on a colour basis, which depends on the process.
The squared matrix element\footnote{Note that the normalisation over spin/colour has been omitted for readability.} is then given by
\begin{equation}\label{eq:ms_loop_squared}
|\mcal|^2 =
\sum_{\lambda,h}\sum_{c,c'}J_c(\lambda,h)\,C_{cc'}\,J_{c'}^*(\lambda,h),
\end{equation}
where $J_c$ corresponds to the projection of the 1-loop amplitude on the $c$ element of the colour basis, $C_{cc'}$ is the scalar product of the elements of the basis (which is not orthogonal)\footnote{\madloop\ performs the same decomposition for the single-pole and double-pole coefficients of a general 1-loop amplitude, but these vanish for loop-induced processes, which provides a nice validity check.}. Here $\lambda$ represents the helicities of the decaying particles, while $h$ represents the helicities of the ``spectator'' particles that do not decay.

Therefore, in the standalone implementation for loop-induced processes, we first evaluate the 1-loop amplitude for every helicity configuration and store its coefficients $J_c(\lambda)$ in the chosen colour basis. The spin-density matrix is then constructed from these stored coefficients by multiplying the amplitudes with helicities $\lambda$ and $\lambda'$ for the particles that decay, while summing over the helicities of all other spectator particles:
\begin{equation}
\rho_{\lambda\lambda'} = \sum_h\sum_{c,c'}J_c(\lambda,h)\,C_{cc'}\,J_{c'}^*(\lambda',h).
\label{eq:ms_rho_loop}
\end{equation}

Comparing Eqs.~\ref{eq:ms_loop_squared} and \ref{eq:ms_rho_loop}, it is obvious that for the construction of the spin-density matrix, every helicity configuration has to be evaluated explicitly and therefore the
\madloop\ helicity filter, which normally identifies vanishing helicity configurations and configurations that are identical up to a phase due to C-symmetries, cannot be used\footnote{\newms\ therefore sets \texttt{HelicityFilterLevel} to zero.}.

We use two types of numerical checks to validate the construction of the density matrix for loop-induced processes. First of all, the trace of the density matrix must be equal to the helicity-summed matrix element. The second test is to reproduce the convolution relation of Eq.~\ref{eq:densityMaster}.

An example process which is used for the first test is vector-boson pair production through the gluon-initiated channel \cite{Glover:1988fe}. The syntax is the same as the one explained in more detail in \cite{Durupt:2025wuk}, i.e. for the standalone mode it is
\begin{lstlisting}[style=scipostcode,language=Python,caption={},label={}]
generate g g > w+ w- [sqrvirt=QCD]
output standalone --density=3,4
\end{lstlisting}
with the density matrix carried by the particles $3$ and $4$, i.e. the two $W$ bosons (which have nine helicity states, 45 independent complex entries).

A given phase-space point gives
\begin{align}
\mathrm{Tr}\,\rho &= 3.10537552440078 \times 10^{-4}\, , \\
|\mcal|^2_{\mgamc}    &= 3.10537552440141 \times 10^{-4}\, ,
\end{align}
a relative agreement of $2 \times 10^{-13}$, which reaches the same accuracy estimate as \madloop\ ($6 \times 10^{-13}$). The same test was repeated for multiple phase-space points and for other Standard-Model processes, such as \code{p p > h j}.

For the second test, we need a process which proceeds via the exchange of a resonance in the $s$-channel. One such process is $g g \to (Z^{*} \to e^{+} e^{-})\, g$ where the $Z$ boson is off-shell. This is generated via
\begin{equation}
\left\{ \begin{aligned} 
  &\textrm{production}: g\, g \to Z^{*} \, g,\\
  &\textrm{decay}: Z^{*} \to e^+ \, e^-,
\end{aligned} \right.
\end{equation}
where the density matrix computed is the one of the $Z$ boson alone (3 helicity states). For a given phase-space point, we compare the full computation of $g \, g \to Z g\, \to e^+\, e^- \, g$ to the result of the convolution of Eq.~\ref{eq:densityMaster}:

\begin{align}
|\mcal|^2_{\text{convolution}} &= 2.9443978783 \times 10^{-6}~\mathrm{GeV}^{-2}\, ,\\
|\mcal|^2_{\mgamc}             &= 2.9443978772 \times 10^{-6}~\mathrm{GeV}^{-2}\, ,
\end{align}
which results in a relative agreement of $3 \times 10^{-10}$.

%% file: applications.tex
\section{New phenomenological applications}\label{sec:applications}

In this section we present a broad range of applications, starting with the study of the top lineshape (Sec.~\ref{subsec:valid_offshell}). We then move to applications that were not possible with \ms: the generation of polarised resonances at NLO including off-shell effects (Sec.~\ref{subsec:polarised}), the study of spin correlations in loop-induced processes (Sec.~\ref{subsec:loop_induced}) and the study of spin correlations for non-positive-definite cross-sections, looking here at the impact of the chromomagnetic operator (Sec.~\ref{subsec:valid_interf}).
Unless otherwise specified, we use the same setup as in the previous section.

\subsection{Off-shell effects: top lineshape}\label{subsec:valid_offshell}

\begin{figure}[htp]
    \centering
    \includegraphics[width=0.7\textwidth]{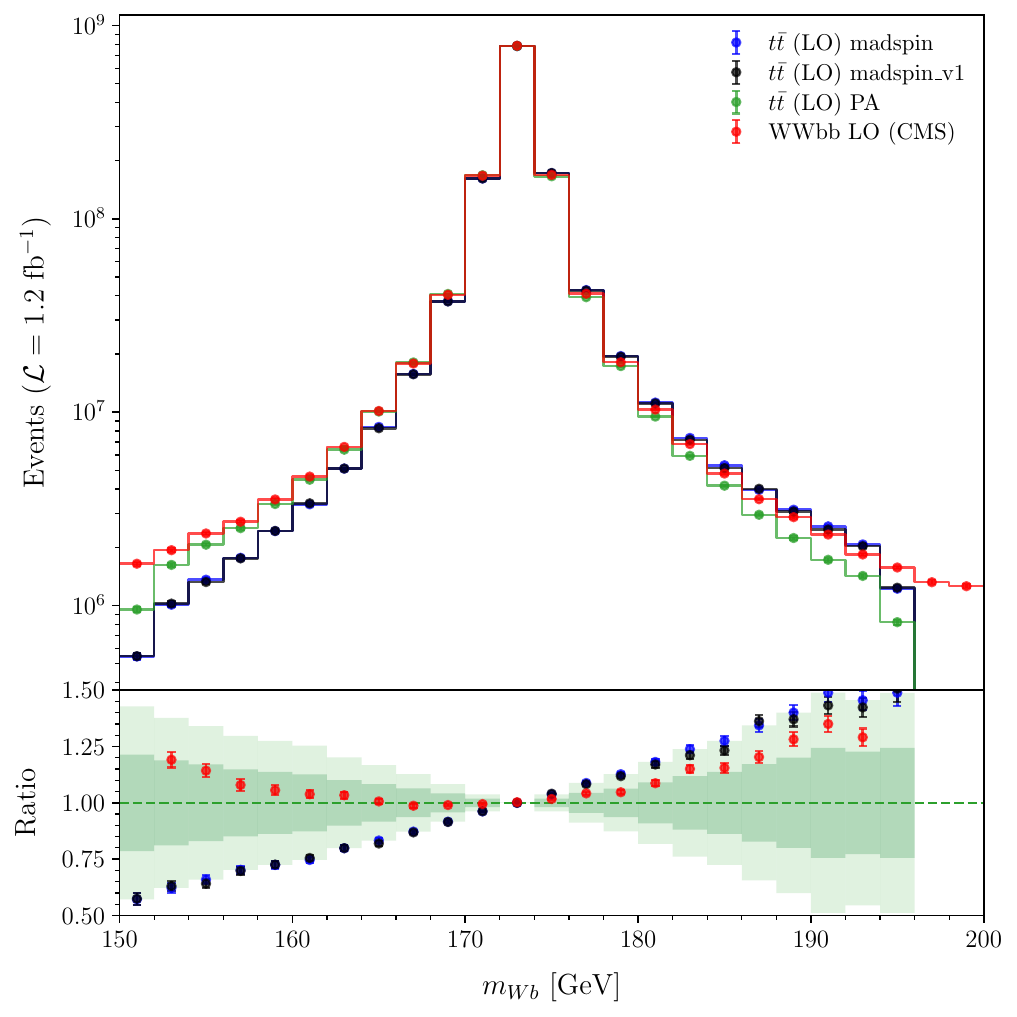}
    \caption{The invariant mass of the top-quark decay products simulated with different off-shell modes of \textsc{MadSpin2} compared to the full $pp\to W^+W^-b\bar{b}$ calculation. The light green band corresponds to the difference between the pole approximation and the FLMW approach (\texttt{madspin} mode) and the dark green band to half of this difference. Both top and anti-top quarks are included in the distribution.} 
    \label{fig:offshell}
\end{figure}

In order to illustrate the differences between the FLMW (\code{spinmode=madspin}) and MPA (\code{spinmode=PA}) modes in the modelling of the off-shell effects, we study the lineshape of the top quark. 

An LO $pp\to t\bar{t}$ sample is generated with \mgamc\ and the top quarks are decayed in \newms\ with both the FLMW and MPA approximations. These are compared to an LO $pp\to W^+W^-b\bar{b}$ sample, which includes an exact description of the off-shell effects~\cite{Denner:2010jp,Bevilacqua:2010qb}.

The samples are generated with $m_{\rm top}=173$ GeV and $\Gamma_{\rm top}=1.4915$ GeV. Since \mgamc\ by default limits the virtuality of the resonances to $|m-m_{\rm pole}|<15\Gamma$, the generated samples have kinematic cutoffs at 151 and 195 GeV, while the $W^+W^-b\bar{b}$ sample does not have any such restriction.\footnote{We generated the $W^+W^-b\bar{b}$ sample both in the complex-mass scheme and without it, and none of the plots of this section show a significant difference between the two schemes.}

It should be noted that the $W^+W^-b\bar{b}$ calculation also includes single-resonant and non-resonant contributions as well as interference effects, which are absent in the $pp\to t\bar{t}$ sample. Away from the pole these effects are expected to dominate and therefore both FLMW and MPA, which rely on the narrow-width approximation, will yield incorrect results. This is clearly illustrated in \cref{fig:offshell}.

As shown in \cref{fig:offshell}, all calculations agree close to the top pole mass and deviate away from the pole. Below the top mass, the MPA is found to be closer to the $W^+W^-b\bar{b}$ calculation than the FLMW one, while the opposite holds above the top mass. 
One can use 
the difference between the FLMW and MPA approximations to estimate a ``lineshape uncertainty'', i.e. an uncertainty due to the modelling of off-shell effects, not too far from the resonance pole.

\begin{figure}
     \centering
     \begin{subfigure}[b]{0.44\textwidth}
         \centering
         \includegraphics[width=\textwidth]{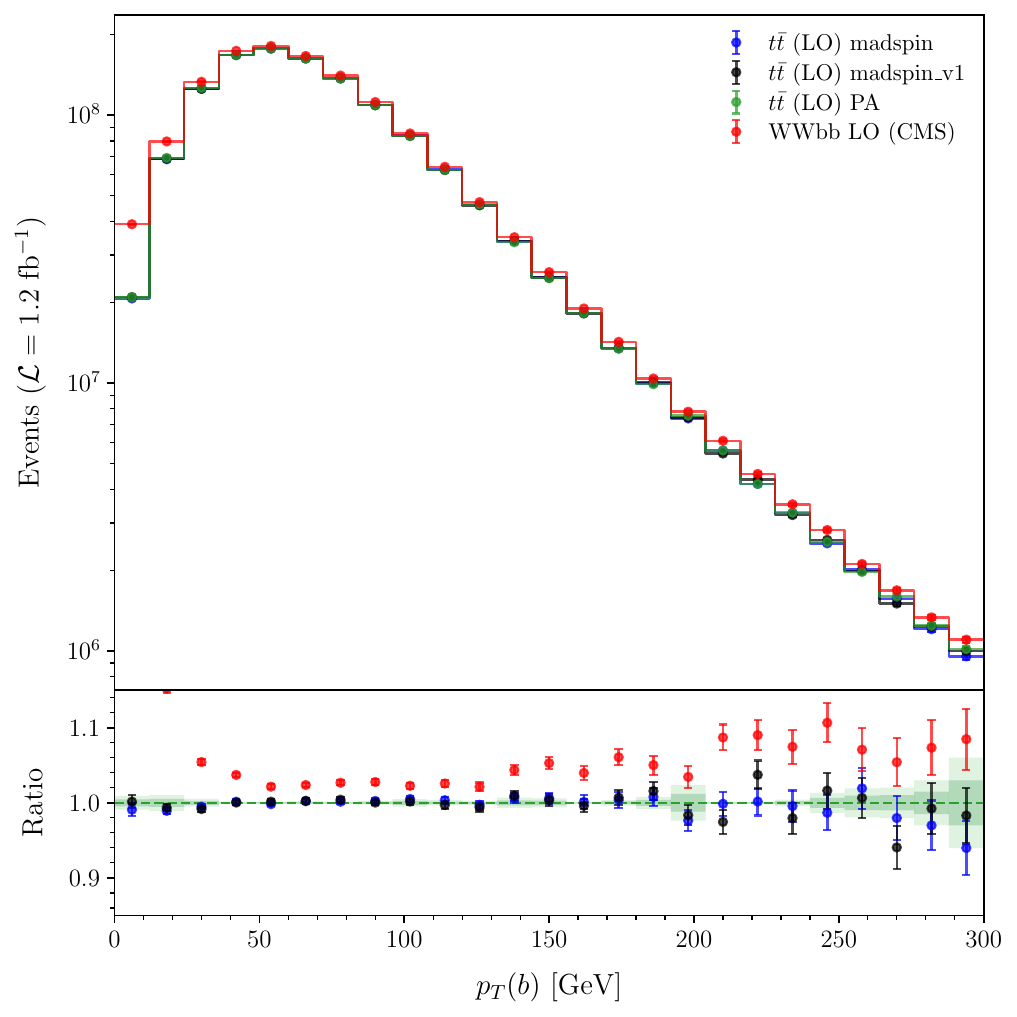}
         \caption{$p_T(b)$ inclusive}
         \label{fig:ptb_inclusive}
     \end{subfigure}
     \hfill
     \begin{subfigure}[b]{0.44\textwidth}
         \centering
         \includegraphics[width=\textwidth]{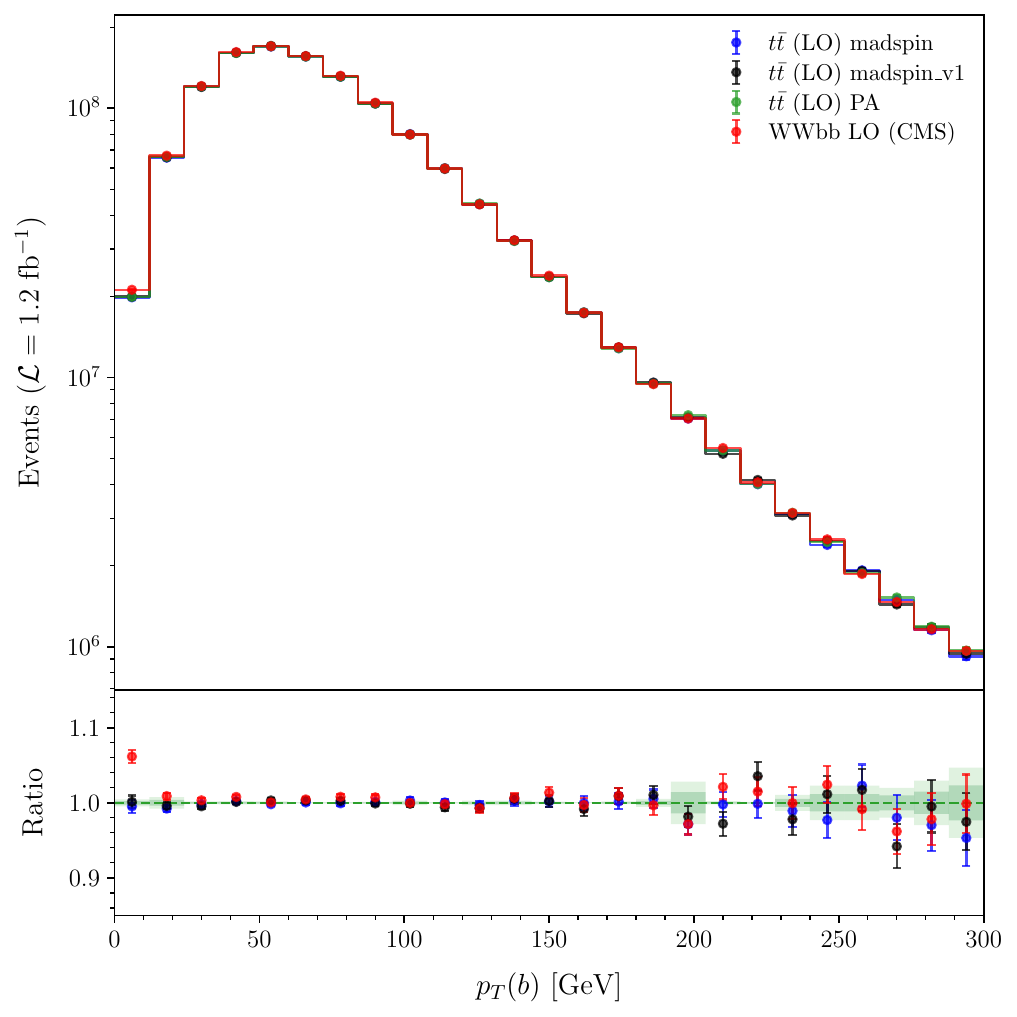}
         \caption{$p_T(b)$ on-shell}
         \label{fig:ptb_onshell}
     \end{subfigure}
     \hfill
     \begin{subfigure}[b]{0.44\textwidth}
         \centering
         \includegraphics[width=\textwidth]{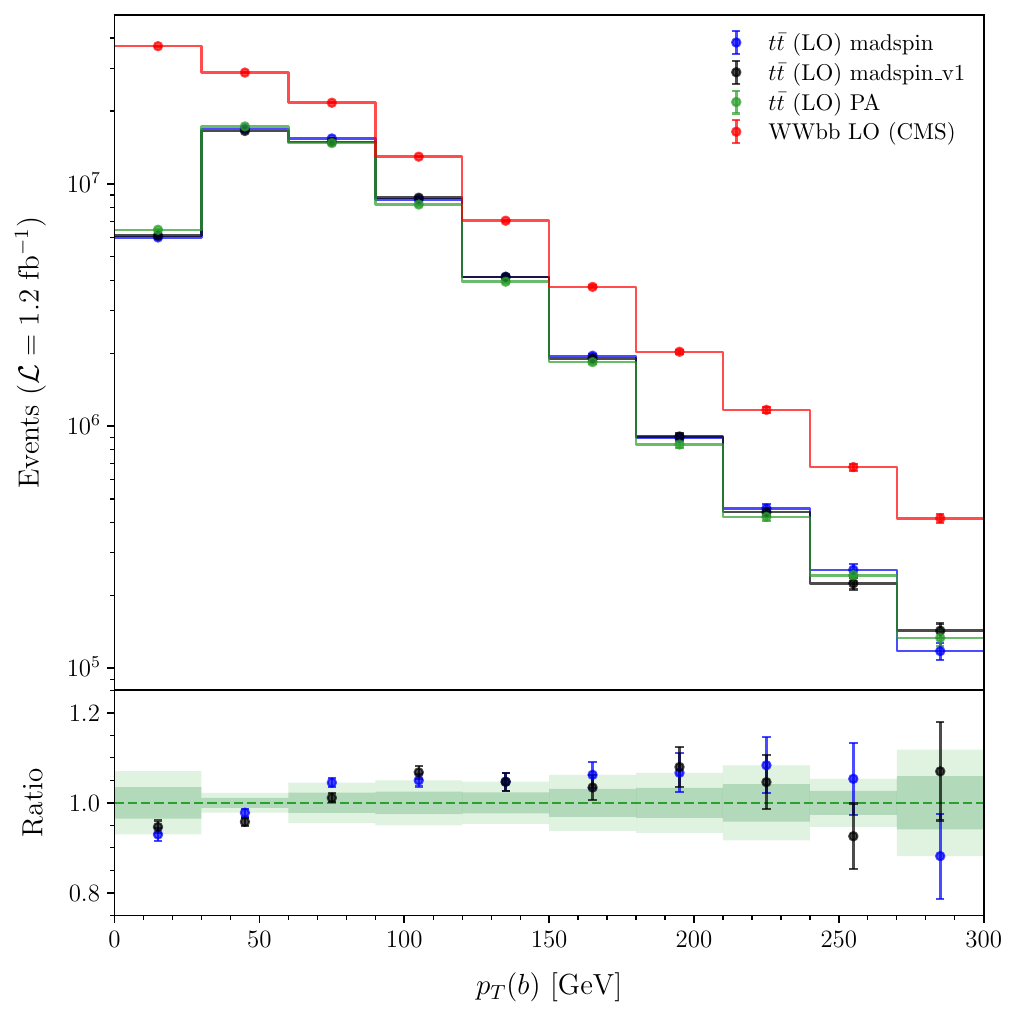}
         \caption{$p_T(b)$  off-shell}
         \label{fig:ptb_offshell}
     \end{subfigure}\hfill
     \begin{subfigure}[b]{0.44\textwidth}
         \centering
         \includegraphics[width=\textwidth]{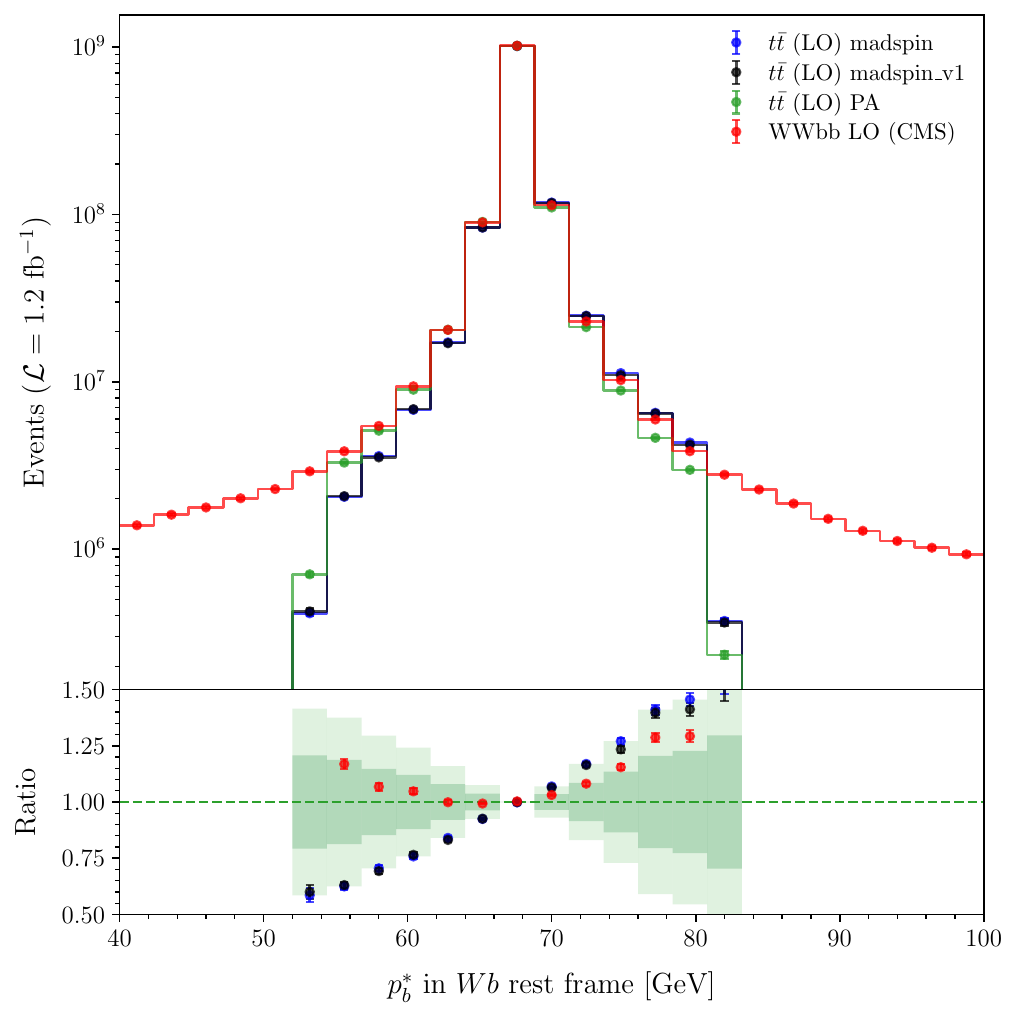}
         \caption{$p_b^*$ inclusive}
         \label{fig:pbstar_inclusive}
     \end{subfigure}
     \hfill
     \begin{subfigure}[b]{0.44\textwidth}
         \centering
         \includegraphics[width=\textwidth]{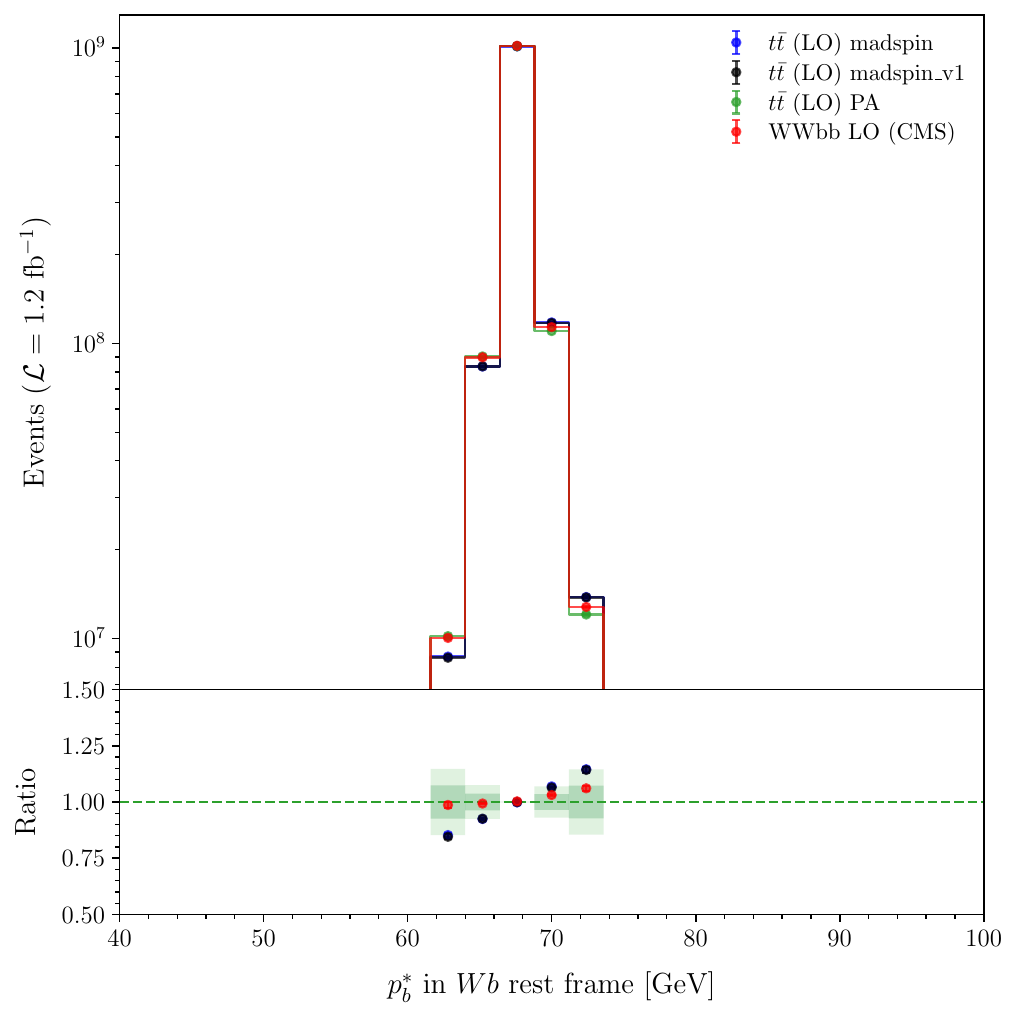}
         \caption{$p_b^*$ on-shell}
         \label{fig:pbstar_onshell}
     \end{subfigure}
     \hfill
     \begin{subfigure}[b]{0.44\textwidth}
         \centering
         \includegraphics[width=\textwidth]{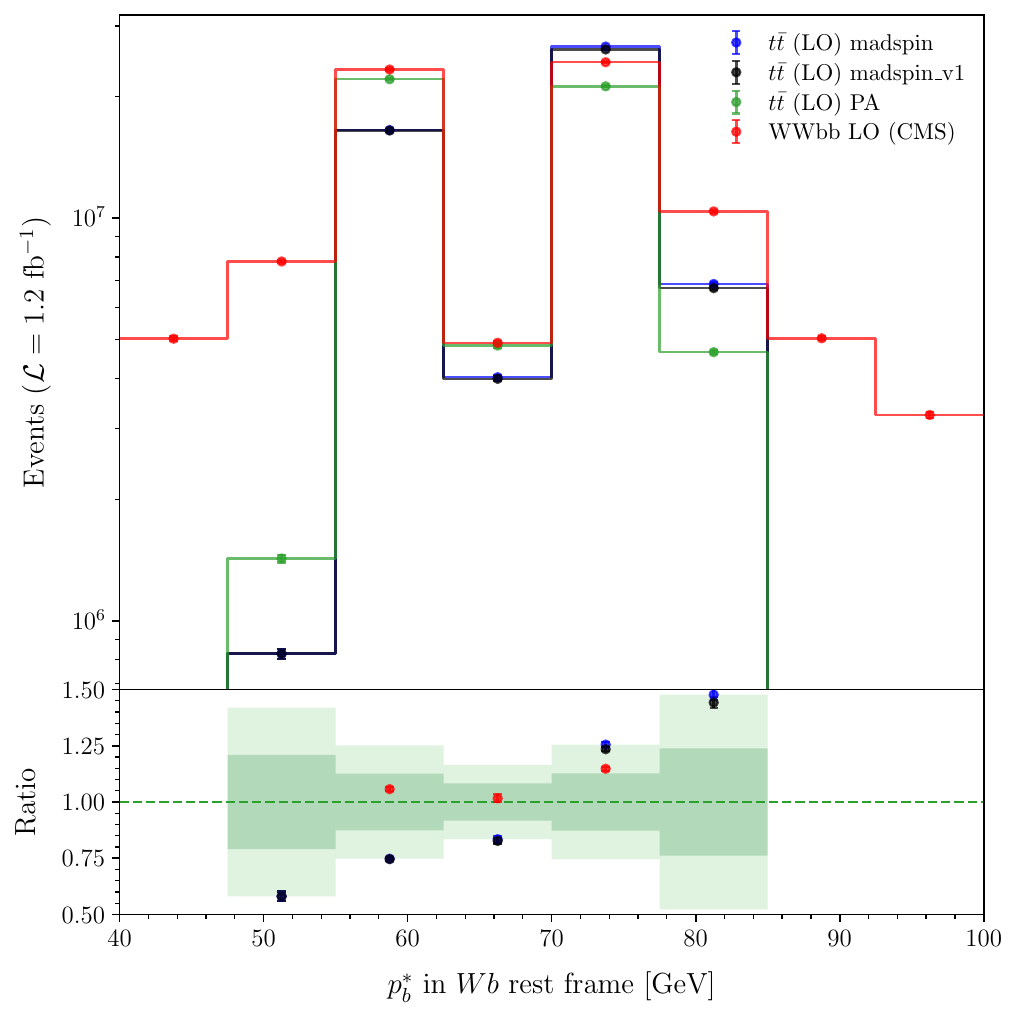}
         \caption{$p_b^*$ off-shell}
         \label{fig:pbstar_offshell}
     \end{subfigure}     
        \caption{Differential distributions of the full $W^+W^-b\bar{b}$ calculation and the FLMW and MPA approximations for different observables and phase spaces. The on-shell (off-shell)  distributions are filled with top and anti-top candidates with $|m_{Wb}-m_{\rm pole}|$ smaller (larger) than 7 GeV.}
\end{figure}

Finally, we note that the largest differences between the FLMW and MPA approximations are visible in quantities that directly probe the mass of the intermediate top. This is clear in the reconstructed $m(Wb)$ distribution, and also in the momentum of the $b$-quark in the $Wb$ rest frame, $p_b^*$ (\cref{fig:pbstar_inclusive,fig:pbstar_onshell,fig:pbstar_offshell}). Lab-frame observables, such as $p_T(b)$ (\cref{fig:ptb_inclusive,fig:ptb_onshell,fig:ptb_offshell}), average over the top mass distribution and depend more on the top-quark boost and recoil, and are therefore much less affected by changes in the modelling of the top virtuality. As a result, differences in the top lineshape only have a limited impact on more inclusive kinematic observables.

\subsection{Polarised $ZZ$ production at NLO including off-shell effects}\label{subsec:polarised}

Studying specific polarisation modes of electroweak bosons is particularly important to have a better understanding of electroweak symmetry breaking \cite{Ballestrero:2017bxn,Denner:2021csi}. Indeed, in the Standard Model, the electroweak bosons acquire a mass and a longitudinal polarisation through this mechanism. This means that the longitudinal polarisation of the $Z$ boson (and its mass) is an observable that we can use to detect deviations from the SM and thus find hints of BSM models, and has recently become experimentally accessible \cite{ATLAS:2023zrv}.
Dedicated polarised predictions for di-boson production, based on the double-pole approximation \cite{Denner:2000bj,Denner:2019vbn}, have been obtained at NLO and NNLO accuracy with several codes, such as \textsc{MoCaNLO} \cite{Denner:2020bcz,Denner:2020eck,Denner:2021csi}, \textsc{Stripper} \cite{Czakon:2010td,Czakon:2014oma,Czakon:2019tmo,Poncelet:2021jmj,Pellen:2021vpi,Pellen:2022fom}, \textsc{MulBos} \cite{Le:2022lrp,Le:2022ppa,Dao:2023kwc,Dao:2024ffg} and \textsc{BBMC} \cite{Denner:2022riz,Denner:2023ehn,Denner:2024tlu}; a systematic comparison of polarised $ZZ$ predictions, including \mgamc, can be found in \cite{Carrivale:2025mjy}.

While \mgamc\ does not allow one to generate polarised samples at NLO accuracy\footnote{This feature will be present in \textsc{MadGraph7}.}, \newms\ allows one to obtain such quantities via a reweighting mechanism (see App.~\ref{subsec:polarised_syntax}). To illustrate this new capability, we study polarised $ZZ$ production at the LHC at NLO precision, specifically the process
\begin{equation}
p \, p \to Z \, Z \to e^+ \, e^- \, \mu^+ \mu^-,
\end{equation}
where the LO contribution comes from the $q \bar q$ channel, while the $q g$ and $\bar q g$ channels are part of the NLO corrections. The NNLO loop-induced channel $gg$ is discussed in Sec.~\ref{subsec:valid_ms_loop_induced_density}.

\begin{figure}
     \centering
     \begin{subfigure}[b]{0.5\textwidth}
         \centering
         \includegraphics[width=\textwidth]{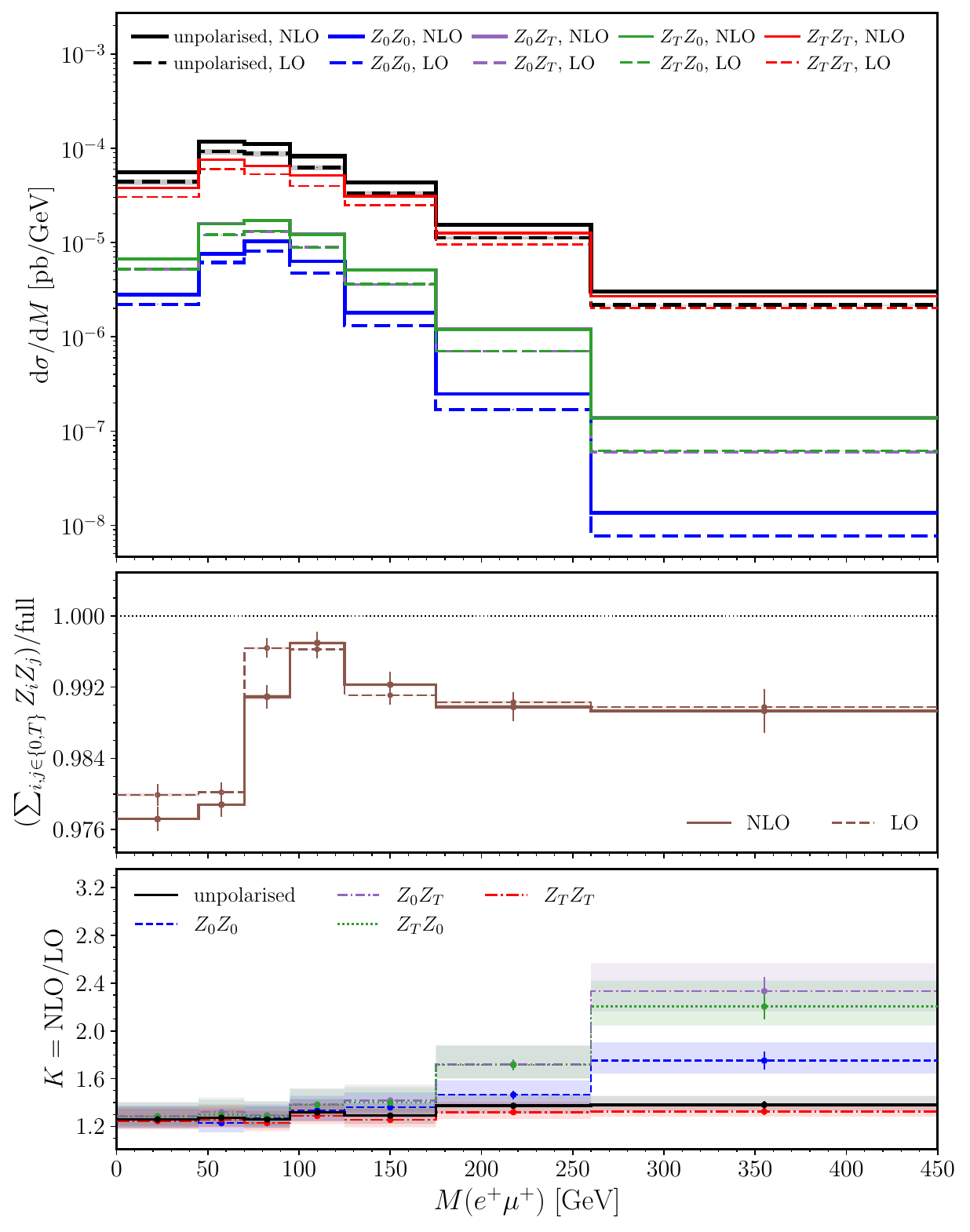}
         \caption{$M_{e^+ \mu^+}$}
         \label{fig:Mepmup_NLO}
     \end{subfigure}%
     \begin{subfigure}[b]{0.5\textwidth}
         \centering
         \includegraphics[width=\textwidth]{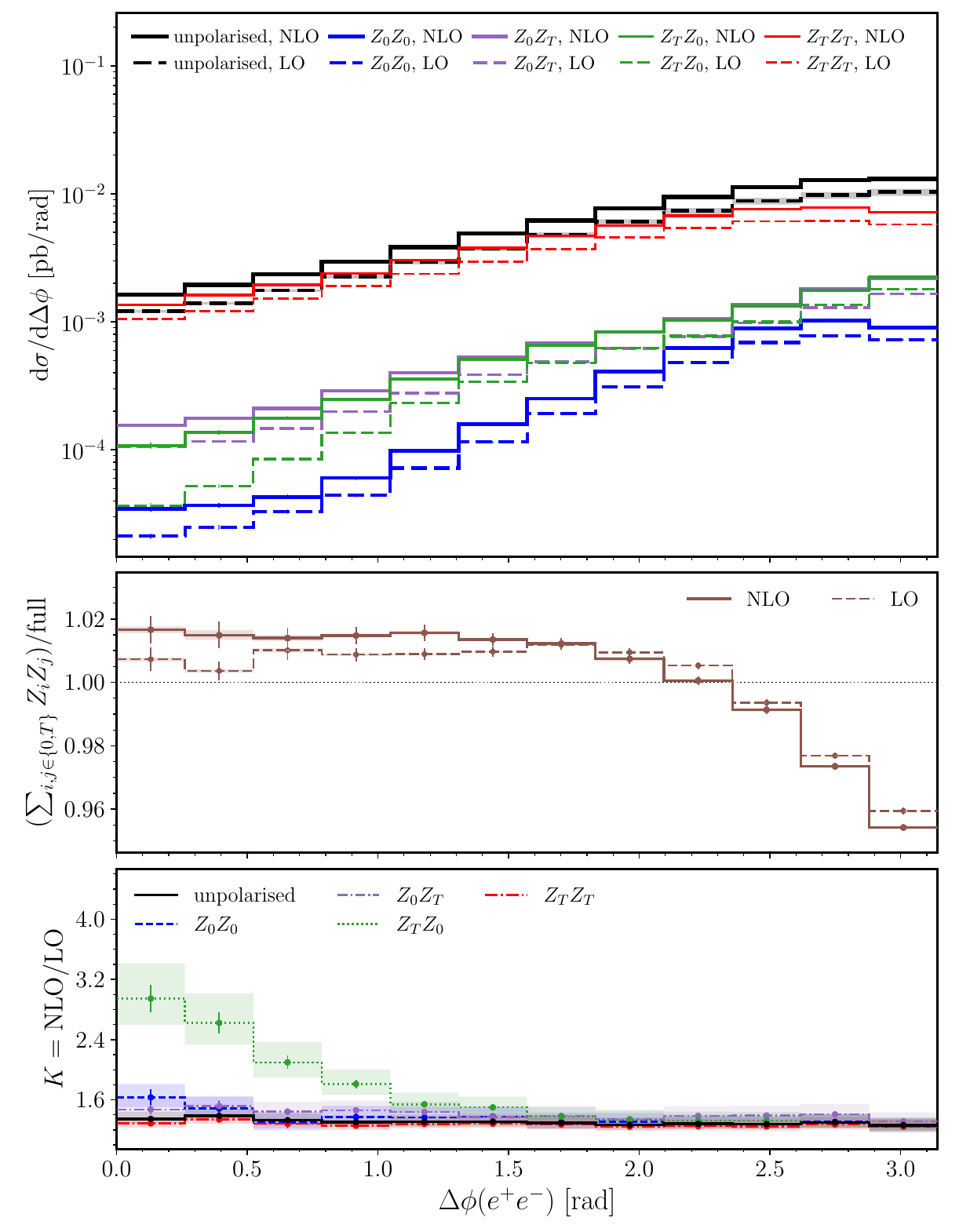}
         \caption{$\Delta\phi(e^+ e^-)$}
         \label{fig:DeltaPhi_NLO}
     \end{subfigure}
        \caption{Differential cross-section for $p \, p \to Z \, Z \to e^+\, e^- \, \mu^+ \, \mu^-$ at LO/NLO in the unpolarised case and in each possible helicity state. The first $Z$ decays to electrons while the second decays to muons. The left plot (a) is shown as a function of the invariant mass of the positively charged leptons $M_{e^+ \mu^+}$, while the right plot (b) is shown as a function of the azimuthal angle between the two electronic leptons $\Delta\phi(e^+ e^-)$. The second row compares the unpolarised result to the one reconstructed by summing the different polarised components (whose ratio should not be exactly one), and the last row presents the $K$-factor for each contribution.}
        \label{fig:NLO_polarisations_application}
\end{figure}

Fig.~\ref{fig:NLO_polarisations_application} splits the $Z$ polarisation states into two contributions: the longitudinal and the transverse one (the sum of the left/right polarisations), which for $ZZ$ leads to four possible contributions: $\{Z_0 Z_0, Z_0 Z_T, Z_T Z_0, Z_T Z_T\}$. Each sample is composed of $250$k unweighted events generated at fixed scale ($M_Z$), with the polarisation defined in the $ZZ$ rest frame, the quantisation axis of each boson being its own flight direction in that frame (helicity basis). The same PDF set is used for the LO and NLO samples, so that the quoted $K$-factors reflect the change of perturbative order only, and not a change of PDF. 
The contribution of each component is plotted together with the non-polarised process as a function of the invariant mass of the positively charged leptons $M_{e^+ \mu^+}$ (\ref{fig:Mepmup_NLO}) and as a function of the azimuthal angle between the two leptons of the same decay $\Delta \phi(e^+e^-)$ (\ref{fig:DeltaPhi_NLO}). The contribution of each helicity state is also represented relative to the unpolarised result both at LO and NLO precision. We also plot the sum of all these contributions relative to the unpolarised result; this serves as a check of the importance of the helicity interference terms, which are responsible for at most 2\% of the differential cross-section per bin for those two observables.

While the inclusive $K$-factor is relatively flat in the unpolarised case, several subleading polarisation channels exhibit a nontrivial dependence on the kinematics. In the invariant-mass distribution, the $K$-factor grows with increasing diboson invariant mass whenever at least one longitudinally polarised $Z$ boson is present, most notably in the $Z_TZ_0$ channel. For the azimuthal-angle distribution, a large local enhancement is observed only for $Z_TZ_0$ when the transverse $Z$ decays into electrons. The origin of this effect lies in the small-$\Delta\phi(e^+e^-)$ region, where the electrons are nearly collinear. Such configurations favour symmetric energy sharing and hence a longitudinally polarised electron-side $Z$. Since this boson is transverse in the $Z_TZ_0$ contribution, the LO rate is strongly suppressed. NLO radiation induces a recoil of the $ZZ$ system that boosts both bosons and repopulates this phase-space region, leading to a substantially enhanced $K$-factor relative to the suppressed LO prediction.

\subsection{Loop-induced processes}\label{subsec:loop_induced}

The computation of spin-density matrices for loop-induced processes, validated in
Sec.~\ref{sec:validation} through two independent tests, enables the use of \newms\
for this class of processes.

An interesting phenomenological application is the production of a pair of weak bosons
at the LHC through the gluon-initiated channel \cite{Glover:1988fe,Caola:2015psa,Javurkova:2024bwa}
\begin{equation}
    g \, g \to Z \, Z,
\end{equation}
with the $Z$ bosons decaying leptonically, $Z \to e^+e^-$ and $Z\to \mu^+\mu^-$.
The matrix element of the full process
\begin{equation}
    g \, g \to e^+ \, e^- \, \mu^+ \, \mu^-
    \label{eq:full_process_62}
\end{equation}
is computationally expensive to evaluate, motivating the use of \newms.

To illustrate the importance of spin-correlation effects, we compare the four
\texttt{spinmode} options available for loop-induced processes (\texttt{none},
\texttt{onshell}, \texttt{PA} and \texttt{madspin}; see \cref{sec:madspin2} and
\cref{tab:madspin_modes_combined}) against the full computation of \cref{eq:full_process_62} performed with \mgamc, hereafter referred to as the
reference.

The samples generated with \newms\ use a fixed scale ($\mu_R=\mu_F=M_Z$) and include a cut on the transverse momentum of the
$Z$ bosons, $p_T(Z) > 1\,\textrm{GeV}$, which regulates the integrable singularity
of the fermionic box loop integral in the $p_T(Z)\to 0$ limit \cite{Campbell:2010ff, Hirschi:2015iia}, while their virtuality is restricted to the pole region,
$|m_Z - m_{Z,\rm pole}| < 15\,\Gamma_Z$. For a consistent comparison, the same
$p_T$ cut is applied to the reference sample, together with the requirement that
the invariant masses of the two lepton pairs fall in the same on-shell regions. 

\begin{figure}
     \centering
     \begin{subfigure}[b]{0.45\textwidth}
         \centering
         \includegraphics[width=\textwidth]{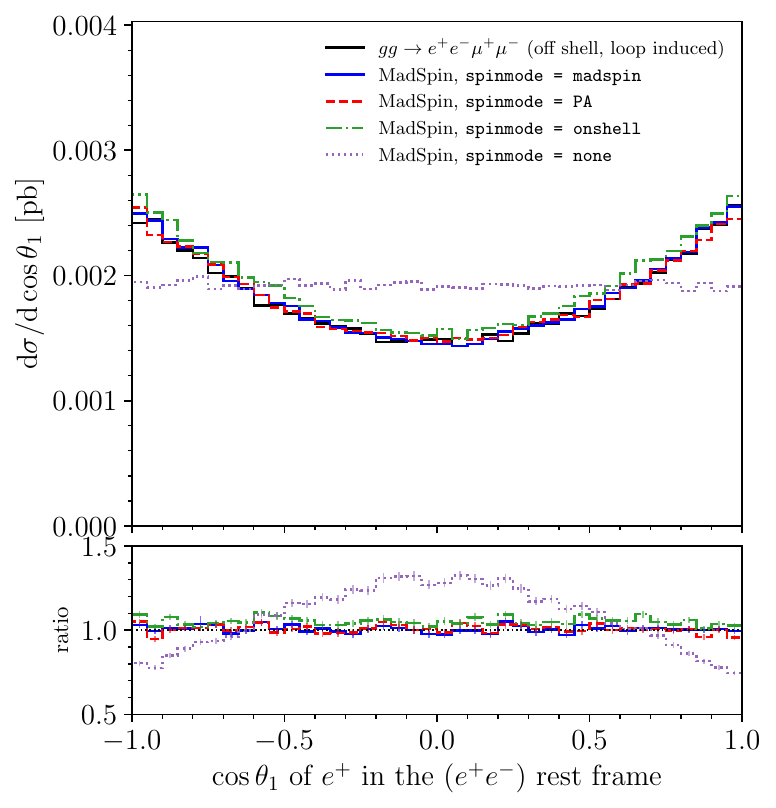}
         \caption{$\cos\theta_1$}
         \label{fig:theta1_LI}
     \end{subfigure}
     \hfill
          \begin{subfigure}[b]{0.45\textwidth}
         \centering
         \includegraphics[width=\textwidth]{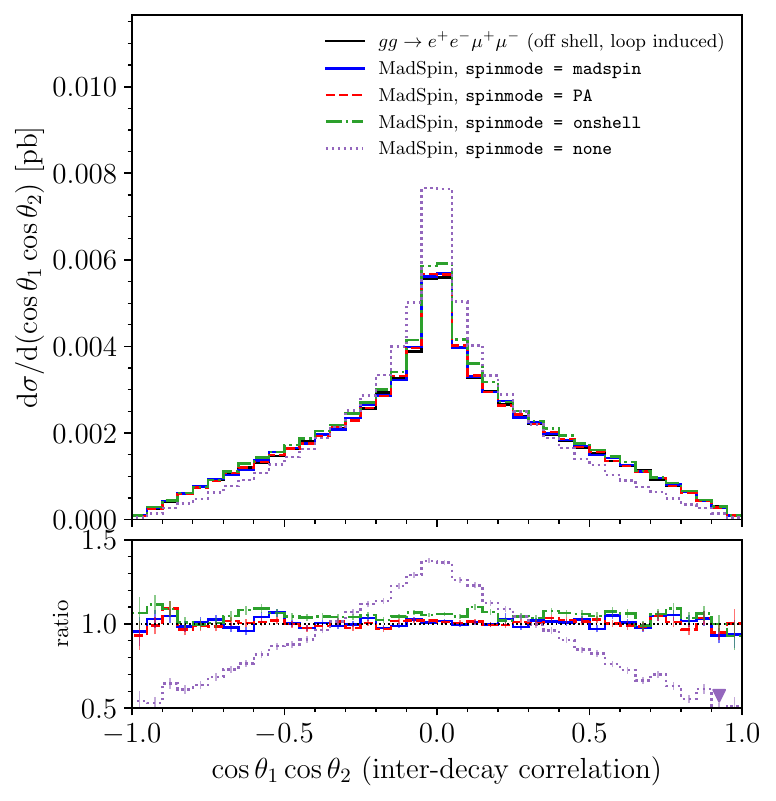}
         \caption{$\cos\theta_1\cos\theta_2$ }
         \label{fig:cos1cos2_LI}
     \end{subfigure}
        \caption{Distributions of two different observables, (a) $\cos\theta_1$ and (b) $\cos\theta_1\cos\theta_2$, for the process $g\, g \to Z\, Z \to e^+\, e^-\, \mu^+\,\mu^-$. The solid black line represents the reference (the full matrix element computed by \mgamc); the other curves are the results given by \newms\ with the different \texttt{spinmode} options available for loop-induced processes.}
        \label{fig:LI_application}
\end{figure}

Fig.~\ref{fig:theta1_LI} compares the various \newms\ \texttt{spinmode} options for the cosine of the helicity angle $\theta_1$, where $\theta_1$ is the angle between the direction of the $e^+$ in the $(e^+e^-)$ rest frame and the direction of the $(e^+e^-)$ system in the four-lepton rest frame. The distribution is flat without spin correlations (\code{spinmode=none}) and proportional to $1+\cos^2\theta_1$ otherwise. Note that the \code{spinmode=onshell} curve does not match the other curves, as it has a larger cross-section, not being impacted by the cut on the invariant mass.\footnote{The \code{none} curve also has a slightly larger cross-section for the same reason, but this is not visible in these plots.}

In Fig.~\ref{fig:cos1cos2_LI}, we compare the predictions for the
product of the cosines of the lepton helicity angles, where $\theta_2$ is defined analogously for the $\mu^+$ and the $(\mu^+\mu^-)$ system. As in the top-quark case (see \cref{subsec:validation-polarised}), the average value of the distribution is associated with a spin-correlation coefficient, here called $C_{kk}$:
\begin{equation}
C_{kk} = \frac{(g^2_V+g^2_A)^2}{g_V^2g_A^2}\langle\cos\theta_1\cos\theta_2\rangle,
\end{equation}
where $g_V$ and $g_A$ are the vector and axial coupling of the $Z$. 
Our simulations were performed with  $\alpha_{\rm EW}^{-1}=132.507$, $G_F=1.16639\times10^{-5}$ and
$m_Z=91.188$~GeV, meaning that $\frac{(g^2_V+g^2_A)^2}{g_V^2g_A^2} = 83.15$. 
One can relate $C_{kk}$ to the integrated density matrix:
\begin{equation}
C_{kk} = \rho_{++} + \rho_{--} -\rho_{+-} - \rho_{-+},
\end{equation}
which is a pure transverse quantity.

The computed values of this observable are reported in Table~\ref{tab:spincoeff}. For comparison, we also show the corresponding results for the NLO processes discussed in the previous section, evaluated with the same event selection as the loop-induced sample. As a first observation, in the absence of spin correlations (\code{spinmode=none}), the prediction is incompatible with the reference result and instead remains consistent with zero, as expected for an isotropic spin-density matrix. 

A second noteworthy feature is the markedly different behaviour of the loop-induced and NLO production mechanisms, including the opposite sign of the spin-correlation coefficient. In the $q\bar q$ channel, chirality conservation along the open quark line favours the production of transverse $Z$-boson pairs with opposite helicities, leading to a negative value of $C_{kk}$. By contrast, in the loop-induced $gg$ channel, the closed quark loop relaxes this constraint, while the dominant $J_z=0$ gluon configuration favours equal-helicity transverse states, resulting in a positive value of $C_{kk}$.

Finally, the \code{madspin} prediction for the loop-induced process agrees with the reference result at the level of roughly one standard deviation. While this agreement is somewhat marginal, it should be noted that the different \newms\ predictions are evaluated from the same underlying event sample and are therefore statistically correlated. The comparison is somewhat less satisfactory for the NLO sample, where deviations of up to $1.5\sigma$ are observed. This discrepancy may not be purely statistical. Indeed, the FLMW procedure implemented in \newms\ restores spin correlations at leading-order accuracy on top of an NLO event sample, whereas the reference prediction includes the complete NLO spin-correlation structure.

\begin{table}[htbp] 
\centering \small \setlength{\tabcolsep}{4.5pt} 
\begin{tabular}{lcc} 
\toprule sample & $C_{kk}$ & pull \\ \midrule \multicolumn{3}{@{}l}{\emph{\mg\ reference}}\\ 
$gg\to ZZ \to e^+e^-\mu^+\mu^-$, loop induced & $+0.380\pm0.072$ & -- \\ 
$pp\to ZZ \to e^+e^-\mu^+\mu^-$ at NLO & $-0.645\pm0.080$ & -- \\ 
\midrule \multicolumn{3}{@{}l}{\emph{$gg\to ZZ$, loop-induced}}\\ 
\code{madspin} & $+0.460\pm0.072$ & $+0.79$ \\ \code{PA} & $+0.491\pm0.072$ & $+1.09$ \\ 
\code{onshell} & $+0.431\pm0.072$ & $+0.50$ \\ \code{none} & $-0.034\pm0.062$ & $-4.36$ \\ 
\midrule \multicolumn{3}{@{}l}{\emph{$q\bar q\to ZZ$ at NLO}}\\ 
\code{madspin} & $-0.540\pm0.078$ & $+0.94$ \\ \code{PA} & $-0.524\pm0.079$ & $+1.08$ \\ 
\code{onshell} & $-0.482\pm0.078$ & $+1.46$ \\ 
\code{none} & $-0.028\pm0.071$ & $+5.77$ 
\\ \bottomrule 
\end{tabular} \caption{Spin coefficient $C_{kk}$ of the two $ZZ$ production mechanisms, together with the pull with respect to the corresponding \mg\ reference prediction. The values are unbinned moments over $200\,000$ events per sample, on the common selection described in the text.} \label{tab:spincoeff} \end{table}

\begin{figure}
     \centering
     \begin{subfigure}[b]{0.45\textwidth}
         \centering
         \includegraphics[width=\textwidth]{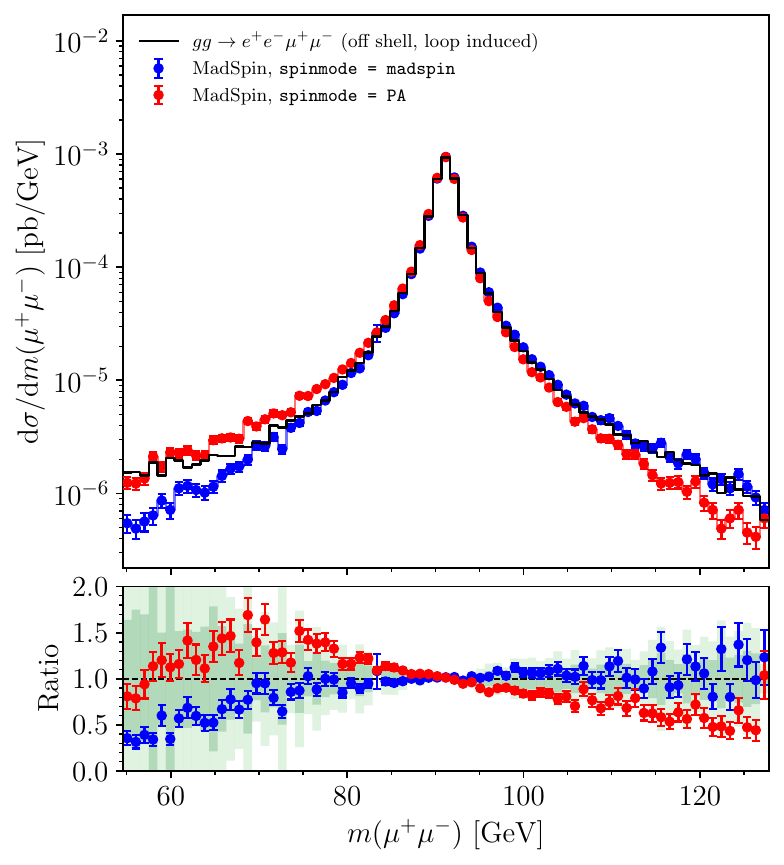}
         \caption{$M_{\mu^+\mu^-}$ }
         \label{fig:MZ_LI}
     \end{subfigure}
     \hfill
     \begin{subfigure}[b]{0.46\textwidth}
         \centering
         \includegraphics[width=\textwidth]{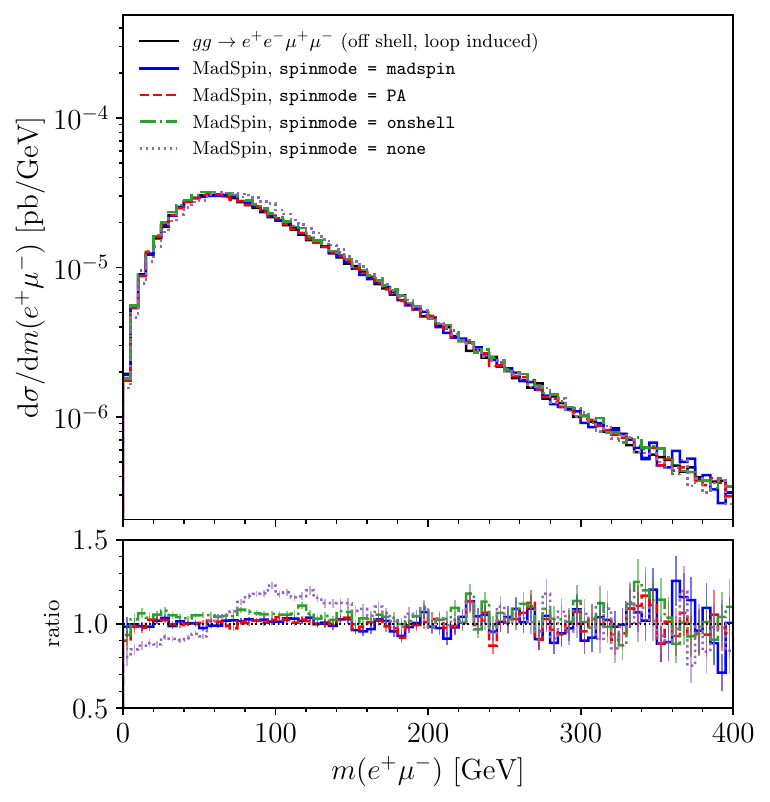}
         \caption{$M_{e^+\mu^-}$ }
         \label{fig:Memu_LI}
     \end{subfigure}

        \caption{Invariant-mass distributions of (a) the muon pair, $M_{\mu^+\mu^-}$, and (b) the positron--muon pair, $M_{e^+\mu^-}$, for the process $g\, g \to Z\, Z \to e^+\, e^-\, \mu^+\,\mu^-$. The solid black line represents the reference (the full matrix element computed by \mgamc); the other curves are the results given by \newms\ with the different \texttt{spinmode} options available for loop-induced processes.}
        \label{fig:LI_MXX}
\end{figure}

In Fig.~\ref{fig:LI_MXX}, we present the invariant-mass distributions of the muon pair (which corresponds to the Breit-Wigner lineshape) and of the positron--muon pair. In the first case, both the \code{none} and \code{onshell} curves are delta functions at the $Z$ peak due to the absence of reshuffling. The difference between \code{PA} and \code{madspin} is related to the evaluation of the matrix element with off-shell or on-shell momenta. In this case, \code{madspin} is a better approximation than \code{PA}, even if it undershoots the reference curve at low invariant mass. The invariant mass of the positron--muon pair is more consistent among the modes, with the exception of the sample without spin correlations, which has a different shape (by up to 20\%).

\subsection{Spin correlations in amplitude interference: SMEFT}\label{subsec:valid_interf}

In order to test the impact of spin correlations in processes with interference, we use the \texttt{SMEFTsim} model~\cite{Brivio:2020onw,Brivio:2017btx} that encodes the SMEFT~\cite{Grzadkowski:2010es,Brivio:2017vri} operators and focus on the effect of the top chromomagnetic dipole moment operator, defined by
\begin{eqnarray}
    &&\frac{c_{tG}}{\Lambda^2}\mathcal{O}_{tG}+\text{h.c.},\\
    \mathcal{O}_{tG}&=&g_s\bar{L}_3\sigma^{\mu\nu}T^At_R\tilde{H}G_{\mu\nu}^A,    
\end{eqnarray}
where $c_{tG}$ is the Wilson coefficient associated with the $\mathcal{O}_{tG}$ operator, $\Lambda$ denotes the energy scale at which the SMEFT expansion breaks down, $g_s$ is the strong coupling constant, $T^A$ are the $SU(3)$ generators, $L_3$ is the third generation left-handed quark doublet, $\tilde{H}$ is the Higgs doublet and $G_{\mu\nu}^A$ is the gluon field-strength tensor. After electroweak symmetry breaking, this operator induces an anomalous $gt\bar{t}$ vertex, parametrised by~\cite{Aguilar-Saavedra:2008nuh}
\begin{equation}\label{eq:anomalous_vgtt}
    V_{gt\bar{t}}=-g_s\bar{t}\frac{i\sigma^{\mu\nu}q_{\nu}}{m_t}(d_V^g+i\gamma_5d_A^g)T^AtG^A_{\mu},
\end{equation}
where $q_\nu$ is the gluon momentum and $d_V^g,d_A^g$ are the anomalous couplings related to the top chromomagnetic and chromoelectric dipole moment respectively, given by
\begin{eqnarray}
    d_V^g&=&\sqrt{2}\text{Re}[c_{tG}]\frac{vm_t}{\Lambda^2}\\
    d_A^g&=&\sqrt{2}\text{Im}[c_{tG}]\frac{vm_t}{\Lambda^2}.
\end{eqnarray}
While the latter induces CP-violating observables, the former is CP-even and couples the spin of the top quarks to the gluon momentum via the $\sigma_{\mu\nu}$ operator, thereby modifying the SM spin correlations.

The dominant effect of the $\mathcal{O}_{tG}$ operator arises from the interference term \begin{equation}\label{eq:CMDM_interf} 2\frac{c_{tG}}{\Lambda^2}\,\mathrm{Re}\!\left(\mathcal{M}_{\text{SM}}^{*}\mathcal{M}_{tG}\right), 
\end{equation}
where $\mathcal{M}_{\rm SM}$ denotes the SM amplitude for the $pp\to t\bar t$ process, while $\frac{c_{tG}}{\Lambda^2}\mathcal{M}_{tG}$ denotes the corresponding amplitude obtained by replacing the SM $gt\bar t$ vertex with the anomalous vertex of Eq.~\eqref{eq:anomalous_vgtt}; the factor 
$\frac{c_{tG}}{\Lambda^2}$ has been extracted to make its dependence explicit.

Measurements of spin correlations in $t\bar{t}$ production can thus constrain this operator~\cite{CMS:2019nrx,CMS:2019kzp,ATLAS:2019zrq}. In this procedure, it is obviously of crucial importance that the (amplitude) interference term in Eq.~(\ref{eq:CMDM_interf}) be generated keeping the off-diagonal terms in the spin-density matrices (i.e. keeping the helicity interference term). Since \mg\ is not able to compute the equivalent matrix element, \ms\ was therefore restricted to \code{spinmode=none}, the NSA.

Since we skipped such a numerical validation in Sec.~\ref{sec:validation}, we first compare the matrix element calculated in \newms\ against the matrix element calculated by \mgamc\ using
\begin{lstlisting}[style=scipostcode,language=Python,caption={},label={}]
generate g g > t t~ > b b~ e+ ve e- ve~ NP=1 NP^2==1
\end{lstlisting}
for one phase-space point and find
\begin{eqnarray} 2\operatorname{Re}\!\left(\mathcal{M}^{*}_{SM}\mathcal{M}_{tG}\right)\Big|_{\rm MadSpin2} &=& -0.02695274209\ \text{GeV}^{-4}\\ 2\operatorname{Re}\!\left(\mathcal{M}_{SM}^*\mathcal{M}_{tG}\right)\Big|_{\rm MG5aMC} &=& -0.02695274333\ \text{GeV}^{-4} \end{eqnarray}
which shows an agreement with a relative precision of $10^{-8}$.

Then, to estimate the importance of spin correlations in such measurements, and also validate the new implementation, we generate the interference term alone \eqref{eq:CMDM_interf} for the $pp\to t\bar{t}$ process using the following commands
\begin{lstlisting}[style=scipostcode,language=Python,caption={},label={}]
import model SMEFTsim_topU3l_MwScheme_UFO
generate p p > t t~ NP=1 NP^2==1
\end{lstlisting}
setting all Wilson coefficients to zero, except for $c_{tG}$, which is set to one. The off-shell effects, found to have a negligible impact, are not reported here; we therefore compare the top decay in the
\texttt{onshell} mode with the \texttt{none} mode to quantify the importance of spin correlations.

\begin{figure}[htp]
    \centering
    \includegraphics[width=0.5\textwidth]{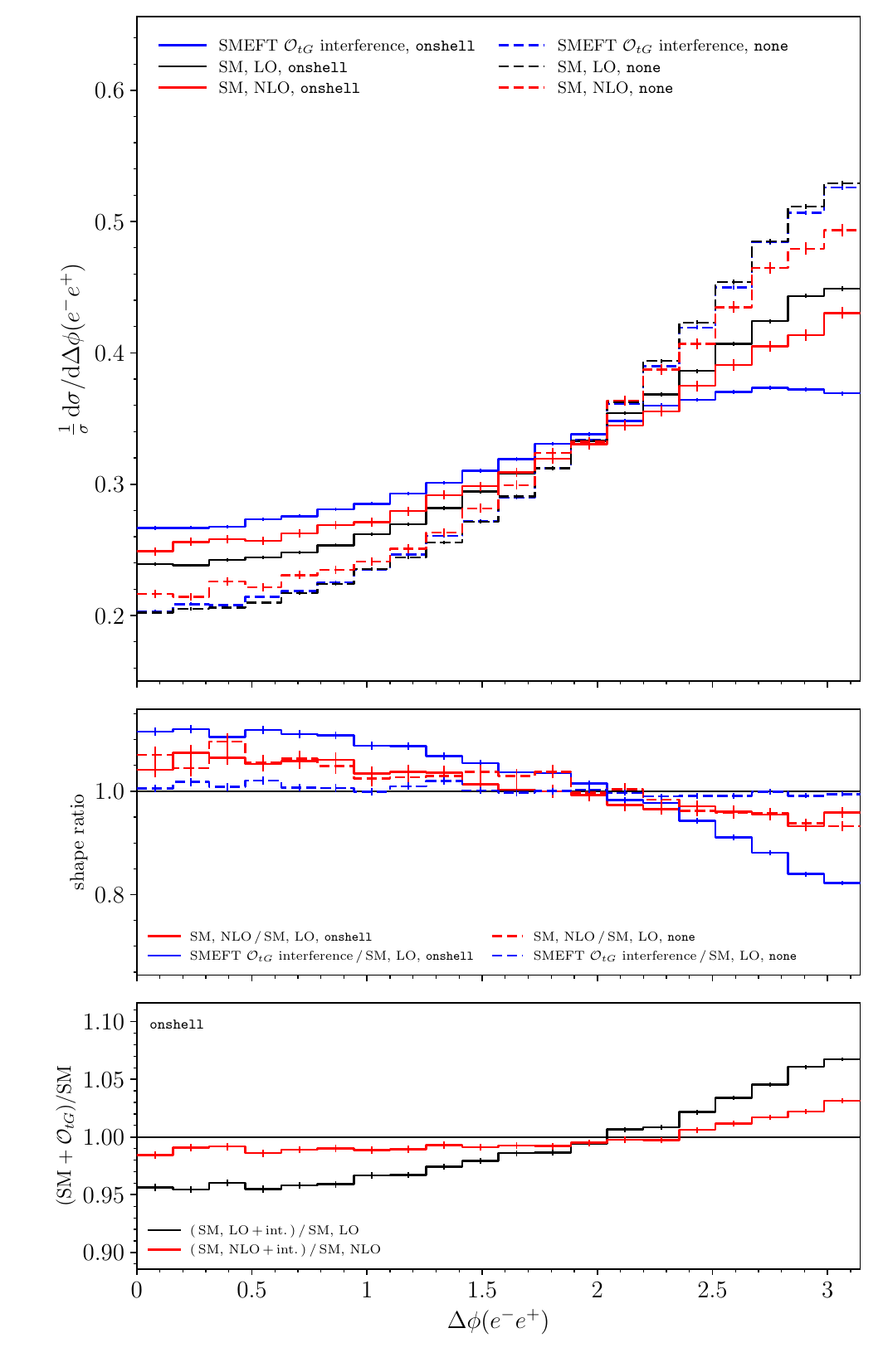}
    \caption{$\Delta\phi(e^-e^+)$ of the electron and positron from the $t\bar{t}$ decay in the $pp\to t\bar{t}$ process, normalised to unity. The SMEFT amplitude-level interference $2Re\left(\mathcal{M}_{tG}^*\mathcal{M}_{\text{SM}}\right)$ is shown with (\texttt{onshell}) and without (\texttt{none}) spin correlations, together with the SM prediction at LO and NLO accuracy. The first subplot shows the shape ratios to the SM LO prediction; the second subplot shows the ratio of the LO (NLO) SM prediction augmented by the interference contribution (for $c_{tG}=1$) to the corresponding SM prediction.
    }
    \label{fig:valid_interf}
\end{figure}

A comparison of the two simulations is presented in \cref{fig:valid_interf}, which shows the $\Delta\phi$ distribution between the two electrons from the $t\bar{t}$ decay, normalised to unity. This observable is particularly sensitive to spin-correlation effects (see Sec.~\ref{subsec:validation-polarised} for more details). The inclusion of spin correlations modifies the distribution by up to $25\%$, demonstrating that an accurate treatment of spin-correlation effects is essential for such measurements.

The same figure also includes the Standard Model prediction at both LO and NLO accuracy, together with the corresponding shape ratios shown in the first subplot. In the absence of spin correlations, the SMEFT interference and Standard Model distributions are identical and their ratio is exactly one --up to statistical fluctuations--. This implies that the observable would have no discriminating power in the absence of spin-correlation effects.

A second noteworthy feature of this subplot is that the NLO-to-LO ratio is not constant across the distribution. However, this ratio is identical when comparing those curves with and without spin correlations, indicating that the effect originates from a purely kinematical distortion induced by the real-emission contribution. Finally, the SMEFT interference shape differs from the LO Standard Model prediction by up to $10\%$, which is twice as important as the NLO corrections (at most $5\%$).

In the bottom subplot, we present a more phenomenologically relevant ratio. Here, the interference contribution is added either to the LO or to the NLO Standard Model prediction, and the result is divided by the corresponding LO or NLO distribution. These are the only curves in the figure that depend explicitly on the value of $c_{tG}$, which is set to unity for this plot. This ratio therefore illustrates the level of experimental precision required for the observable to provide sensitivity to the operator coefficient.

While the effect reaches about $5\%$ when the interference contribution is added to the SM LO prediction, the presence of a larger-than-one $K$-factor reduces the effect to approximately $2\%$ when the same contribution is added on top of the NLO prediction. However, one should note that the interference term is itself computed at LO. If one assumes that the interference term receives a similar $K$-factor (bin by bin), then the effect goes back to its LO size.
The computation of such a $K$-factor is left to a dedicated phenomenological study, as well as the assessment of the scale/PDF uncertainties on such observables.

%% file: performance.tex
\section{Computational performance}
\label{sec:performance}

The \newms\ code features improved computational performance with respect to \ms, stemming from the parallelisation of the different computational phases (decay-event generation, unweighting) and from the simplified evaluation of the matrix elements via spin-density matrices described in Sec.~\ref{sec:madspin2}.

In order to illustrate how these improvements impact the computational performance, we analyse here the impact of parallelisation (Sec.~\ref{subsec:parallelism}) and study the time fractions spent in different computational phases for various run modes and physics processes (Sec.~\ref{subsec:detail_time_estimates}).

\subsection{Technical setup}

The benchmark runs were performed on a heterogeneous machine equipped with an Apple M2 system-on-chip comprising 4 performance and 4 efficiency CPU cores using an ARM64 architecture and 16 GB of unified memory. No other computationally intensive tasks were running concurrently.

To obtain stable wall-time estimates, each process/mode combination was run several times in a random order, including a warm-up run (which was discarded) and a cool-down period between runs to avoid thermal throttling. The inter-quartile range (IQR) of the total wall-clock time was evaluated after each run, and additional runs were performed until the IQR dropped below 5\% of the median. In the following, the reported wall-clock times always correspond to the median of all runs.

\subsection{Computational improvement from parallelisation}\label{subsec:parallelism}

One of the technical improvements in \newms\ is the ability to split the generation of decay events and unweighting over several cores. This is steered via the \texttt{nb\_core} setting. To be precise, \texttt{nb\_core}, hereafter denoted by $p$, defines the number of processes assigned to each parallel operation and not the number of machine cores used, since the code can oversubscribe or underuse cores depending on how many decay channels are generated.

In order to study how the execution time scales with $p$, \newms\ was run on $pp\to t\bar{t}t\bar{t}$ production events with a fully leptonic top decay chain, in the \texttt{madspin} mode with \texttt{unweighting=joint} (see App.~\ref{seq:commands}).

The wall-clock times for decay-event generation, unweighting and total end-to-end runtime were registered as a function of $p$. The time estimates were obtained for two workloads: $10^4$ and $10^5$ events. For each $p$, \newms\ was run several times with different random-number seeds to ensure that the time estimates cover a representative range of phase-space configurations. The reported times correspond to the medians of all runs and the error bars correspond to the respective IQR.

As explained before, for the generation of decay events, we generate unweighted events using \mg; in that step, each type of decaying particle is run in parallel, each requesting multiple cores in an oversubscribed way (because they are highly inefficient). The reported time estimate corresponds to the slowest of them.

For each $p$ the speedup relative to the one-core measurement is defined from the plotted medians
\begin{equation}
    S(p)\equiv \frac{T_1}{T_p},
\end{equation}
where $T_p$ corresponds to the time using $p$ processes and $T_1$ to the time using 1 process.

For a homogeneous machine, the runtime is expected to follow Amdahl's law \cite{Amdahl}
\begin{equation}
    S^{-1}(p)=f+\frac{1-f}{p},
\end{equation}
where $f$ is the non-parallelisable fraction. This law assumes that all cores provide equal throughput and that the amount of computational work is unchanged. 

Since the machine which was used for benchmarking contains performance and efficiency cores of different throughput, a modified Amdahl law was used to model the scaling behaviour
\begin{equation}\label{eq:modified_amdahl}
    S^{-1}(p)=f+\frac{(1-f)}{C_k(p,r)}, 
\end{equation}
where $C_k(p,r), k\in\{U,G\}$ corresponds to an effective number of cores, that depends on the number of processes $p$ and the ratio of the throughput of an efficiency core relative to that of a performance core, $r$.

The non-parallelisable fractions can differ between computational phases and workloads, since different algorithms and parallelisation overheads are involved. 

\begin{figure}[htbp]
     \centering
     \begin{subfigure}[b]{0.45\textwidth}
         \centering
         \includegraphics[width=\textwidth]{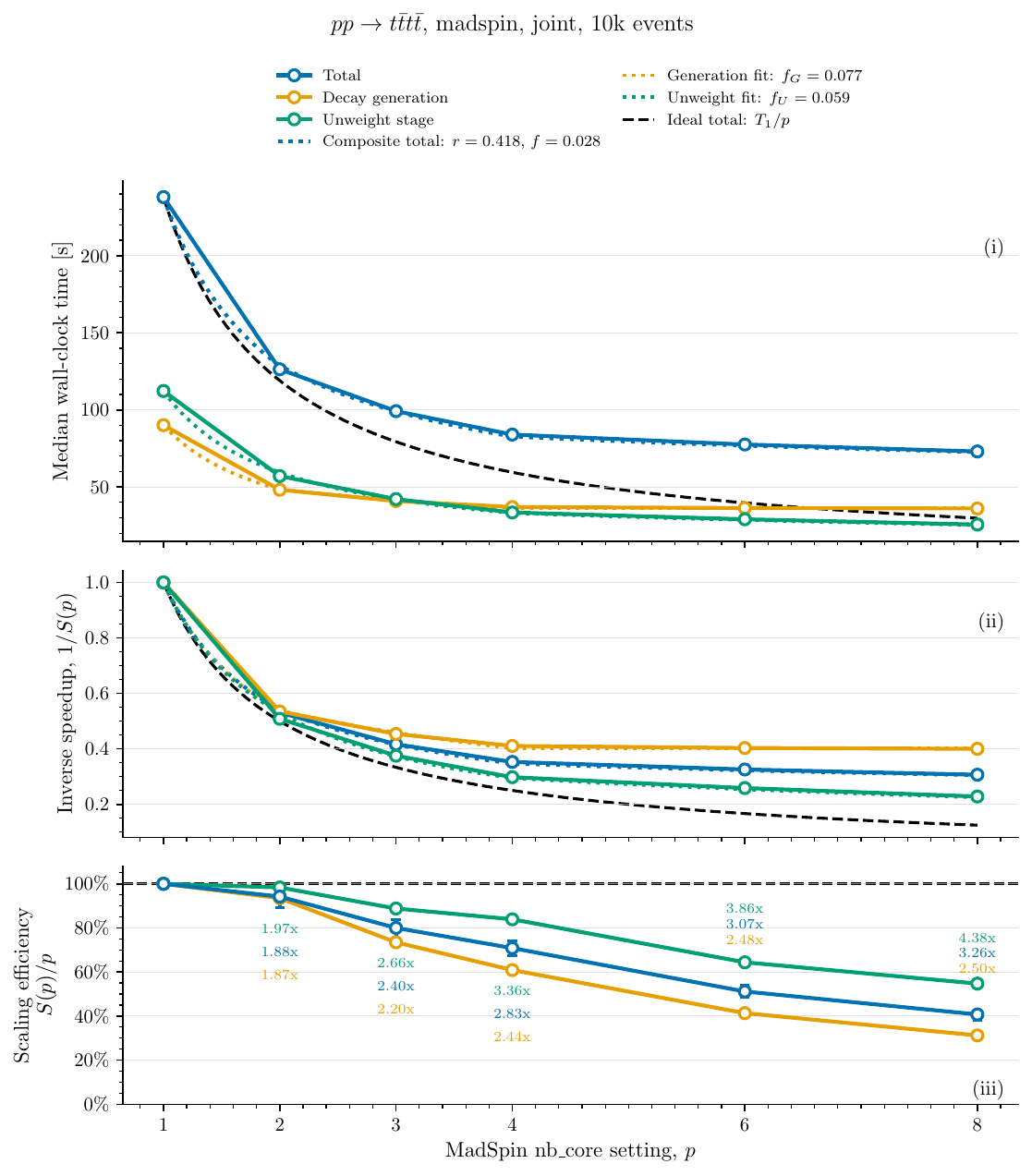}
         \caption{$t\bar{t}t\bar{t}$ with $W$ decays $10^4$ events}
         \label{fig:core_scaling_10k}
     \end{subfigure}
     \hfill
     \begin{subfigure}[b]{0.45 \textwidth}
         \centering
         \includegraphics[width=\textwidth]{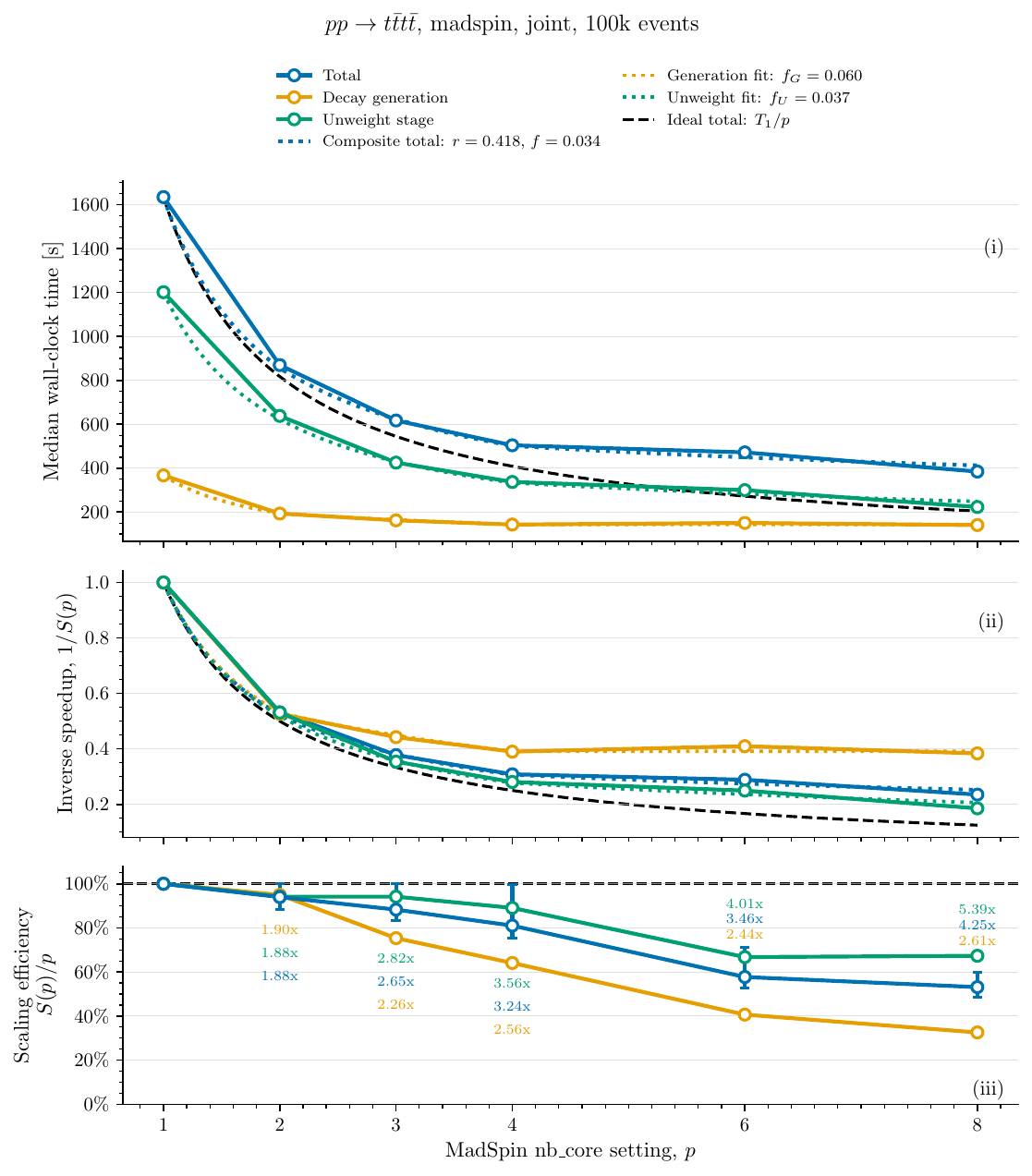}
         \caption{$t\bar{t}t\bar{t}$ with $W$ decays  $10^5$ events}
         \label{fig:core_scaling_100k}
     \end{subfigure}
     \begin{subfigure}[b]{0.45 \textwidth}
         \centering
         \includegraphics[width=\textwidth]{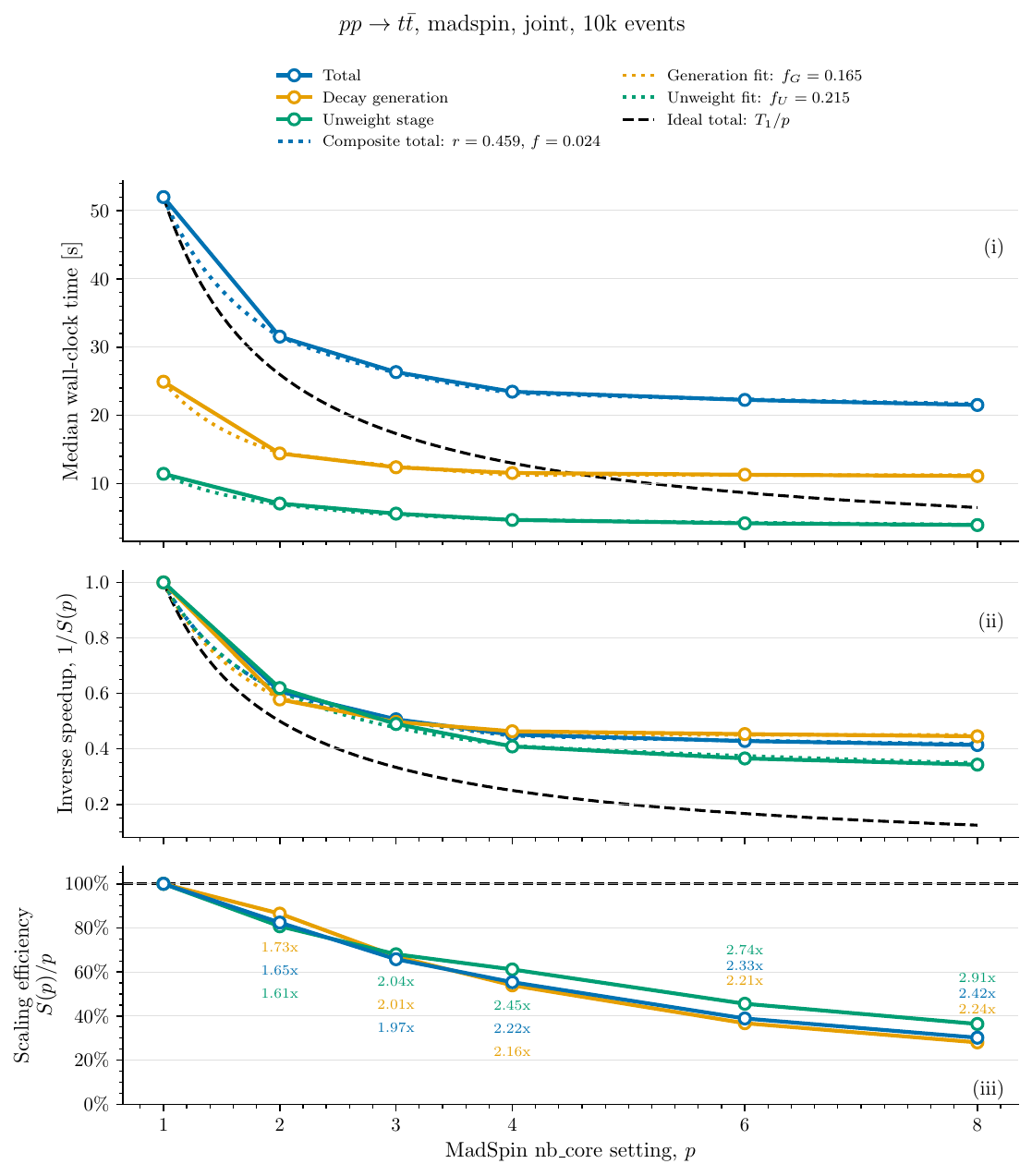}
         \caption{$t\bar{t}$ no $W$ decays  $10^4$ events}
         \label{fig:tt_2body_scaling}
     \end{subfigure}
        \caption{Scaling of wall-clock time for decay-event generation, unweighting and total as a function of the \texttt{nb\_core} setting, for the $pp\to t\bar{t}t\bar{t}$ process with a fully leptonic decay chain (\cref{fig:core_scaling_10k}, \cref{fig:core_scaling_100k}) and for the $pp\to t\bar{t}\to W^+W^-b\bar{b}$ process (\cref{fig:tt_2body_scaling}). The points on the first pane correspond to the median wall-clock times and the coloured dashed lines to the respective modified Amdahl fits. The dashed black line corresponds to the ideal scaling $T_1/p$ for the total run time. The second pane shows the inverse speedup ($1/S(p)$, where $S(p)\equiv T_1/T_p$) and the last pane the scaling efficiency $S(p)/p$. The numbers in the last pane correspond to the speedup at each core multiplicity. The pink dashed-dotted line corresponds to the estimated total time assuming the decay-event generation would have the same scaling behaviour as the unweighting. The error bars correspond to the IQR.}
\end{figure}

As will be demonstrated in \cref{subsec:detail_time_estimates}, the speedup from parallelisation is significantly worse for the decay-event generation than the unweighting phase. For small workloads (typically used for parallel event generation on computing clusters or on the grid), decay-event generation accounts for a large fraction of the total runtime and therefore constitutes a bottleneck for further speedup of the code.

We therefore expect that combining the improvements in \newms\ with improvements in the generation of events, which were recently achieved in \textsc{MadNIS/MadSpace} \cite{Heimel:2022wyj,Heimel:2023ngj,Heimel:2024wph,DeCrescenzo:2026tsp}, will lead to a further significant reduction of the computation time. This will be achieved with the release of \textsc{MadGraph7}.

We also observe that for simpler processes, like $pp\to t\bar{t}\to W^+W^-b\bar{b}$, the overall speedup scales more weakly with the number of processes, given that the time spent in the parallel process orchestration is significantly larger (cf. \cref{fig:timing_unweight_breakdown}).

\subsection{Computational performance for different physics processes}\label{subsec:detail_time_estimates}

In order to illustrate the computational performance of \newms, a set of benchmark processes are used corresponding to different multiplicities and spins of decaying resonances and different complexity of the production matrix elements. The following processes are generated with all available \texttt{spinmode} settings:
\begin{enumerate}
    \item $WH$, with $H\to b\bar{b}$, while the $W$ is left undecayed. This is the simplest process which does not feature any spin correlations and is used to benchmark the overhead of the new framework and of the decay-event generation.
    \item $t\bar{t}$ with both tops decaying, which features two spin-$1/2$ decays and a relatively simple production process,
    \item $ZZ$, with both $Z$ bosons decaying to $d\bar{d}$, which compared to $t\bar{t}$ isolates the effect of a higher-dimension spin-density matrix.
    \item $t\bar{t}Z$, a process which includes both fermions and vector bosons,
    \item $t\bar{t}t\bar{t}$, a complicated process with four fermions, which is very costly to generate in \ms,
    \item $W^+W^-W^+W^-$, a complicated process with four vector bosons, costly amplitudes and the largest spin-density space. It exposes costs that are negligible in less complicated processes.
\end{enumerate}

The $t\bar{t}, t\bar{t}Z$ and $t\bar{t}t\bar{t}$ processes were generated twice, once with a one-step decay, e.g. $t\to bW^+$, and once with a two-step decay chain, e.g. $t\to bW^+, W^+\to e^+\nu_e$.  All runs were performed with \texttt{nb\_core=6} and identical settings, including the same random-number seeds. The total wall-time for each process and each \texttt{spinmode} is shown in Table~\ref{tab:madspin-wall-times}.

\begin{table*}[h]
  \centering
  \scriptsize
  \caption{Total \newms\ wall-time in seconds. Each entry is the
  median over the retained runs. The speedup
  $S=T_{\rm ref}/T$ is shown in parentheses; entries in bold indicate a speedup ($S>1$).
  The reference is \texttt{onshell\_v1} for the on-shell modes and
  \texttt{madspin\_v1} for the \texttt{madspin} and \texttt{PA} modes. For every entry, the observed interquartile range
  is below 1\% of the median. ``joint'' corresponds to \texttt{unweighting=joint} and ``seq.'' to \texttt{unweighting=sequential} (see App.~\ref{app:sequential}). The reported wall-time corresponds to $10^4$ events.}
  \label{tab:madspin-wall-times}
  \resizebox{\textwidth}{!}{%
    \begin{tabular}{ccccccccc}
      \toprule
      \multirow{2}{*}{Process} &
      \multicolumn{3}{c}{On-shell} &
      \multicolumn{3}{c}{MadSpin} &
      \multicolumn{2}{c}{Pole approximation} \\
      \cmidrule(lr){2-4}
      \cmidrule(lr){5-7}
      \cmidrule(lr){8-9}
      &
      \texttt{onshell\_v1} &
      \texttt{onshell} joint &
      \texttt{onshell} seq. &
      \texttt{madspin\_v1} &
      \texttt{madspin} joint &
      \texttt{madspin} seq. &
      \texttt{PA} joint &
      \texttt{PA} seq. \\
      \midrule

      $WH$ &
      $14.0$ &
      $13.4\;(\mathbf{1.05}\times)$ &
      $13.5\;(\mathbf{1.04}\times)$ &
      $10.4$ &
      $15.0\;(0.69\times)$ &
      $20.2\;(0.51\times)$ &
      $14.7\;(0.71\times)$ &
      $19.7\;(0.53\times)$ \\

      $t\bar t$ &
      $20.5$ &
      $21.6\;(0.95\times)$ &
      $19.6\;(\mathbf{1.05}\times)$ &
      $17.8$ &
      $22.4\;(0.80\times)$ &
      $28.2\;(0.63\times)$ &
      $22.3\;(0.80\times)$ &
      $26.8\;(0.66\times)$ \\

      $t\bar t\;(\mathrm{dec.\ chain})$ &
      $24.3$ &
      $25.8\;(0.94\times)$ &
      $23.5\;(\mathbf{1.03}\times)$ &
      $17.1$ &
      $26.9\;(0.64\times)$ &
      $33.3\;(0.51\times)$ &
      $27.0\;(0.63\times)$ &
      $31.4\;(0.54\times)$ \\

      $ZZ$ &
      $32.3$ &
      $29.9\;(\mathbf{1.08}\times)$ &
      $28.5\;(\mathbf{1.14}\times)$ &
      $16.7$ &
      $41.8\;(0.40\times)$ &
      $40.8\;(0.41\times)$ &
      $39.2\;(0.43\times)$ &
      $38.7\;(0.43\times)$ \\

      $t\bar t Z$ &
      $43.8$ &
      $37.4\;(\mathbf{1.17}\times)$ &
      $37.2\;(\mathbf{1.18}\times)$ &
      $50.4$ &
      $43.9\;(\mathbf{1.15}\times)$ &
      $49.5\;(\mathbf{1.02}\times)$ &
      $41.4\;(\mathbf{1.22}\times)$ &
      $46.6\;(\mathbf{1.08}\times)$ \\

      $t\bar t Z\;(\mathrm{dec.\ chain})$ &
      $49.9$ &
      $46.5\;(\mathbf{1.07}\times)$ &
      $45.0\;(\mathbf{1.11}\times)$ &
      $45.2$ &
      $58.7\;(0.77\times)$ &
      $58.4\;(0.77\times)$ &
      $53.3\;(0.85\times)$ &
      $55.3\;(0.82\times)$ \\

      $t\bar t t\bar t$ &
      $330.2$ &
      $28.7\;(\mathbf{11.52}\times)$ &
      $30.0\;(\mathbf{11.01}\times)$ &
      $1820.7$ &
      $40.5\;(\mathbf{44.92}\times)$ &
      $49.0\;(\mathbf{37.17}\times)$ &
      $32.5\;(\mathbf{56.09}\times)$ &
      $40.3\;(\mathbf{45.19}\times)$ \\

      $t\bar t t\bar t\;(\mathrm{dec.\ chain})$ &
      $59.2$ &
      $38.4\;(\mathbf{1.54}\times)$ &
      $37.9\;(\mathbf{1.56}\times)$ &
      $690.1$ &
      $61.7\;(\mathbf{11.18}\times)$ &
      $58.9\;(\mathbf{11.72}\times)$ &
      $47.5\;(\mathbf{14.54}\times)$ &
      $49.8\;(\mathbf{13.87}\times)$ \\

      $W^+W^-W^+W^-$ &
      $141.5$ &
      $134.1\;(\mathbf{1.05}\times)$ &
      $118.8\;(\mathbf{1.19}\times)$ &
      $609.1$ &
      $579.1\;(\mathbf{1.05}\times)$ &
      $180.7\;(\mathbf{3.37}\times)$ &
      $166.8\;(\mathbf{3.65}\times)$ &
      $132.3\;(\mathbf{4.60}\times)$ \\

      \bottomrule
    \end{tabular}%
  }
\end{table*}

Table~\ref{tab:madspin-wall-times} shows that the gain from the new implementation grows with the complexity of the process. For the most demanding benchmarks the speedup is substantial --- reaching a factor 56 for $t\bar{t}t\bar{t}$ in the \texttt{PA} mode and a factor 4.6 for $W^+W^-W^+W^-$ --- since the cost of evaluating the full matrix element and the low unweighting efficiency dominate the runtime of the legacy modes. For simple processes ($WH$, $t\bar{t}$, $ZZ$), the off-shell density-matrix modes instead show a mild overhead (speedups of 0.4--0.8) with respect to \texttt{madspin\_v1}, whose direct evaluation of the (cheap) full matrix element is hard to beat once the fixed costs of generating the initial decay-event pool and of orchestrating the parallel processes are taken into account; the on-shell modes remain at par or slightly faster. A detailed breakdown of the time spent in each computational phase, which substantiates these observations, is given in App.~\ref{app:profiling}.

%% file: conclusion.tex
\section{Conclusion and outlook}\label{sec:conclusions}

In this paper we applied a new method for simulating resonance decays based on the spin-density matrix formalism. The implementation is included in the \newms\ package that ships with \mgamc\ v3.8.x and will later be incorporated in the \textsc{MadGraph7} framework. This lifts many of the limitations of the previous \ms\ code: the FLMW method can now be applied to arbitrary $N$-body decays and to a wider range of processes (non-positive-definite samples, including pure interference, and loop-induced samples).

The new code also introduces the simulation of off-shell effects based on the multiple pole approximation, as well as a sequential accept/reject method for the unweighting. \newms\ also provides the ability to simulate a process with different options for modelling off-shell effects, thereby allowing one to estimate not only the impact of such effects but also the differences between the involved approximations.

\newms\ also provides an on-the-fly decomposition of the production--decay
density-matrix contraction in resonance-helicity space. Additional event
weights can be assigned to definite-helicity contributions and to
off-diagonal helicity-interference contributions. This can reduce the number of samples needed to study polarisation effects and allow one to study NLO polarised samples for the first time in \mg.

This facilitates a series of phenomenological applications that were not possible in the previous version, such as the study of SMEFT processes with proper $\frac{1}{\Lambda^2}$ truncation or the inclusion of spin correlations in searches for BSM resonances that are produced via loop-induced production and decay to particles of non-zero spin (e.g.\ the $gg\to A\to Z(\ell^+\ell^-)H$ process in models with additional Higgs multiplets~\cite{Argyropoulos:2024yxo}).

\newms\ is found to be faster than \ms\ for processes involving complicated matrix elements and several resonance decays. \newms\ also introduces the ability to launch several parallel processes, which further reduces the computation time. 

Some limitations nevertheless remain. Most importantly, the momentum reshuffling used in the simulation of off-shell effects is performed at fixed $\sqrt{\hat{s}}$, which leads to an incomplete coverage of the phase space (a dead zone) below the production threshold when no additional stable particles are present to absorb the recoil. This limitation is particularly relevant for $2\to 1$ processes. The new reshuffling framework is, however, quite flexible and we plan to investigate different reshuffling strategies in the future.

The spin-density-matrix implementation provides an essential foundation for improving the perturbative accuracy of resonance decays. Achieving NLO-accurate production spin correlations requires replacing the spin-summed NLO ingredients \cite{Frixione:1995ms,Frederix:2009yq} with the corresponding spin-density-matrix counterparts, which are currently unavailable. Such an extension would yield NLO-accurate spin information in the production of resonances. It would also provide a more robust framework for studies based on quantum-information observables \cite{Durupt:2025wuk,Frederix:2021zsh,Bernreuther:2015yna}.

The inclusion of NLO QCD corrections to the decay amplitudes and of real radiation in the decays constitutes a natural next step. Such an extension would, however, require a consistent NLO expansion of the resonance width, together with the introduction of a resonance-aware NLO+PS matching procedure \cite{Jezo:2015aia,Frixione:2023hwz}.

%% file: appendix.tex
\section{\newms\ commands}
\label{seq:commands}

\subsection{Decay command}

The \texttt{decay} command allows one to specify which particles should decay in \newms, and how. Here is one example:
\begin{lstlisting}[style=scipostcode,language={},caption={},label={}]
define mylep = e+ mu+           # multiparticle labels
decay t > w+ b, w+ > mylep vl   # a chain: comma-separated sub-decays
decay w+ > all all              # MG5's own labels work too
\end{lstlisting}

\noindent
Only final-state particles of the production events are decayed; a \code{decay}
line for a species that never appears is ignored.  A card with no \code{decay}
line at all is refused outright (\emph{Nothing to decay \ldots\ Please specify
some decay}).
Everything after the particle
name is handed to \mgamc, so the usual process syntax applies --- including
coupling-order restrictions (\code{QED=1}) and subdecays.

\paragraph{Several channels for one particle.}  Repeating \code{decay} for the
same species is what drives the multi-channel logic.
\begin{lstlisting}[style=scipostcode,language={},caption={},label={}]
decay w+ > l+ vl
decay w+ > j j       # Meaning depends on the number of W+ in the sample
\end{lstlisting}
Defining $n_{\rm part}$ as
the number of particles of that species in one event and $n_{\rm dec}$ as the
number of \code{decay} lines given for it, we have the following cases.

\begin{center}
\small
\begin{tabular}{@{}p{2.9cm}p{3.5cm}p{7.4cm}@{}}
\toprule
Case & Channel choice & Branching ratio folded into the weight\\
\midrule
$n_{\rm dec}=1$ & the only channel & $\Gamma_{\rm chan}/\Gamma_{\rm tot}$\\
$n_{\rm dec}=n_{\rm part}>1$ & \textbf{positional}: the $i$-th particle takes
  the $i$-th line & $\bigl(\prod_i\Gamma_i\bigr)/\Gamma_{\rm tot}^{\,n_{\rm part}}
  \times n_{\rm part}!/\prod_c m_c!$ --- the combinatorial factor counts the
  \emph{distinct} ways of assigning the channels to the identical parents, so two
  identical \code{decay} lines ($m_c=2$) are not counted twice\\
otherwise & \textbf{random}, $\propto$ the channels' cross sections & $\bigl(\sum_i\Gamma_i/\Gamma_{\rm tot}\bigr)^{n_{\rm part}}$\\
\bottomrule
\end{tabular}
\end{center}

\paragraph{Semi-leptonic samples, distinct parents ($t\bar t$).}  Here each
particle draws its channel independently, so simply giving \code{t} and
\code{t\textasciitilde} a leptonic and a hadronic line each would also produce
the fully leptonic and fully hadronic combinations.  The correlation is expressed
with the \code{@} suffix, which sorts the \code{decay} lines into \emph{groups}:
lines carrying the same tag are used together, so only the listed combinations
exist.

\begin{lstlisting}[style=scipostcode,language={},caption={},label={}]
decay t  > w+ b,  w+ > l+ vl   @1    # group 1: the lepton comes
decay t~ > w- b~, w- > j j     @1    #          from the top
decay t  > w+ b,  w+ > j j     @2
decay t~ > w- b~, w- > l- vl~  @2    # group 2: from the antitop
\end{lstlisting}

\noindent
The tags are honoured both by the legacy \code{madspin\_v1} path and by the
three density modes, by two different mechanisms.  Under \code{madspin\_v1} one
full matrix element is generated per group, so the sample carries both charge assignments with
their relative rates taken from the matrix elements.  In the density modes each
event instead draws one group, with probability $\mathrm{BR}_g/\sum_{g'}
\mathrm{BR}_{g'}$, and every particle then takes that group's channel; grouping
there forces the \texttt{joint} accept/reject (\opt{unweighting} is ignored, and logged
in the log file).  In both cases the sample contains no fully leptonic and no fully
hadronic event.  

\paragraph{Sample with events that do not all decay the same way.}  When the production
events do not all contain the same decaying species (as for a sample mixing
$Z+{\rm jet}$ and $W+{\rm jet}$), no single global branching
ratio exists.  In the density modes \newms\ then equalises them: it drops events,
per species, with
probability $1-{\rm BR}_{\rm pdg}/{\rm BR}_{\rm max}$ so the output stays
unweighted, and corrects the banner cross section afterwards from the number of
events actually kept.  

\begin{landscape}
\subsection{\newms\ options}
\label{sub:manual}
The following table describes all the various options (\code{set} command in the interface/cards) that can be used to tune the way \newms\ works.
The \textbf{Modes} column indicates in which \opt{spinmode} the option has an
effect.  ``all'' means every mode including \code{none}; \emph{density} is a
shorthand for the three density-matrix modes \code{PA}, \code{onshell} and
\code{madspin}/\code{full}.  An option set outside its modes is usually accepted
by the parser and then simply never read --- with no warning.

\begingroup
\small
\setlength{\tabcolsep}{5pt}
\begin{longtable}{@{}p{5.0cm}p{2.7cm}p{3.9cm}p{12.0cm}@{}}
\toprule
\textbf{Option} & \textbf{Default} & \textbf{Modes} & \textbf{Meaning}\\
\midrule
\endfirsthead
\multicolumn{4}{@{}l}{\emph{\small Quick reference (continued)}}\\[2pt]
\toprule
\textbf{Option} & \textbf{Default} & \textbf{Modes} & \textbf{Meaning}\\
\midrule
\endhead
\multicolumn{4}{@{}l}{\emph{Physics model of the decay}}\\*
\midrule
\opt{spinmode}        & \code{madspin} & all & Approximation used: \code{none}, \code{onshell}, \code{PA}, \code{madspin}/\code{full}, \code{madspin\_v1}, \code{onshell\_v1} (cf. Table~\ref{tab:madspin_modes_combined}).\\
\opt{BW\_cut}         & $-1$ (banner)  & \code{PA}, \code{madspin}/\code{full}, \code{madspin\_v1} & Number of widths a resonance may be off-shell.  Read from \code{bwcutoff} in the banner when left at $-1$ (and set to $15$ when there is no banner).\\
\opt{beampol}         & $[0,0]$ (run card) & all & Polarisation of each beam \emph{in percent}, $-100\ldots100$, exactly the run-card \code{polbeam1}/\code{polbeam2} convention; $0$ is unpolarised.\\
\midrule
\multicolumn{4}{@{}l}{\emph{partial polarization}}\\
\midrule
\opt{pure\_interference} & \code{''} & density & Which polarisation interference term to select (see App.~\ref{subsec:polarised_syntax}).\\
\opt{keep\_weight\_for\_polarization\_vector} & \code{[]} & density & Add the corresponding spin-state partial weight to the LHEF file (see App.~\ref{subsec:polarised_syntax}). The default (\code{[]}) corresponds to not including any extra weight. \\
\opt{keep\_weight\_for\_polarization\_fermion} & \code{[]} & density & Same for the spin-1/2 decaying particles.\\
\opt{frame\_id}       & 6 & all & Frame defining the helicity basis of the density matrices: a bit mask over the external legs (see App.~\ref{subsec:polarised_syntax}). Taken from the run card unless the MadSpin card sets it explicitly.\\

\tabgroup{Unweighting and maximum weight}
\opt{decay\_output} & \code{auto} & density & Choose between unweighted output (default in general) and weighted output (default for an interference-only setup).\\
\opt{Nevents\_for\_max\_weight} & 0 (banner) & all but \code{none} & Number of production events probed to estimate the bound.\\
\opt{max\_weight\_ps\_point} & 500     & all but \code{none} & Decay sets drawn per probed production event.\\
\opt{nb\_sigma}       & 0 (banner)     & all but \code{none} & Number of standard deviations added to the mean in the bound.\\
\opt{seed}            & 0 (random)     & all & Random seed; also seeds the forked workers deterministically.\\
\pagebreak
\tabgroup{How the accept/reject is organised}
\opt{unweighting}     & \code{auto}    & density & Which of the schemes below unweights the decays.  \code{auto}: \code{joint} for one or two decaying particles, \code{sequential} from three --- \code{sequential} throughout under \code{PA}/\code{onshell}, and \code{sequential} at every multiplicity when the production process carries a polarisation tag.\\
\quad \code{unweighting=joint}    &                & density & One test over the virtualities and every decay at once: the historical scheme.\\
\quad \code{unweighting=sequential} &              & density & Unweight the virtualities first, then one test per decaying particle, redrawing only the particle that was rejected.\\
\quad \code{unweighting=sequential\_global\_retry} &  & density & As \code{sequential}, but a rejected decay redraws the virtualities too.  Needs no running-width factor, at $2$--$3\times$ the cost; a cross-check rather than a default.\\
\opt{sequential\_spin\_order} & \code{2 3 1} & density & Spin order ($2S+1$) deciding which particle is accept/rejected first.  Efficiency only, never physics.\\

\opt{density\_keep\_jacobian} & \code{True} & \code{PA} & Fold the production reshuffling jacobian into the accept/reject weight, instead of applying the reshuffle as a post-acceptance kinematic dressing.\\
\tabgroup{Validation}
\opt{density\_debug}  & \code{False}   & density & Check the density-matrix weight against the full matrix element on every trial.  Debugging only --- very expensive.\\
\opt{density\_tolerance} & $10^{-4}$   & density, when \opt{density\_debug} & Relative tolerance of that check.\\
\opt{sequential\_debug} & \code{False} & density & On every accepted chain, recompute the joint weight and check that the product of the stage weights reproduces it.  A deterministic check of the decomposition, at roughly one joint trial per event.  Debugging only, and only for the up-front-mass schemes (\code{sequential}, \code{sequential\_global\_retry}).\\
\pagebreak
\tabgroup{Decay-event pools, performance, bookkeeping}
\opt{nb\_core}        & 0 (global)     & all but \code{madspin\_v1} & Worker processes for the unweighting, the max-weight scan and the decay generation.\\
\opt{decay\_event\_mult} & 1.0         & density, \code{onshell\_v1} & Multiplicative safety factor on the size of every decay pool.\\
\opt{ms\_dir}         & \code{''}      & all & Reuse (and cache into) a persistent MadSpin working directory.  Also sets \opt{curr\_dir}.\\
\opt{use\_old\_dir}   & \code{False}   & \code{madspin\_v1} & Reuse the existing directory content without the pickle checks (debugging only).\\
\opt{curr\_dir}       & \code{cwd}     & all & Working directory of the run.\\
\tabgroup{Input/output and normalisation}
\opt{input\_format}   & \code{auto}    & \code{none} & \code{lhe}, \code{hepmc}, \code{lhe\_no\_banner}.  Only the bridge mode reads a non-LHE input; every other mode needs an LHE file with its banner.\\
\opt{fixed\_order}    & \code{False}   & \code{onshell}, \code{onshell\_v1} & Handle NLO counter-events (event groups).  Forces the joint accept/reject.  Refused in \code{PA} and \code{madspin}/\code{full}, which reshuffle the production and which the counter-events cannot follow.\\
\opt{cross\_section}  & \{\}           & \code{none} (bridge) & Force the cross section per weight entry after the decay.\\
\opt{new\_wgt}        & \code{cross-section} & \code{none} (bridge) & How the event weight is rescaled: global BR (\code{cross-section}) or per-event BR (\code{BR}).\\
\opt{run\_card}       & \code{''}      & all but \code{madspin\_v1} & Path to (or in-line edition of) a run card applying cuts to the generated decay events.\\
\opt{identical\_particle\_\bk in\_prod\_\bk and\_decay} & \code{average} & density, \code{onshell\_v1} & Policy for the momentum-assignment ambiguity when a species appears in both production and decay: \code{average}, \code{max}, \code{first}, \code{crash}.\\
\midrule
\multicolumn{4}{@{}l}{\emph{\textbf{v1 only} --- read exclusively by the legacy code paths}}\\*
\midrule
\opt{onlyhelicity}    & \code{False}   & \code{madspin\_v1} & Only assign helicities, do not decay.  Setting it \emph{forces} \opt{spinmode} to \code{madspin\_v1}.\\
\opt{max\_running\_process} & 100      & \code{madspin\_v1} & Cap on simultaneously alive Fortran helper processes (open file descriptors).\\
\opt{global\_order\_coupling} & \code{''} & \code{madspin\_v1} & Coupling-order restriction appended to every generated decay process.\\
\bottomrule
\end{longtable}
\endgroup

\end{landscape}

\subsection{Polarisation-aware decay}
\label{subsec:polarised_syntax}

In \newms, it is possible to handle polarised decays. The simplest solution is to generate the production sample polarised (using the \mgamc\ syntax, like \code{\{T\}} for transverse polarisation \cite{BuarqueFranzosi:2019boy}). In that case \newms\ will automatically follow the frame chosen in the generation of the sample to evaluate the production density matrix, and will restrict the convolution of the density matrix to the corresponding elements.

Additionally, when decaying a non-polarised sample, one can ask \newms\ to add to the LHE output additional weights corresponding to the weight the event would have carried if a given polarisation had been selected. To select which additional weights are included within the LHE output, we introduced two options: \code{keep\_weight\_for\_polarization\_vector} and \code{keep\_weight\_for\_polarization\_fermion}; both take a list of polarisation restrictions, like
\begin{lstlisting}[style=scipostcode,language=Python,caption={},label={}]
    set keep_weight_for_polarization_vector [0, T]
    set keep_weight_for_polarization_fermion [L, R]
\end{lstlisting}
which for a process like the associated production of two tops with a $Z$ boson --- with all three particles decayed --- will lead to eight additional weights, since we take the Cartesian product of the possibilities per particle.
Not specifying these options means that no additional weight is written into the final file.
Since the polarisation fractions are not Lorentz invariant, the user has to pick a reference frame when using this option, by using \code{frame\_id}
\begin{lstlisting}[style=scipostcode,language=Python,caption={},label={}]
    set frame_id 6
\end{lstlisting}
where the value is a bit mask over the external legs: bit $n$ selects leg $n$, the 4-momenta of the selected legs are summed, and the event is boosted to the rest frame of that sum. The default, $6=2^1+2^2$, selects the two initial-state legs, i.e.\ the partonic centre-of-mass frame.

Finally, \newms\ also allows one to generate samples for the pure interference terms between helicity states. By construction, the cross-section associated with such a sample is identically zero: integrated over the full decay solid angle each decay density matrix reduces to the identity, which has no overlap with an off-diagonal block, so the contraction vanishes exactly. This cancellation follows from complete integration over the decay phase space, rather than from a convention in the implementation, and only observables that resolve the decay orientation are sensitive to the interference. For this we use a decomposition per particle where one can select either diagonal projectors $D^\pm$, or an off-diagonal pair between two sets of helicities:
\begin{lstlisting}[style=scipostcode,language=Python,caption={},label={}]
    set pure_interference t  = + -
    set pure_interference t~ = - -
    set pure_interference z  = 0 T
\end{lstlisting}
Here the code will return the helicity-flip term for the top, the pure negative-helicity (\code{L}) diagonal block for the anti-top, and the interference between the longitudinal and the transverse states of the $Z$. The two sides of an entry must be either disjoint (an interference block) or identical (a diagonal block); a partial overlap is refused, and at least one particle must carry a genuine interference block. One \code{set} line per particle is required.

\section{Handling several decays from the same production event}
\label{app:sequential}

In this appendix, we describe the new option \code{unweighting} of \newms, which determines how to handle the decay of multiple particles from the same production event.

In \newms, all those decays can be handled in a single accept/reject step,
the historical behaviour, where the
decayed event is associated with the weight
\begin{equation}
w =\frac{|\mcal^{\rm off}_{\rm full}|^2}{|\mcal^{\rm on}_{\rm prod}|^2 \prod_{i} |\mcal^{\rm on}_{D_i}|^2}\, w_{\rm shuffle},
\label{eq:joint}
\end{equation}
 where $\mcal^{\rm off}_{\rm full}$ is evaluated with on-shell momenta for the \code{PA} mode
and with off-shell momenta for the \code{madspin} mode, $|\mcal^{\rm on}_{\rm prod}|^2$ is the amplitude squared of the production events (on-shell), $|\mcal^{\rm on}_{D_i}|^2$ is the amplitude squared for the $i$-th decay (on-shell) and where $w_{\rm shuffle}$
collects the Jacobians of the reshuffling described below (it is absent in the
\code{onshell} mode, which reshuffles nothing). More precisely, it is defined as
\begin{equation}
w_{\rm shuffle} = J^{P}\prod_{k}w^{k}_{\rm shuffle},
\end{equation}
where $J^P$ is the Jacobian associated with the production reshuffling and where $w^{k}_{\rm shuffle}$ is the one associated with a dedicated decay, which can be split into two terms:
\begin{equation}
w^k_{\rm shuffle} = J^{BW}_k\, J^{D}_k,
\label{eq:wshuffle}
\end{equation}
where $J^{BW}_k$ is the Jacobian related to the kinematically available
phase-space for the reshuffling --- which can be impacted by thresholds, etc.
---  and $J^{D}_k$ is the Jacobian associated with the reshuffling of the decay $k$.

The \code{unweighting} option allows one either to keep this mode (with \code{unweighting=joint}) or to split the generation of the final kinematics into a multi-stage accept/reject mechanism which uses a joint conditional probability on the previously accepted decays, with a
different formula for the various modes.

\subsection{\texttt{onshell} mode}

In this mode, the two strategies that matter are \code{joint} and \code{sequential}.\footnote{With no virtuality to draw, \code{sequential\_global\_retry} is accepted but only costs more.} In the sequential mode, each decaying particle has its own accept/reject step to accept or reject that particular decay. The $k$-th decay is therefore associated with the weight:
\begin{equation}
w_k =\frac{|\mcal_{\rm{prod}\, +\,k\,-\rm{decay}}|^2}
          {|\mcal_{\rm prod\, +\,(k-1)\, decay}|^2\ |\mcal_{D_k}|^2},
\label{eq:onshell}
\end{equation}
where $|\mcal_{\rm{prod}\, +\,k\,-\rm{decay}}|^2$ stands for the production density matrix contracted with
the density matrices of the $k$ decays accepted so far, the decays not yet drawn
being replaced by the identity in helicity space --- that is, averaged over their
orientation. This is the quantity that makes the construction telescope: the denominator of
$w_k$ contains the same partially contracted quantity that appears in the numerator of $w_{k-1}$, so these factors cancel when the weights are multiplied.

The expected (and achieved) gain of such a method is that it allows one to
reduce the number of decay events to generate. However, strong correlations between
decays can lead the secondary unweighting to have a low efficiency and
negate such effects. In practice, the sequential mode is typically faster than
the joint approach in the \code{onshell} mode (see Sec.~\ref{sec:performance}) and is therefore the default unweighting mode.

\subsection{Off-shell mode (\texttt{PA} and \texttt{madspin})}

In the off-shell modes, the \code{sequential} unweighting needs to handle the reshuffling and the associated effects (threshold, impact of the Jacobian, off-shell matrix-element weight).
A strict sequential handling of each decay, where each stage generates its own virtuality, is only possible if the full chain of generation (mass+angle) is restarted whenever a decay is rejected, which is quite costly.

In \newms, we use two variations of such a strategy which allow one to increase the efficiency of the method (only one being exact in the infinite statistics limit).
In both cases, we start with a dedicated accept/reject step in which only the virtualities of all decaying particles are generated (following the individual Breit--Wigner distributions); the associated weight is:
\begin{equation}
w_{\rm mass} = \frac{|\mprod^{\rm off}|^2}{|\mprod^{\rm on}|^2}\ J^{P}
           \prod_k J^{BW}_k \prod_k Z_k(m_k).
\label{eq:wmass}
\end{equation}

Here $Z_k(m_k)$ is a factor designed to pre-encode the lineshape distortion induced by the subsequent decay of the $k$-th particle when it has an off-shell mass $m_k$.
\begin{equation}
  Z_k(m_k) \;=\; \int p_{\text{pool}}(\Omega)\, w_k(\Omega, m_k)\,\mathrm{d}\Omega
         \;=\; \frac{\int \mathrm{d}\Phi_{\text{off}}(m_k)\,|\mdec|^2}
                    {\int \mathrm{d}\Phi_{\text{on}}\,|\mdec|^2}
         \;=\; \frac{m_k}{M_k}\,\frac{\Gamma_k(m_k)}{\Gamma_k(M_k)} ,
  \label{eq:zk}
\end{equation}
where $M_k$ is the pole mass of the $k$-th particle. Technically, it is a look-ahead weight associated with the mass effect coming from the decay.

\begin{figure}[htbp]
  \centering
  \begin{tabular}{cc}
    \includegraphics[width=0.47\textwidth]{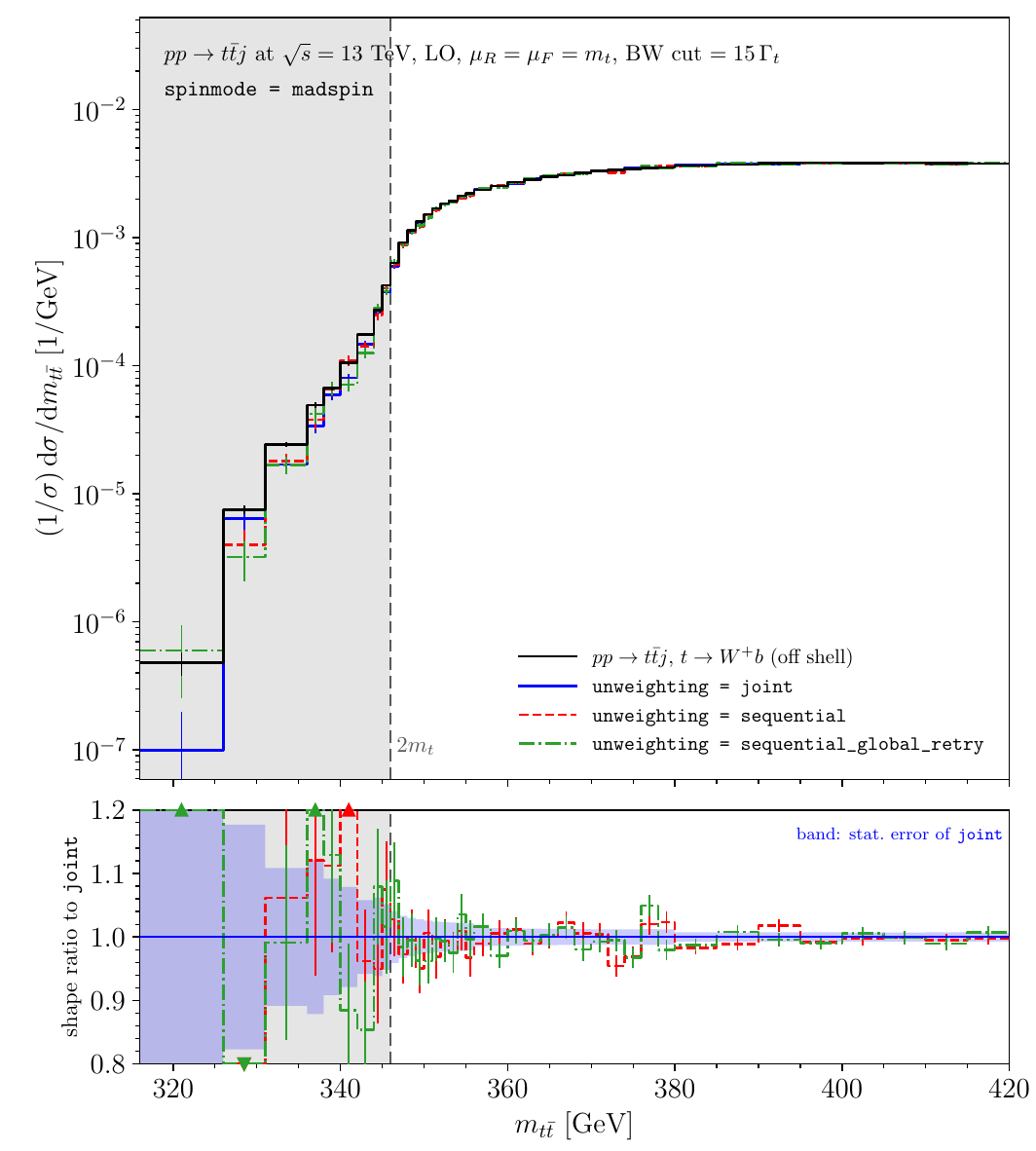} &
    \includegraphics[width=0.47\textwidth]{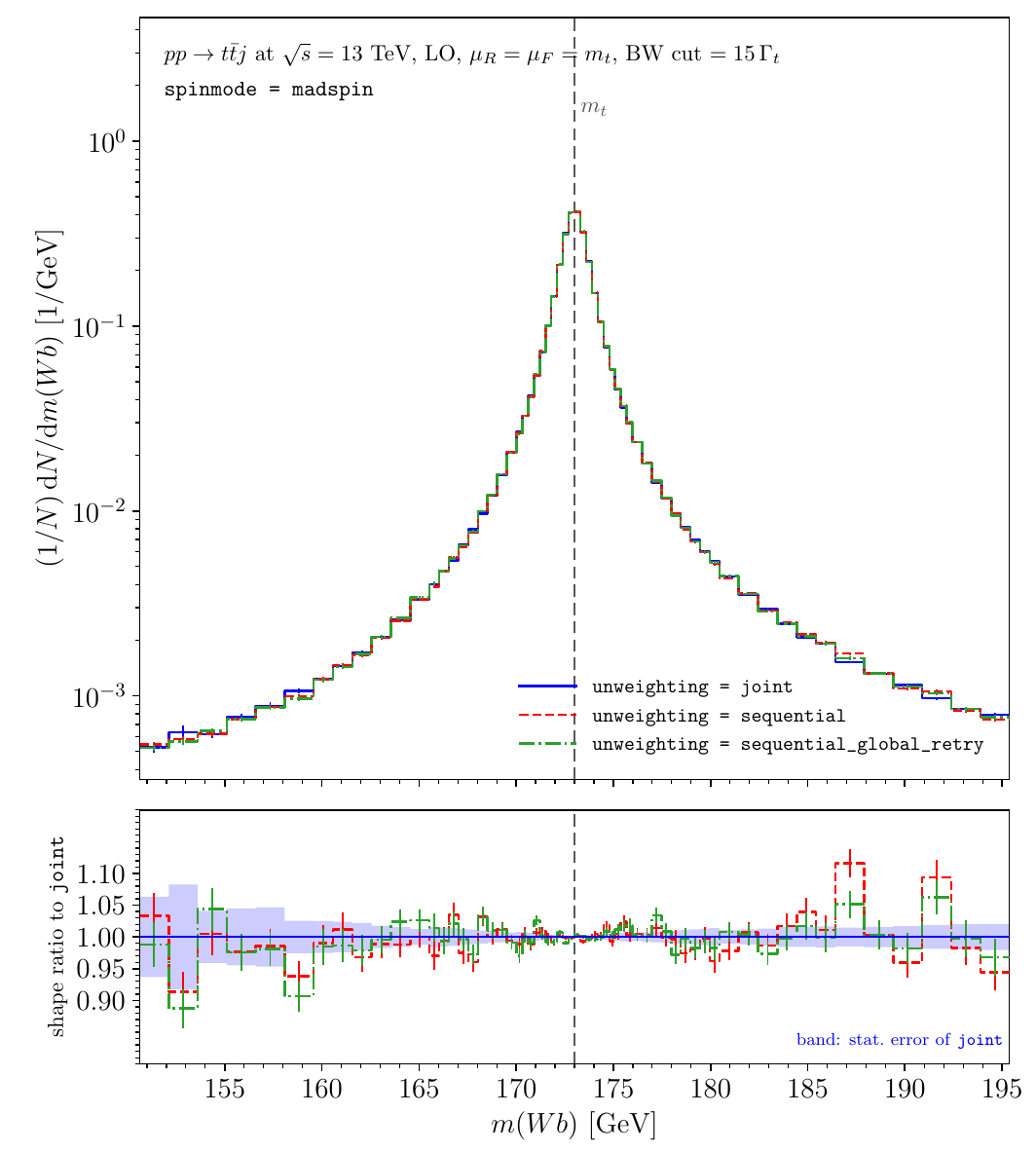} \\[1ex]
  \end{tabular}
  \caption{Invariant mass of the $t\bar{t}$ pair (left) and of the reconstructed top quark (right) in $pp\to t\bar{t}j$ events, comparing the \code{joint}, \code{sequential} and \code{sequential\_global\_retry} unweighting modes in the \code{madspin} spinmode.}
  \label{fig:unweighting-closure}
\end{figure}

Three quantities cancel out of $Z_k$ --- the production event, the other
particles' virtualities, and the decays already accepted --- so $Z_k$ depends on
that particle's virtuality alone and can be tabulated. We measure it from the
same scan that determines the maximum weights, at basically no additional cost, and fit its logarithm as a quadratic in $\ln(m/M)$.
The residual bias of the scheme is exactly the ratio of the tabulated factor to the true one; as can be seen in Fig.~\ref{fig:unweighting-closure}, this bias is not noticeable.

In the mode \code{unweighting=sequential\_global\_retry}, the $Z_k(m_k)$ is only a pre-conditioner to increase the efficiency of the computation. In that mode any subsequent failure to generate one of the successive decays will restart the full procedure and will regenerate a new set of off-shell masses for the particles; any mis-estimation in such a pre-conditioner will therefore be absorbed in the accept/reject procedure and will not create a distortion in the lineshape. On the contrary, \code{unweighting=sequential} will fully trust the $Z_k(m_k)$ value, and a failing decay will only restart the generation of the associated decay, keeping the same virtuality.

After the generation of the various invariant masses, both modes iterate over the decaying particles and associate each decay with a weight for the accept/reject phase.
We stress that such a weight is the same in both modes, but what is accepted/rejected is not the same, as described in the previous paragraph.
This weight for the $k$-th decay is
\begin{equation}
w_k =\frac{|\mcal^{\rm off}_{\rm{prod}\, +\,k\,-\rm{decay}}|^2}
          {|\mcal^{\rm off}_{\rm{prod}\, +\,(k-1)\,-\rm{decay}}|^2\ |\mcal^{\rm on}_{D_k}|^2}\ J^{D}_k .
\label{eq:madspin}
\end{equation}
Neither $J^{BW}_k$ nor $J^{P}$ appears here: both have been paid
in~\eqref{eq:wmass}, which is the purpose of the split.
Note that in the \code{PA} mode the same formula holds, even if all those matrix elements are on-shell.

In Fig.~\ref{fig:unweighting-closure}, we compare two reshuffling-sensitive variables for the associated production of a top quark-antiquark pair with a jet, using the various unweighting strategies. From this plot, we do not observe any bias in the \texttt{sequential} method compared to the \code{joint} method or the \code{sequential\_global\_retry} method, which are unbiased by construction. Rather, all the differences can be explained by the statistical error.

A key advantage of the \code{unweighting=sequential} mode is the fact that, with the virtualities fixed in advance, it is possible to compute a single density matrix for the production event and then convolve that single density matrix (differently) through all the stages of the decay, giving a sizeable computational benefit in terms of speed.

Table~\ref{tab:madspin-unweighting-eps} gives the cost of each scheme for
$p\,p\to t\bar t$.  Reading it per chain attempt separates two things that the
per-event normalisation conflates.  The mass stage is \emph{identical} under
\code{sequential} and \code{sequential\_global\_retry}.  What distinguishes the two schemes is not the cost
of a mass set but how many chain attempts each written event takes: under
\code{sequential\_global\_retry} a decay rejected at any slot discards the mass
set and restarts the chain, so the same stage simply runs about four times as
often.

The last two columns quantify the actual computational cost of a run. $N_{\mathrm{dec}}$ counts
decay events, which are cheap; $N^{P}_{\mathrm{off}}$ counts off-shell
production density matrices, which are not.  Under \code{madspin}, \code{sequential\_global\_retry} consumes
$12\,\%$ \emph{fewer} decay events than \code{joint} while evaluating $3.4$ times
as many production densities.  \code{sequential} is the only scheme that reuses
that object --- one density per drawn mass set, convolved through every decay
stage --- and against \code{joint} it is worth $23\,\%$.  Under \code{PA} and
\code{onshell} the column is identically zero: there the production density is
evaluated once per production event with on-shell momenta and cached, for every
scheme including \code{joint}.

One caveat applies when comparing $\epsilon_m$ across modes.  It measures a bound
against a mean weight, and the bound is not the same kind of object in every
row: under \code{PA} it is a per-event supremum that is never exceeded, while
under \code{madspin} it is extrapolated from a finite probe and occasionally is.

\begin{table}[htbp]
  \centering
  \begin{tabular}{llrrrrr}
    \toprule
    mode & unweighting & $\epsilon_m$ & $\epsilon_t$ & $\epsilon_{\bar t}$
         & $N_{\mathrm{dec}}$ & $N^{P}_{\mathrm{off}}$ \\
    \midrule
    \texttt{PA} & \texttt{joint} & \multicolumn{3}{c}{5.67} & 1\,134\,784 & 0 \\
    \texttt{PA} & \texttt{sequential} & 1.40 & 1.24 & 3.20 & 444\,598 & 0 \\
    \texttt{PA} & \texttt{sequential\_global\_retry} & 1.40 & 3.78 & 3.19 & 697\,292 & 0 \\
    \midrule
    \texttt{madspin} & \texttt{joint} & \multicolumn{3}{c}{4.29} & 858\,196 & 429\,098 \\
    \texttt{madspin} & \texttt{sequential} & 3.31 & 2.39 & 4.24 & 662\,516 & 331\,038 \\
    \texttt{madspin} & \texttt{sequential\_global\_retry} & 3.32 & 4.37 & 3.21 & 757\,657 & 1\,449\,758 \\
    \midrule
    \texttt{onshell} & \texttt{joint} & \multicolumn{3}{c}{3.19} & 638\,504 & 0 \\
    \texttt{onshell} & \texttt{sequential} & -- & 1.10 & 3.19 & 429\,374 & 0 \\
    \texttt{onshell} & \texttt{sequential\_global\_retry} & -- & 3.53 & 3.19 & 671\,733 & 0 \\
    \bottomrule
  \end{tabular}
  \caption{\newms\ unweighting cost for $p\,p\to t\bar t$ with both tops decayed
    leptonically, $100\,000$ production events and one seed shared by every row.
    Each $\epsilon$ is the mean number of points drawn to accept one \emph{per
    chain attempt}, so lower is better and the floor is~1; \texttt{joint} tests
    the virtualities and both decays at once and yields the single figure spanning
    the three columns.  $\epsilon_m$ is undefined for \texttt{onshell}, which
    samples no virtuality.  $N_{\mathrm{dec}}$ counts decay events consumed and
    $N^{P}_{\mathrm{off}}$ off-shell production density matrices evaluated.}
  \label{tab:madspin-unweighting-eps}
\end{table}

\section{Breakdown of the computational performance}
\label{app:profiling}
\begin{figure}[htp]
    \centering
    \includegraphics[width=\textwidth]{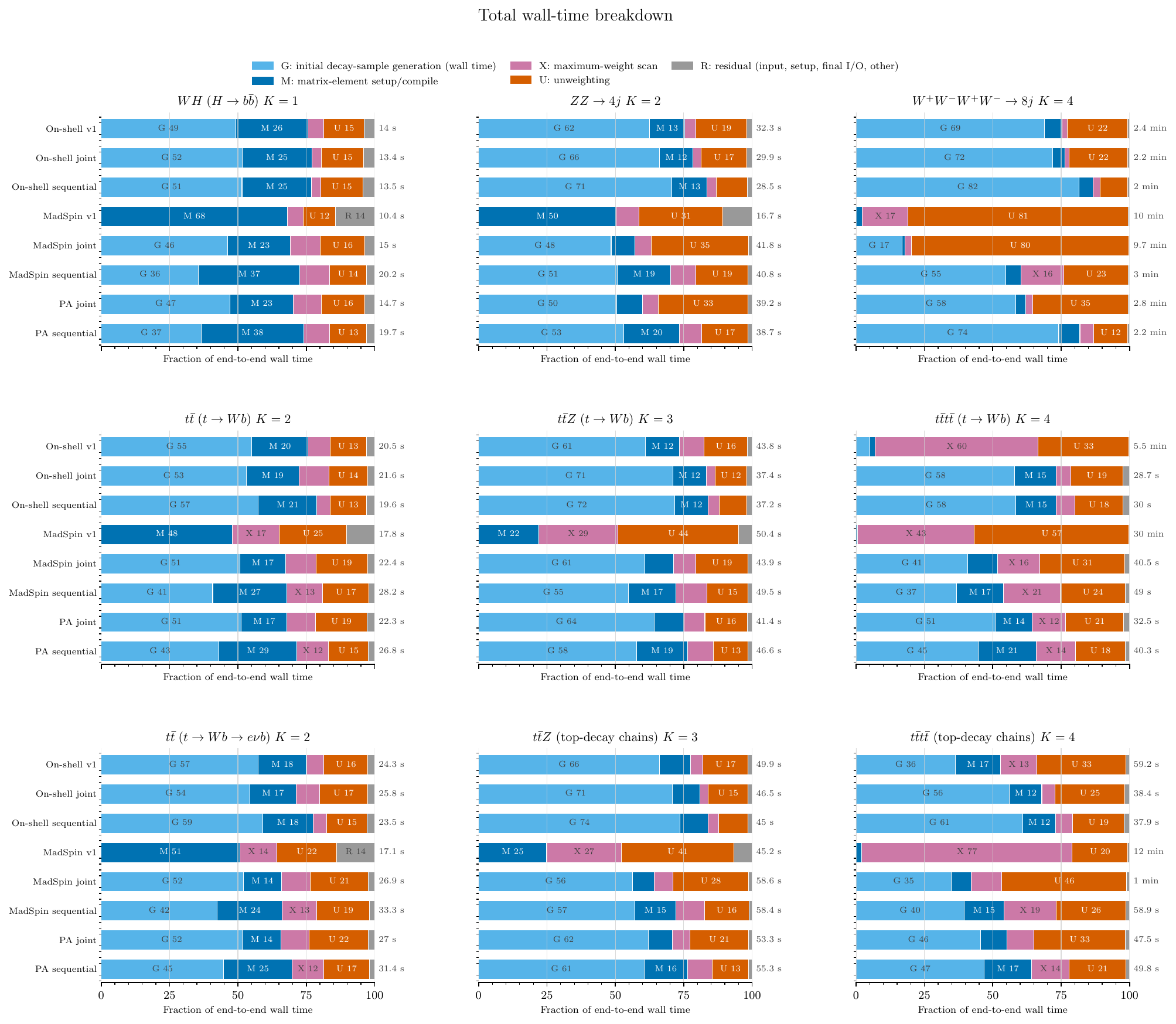}
    \caption{Total and relative wall-time spent in the different computation phases for various physics processes and run modes.}
    \label{fig:timing_phase_breakdown}
\end{figure}

\begin{figure}[htp]
    \centering
    \includegraphics[width=\textwidth]{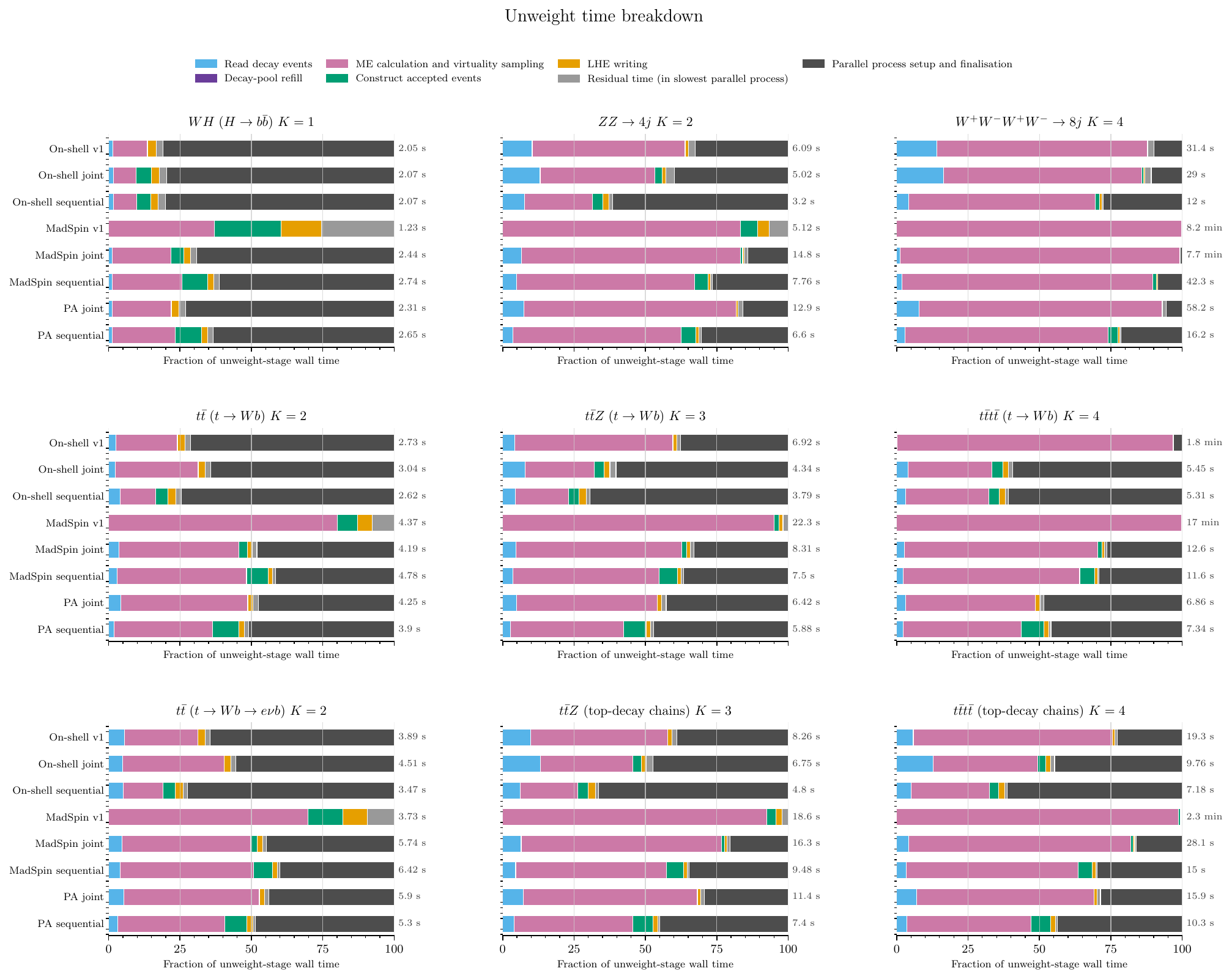}
    \caption{Wall-time spent in the various stages of the unweight step for various physics processes and run modes.}
    \label{fig:timing_unweight_breakdown}
\end{figure}

A breakdown of the time spent in each computational phase (decay-event generation, unweighting, etc.) is shown in \cref{fig:timing_phase_breakdown} and a further breakdown of the time spent in the various stages related to unweighting is shown in \cref{fig:timing_unweight_breakdown}.

The breakdown of the various computation stages is done using the following definitions:
\begin{itemize}
    \item \textbf{G: initial decay-sample generation}, the time required to generate the initial sample of decay events. This includes the matrix-element evaluations, phase-space integration, unweighting and I/O. This separate stage is absent in \texttt{madspin\_v1}, which generates candidate decays directly during the final unweighting.

    \item \textbf{M: matrix-element setup/compile}, the time required to generate, compile and load the process-dependent code used in the subsequent stages. In \texttt{madspin\_v1}, this includes code for the ``full'' (production-and-decay) matrix element, whose computation time grows relatively slowly with process complexity. In the other modes, it includes the code required to evaluate the production and decay spin-density matrices and their contractions. The costly repeated numerical evaluation of these quantities is not included here, but in \textbf{X} and \textbf{U}.

    \item \textbf{X: maximum-weight scan}, the time required to determine the maximum weight that is subsequently used by the rejection algorithm~\cite{Artoisenet:2012st}. Sequential rejection determines separate bounds for its successive rejection steps, while \texttt{joint}  rejection determines a joint bound.

    \item \textbf{U: unweighting}, the time required to produce the unweighted full event sample. This includes the evaluation of the full event weight, the accept/reject step, and, where appropriate (\texttt{madspin}/\texttt{PA} modes), the virtuality sampling and kinematic reshuffling. The unweighting time is affected by the cost of generating a trial event (and thus the time it takes to evaluate a ME, which scales rapidly with the complexity of the decay processes, since larger spin-density matrices need to be evaluated and contracted) and the unweighting efficiency, which typically becomes smaller as the complexity of the decay increases.

    \item \textbf{R: residual}, defined as the time that remains after subtracting the time associated with the phases above from the total wall-clock time. It contains time spent for setting up the parallel processes, splitting and combining the input/output files and other uninstrumented OS operations.
\end{itemize}

With the above definitions, we can make several observations from \cref{fig:timing_phase_breakdown}:
\begin{enumerate}
    \item The generation of a (deliberately oversized) initial pool of decay events dominates the computation time for all but the most complicated processes ($W^+W^-W^+W^-\to 8j$ and $t\bar{t}t\bar{t}\to W^+W^-W^+W^-b\bar{b}b\bar{b}$) which feature a very small unweighting efficiency, and consequently require several trials for an event to be accepted, each of which comes with a costly evaluation of a complicated matrix element. This phase will be significantly accelerated once \newms\ is interfaced to \textsc{MadSpace} in the context of the \textsc{MadGraph7} framework.
    \item The matrix-element setup and compilation step dominates in the \texttt{madspin\_v1} mode for almost all processes, since the  full (production-and-decay) ME has to be evaluated. The exceptions to this are the most complicated processes ($W^+W^-W^+W^-$ and $t\bar{t}t\bar{t}$), where the time for unweighting dominates.
    \item The time spent in the unweight step dominates for the most complicated processes that feature several decays and/or a very small unweighting efficiency, consequently requiring several trial events that involve costly ME evaluations.
    \item The \texttt{sequential} unweighting is generally faster in the \texttt{onshell} mode, because a rejected decay does not require the complete set of decay configurations to be redrawn. The advantage is therefore largest for processes with several decays or a low unweighting efficiency. For simple processes that already have a high unweighting efficiency in the \texttt{joint} mode, \texttt{sequential} unweighting does not lead to any speedup since any potential gain in unweighting is eliminated by the additional per-decay bookkeeping that the \texttt{sequential} mode requires.
    \item In the \texttt{madspin} mode, the relative performance of the two unweighting modes is process dependent. The \texttt{sequential} unweighting first accepts or rejects the complete set of resonance virtualities and subsequently unweights each decay separately, reusing the accepted masses, reshuffled production kinematics and production density matrix. This introduces additional setup and maximum-weight-estimation costs, which dominate for simpler processes. However, for more demanding processes, avoiding the repeated rejection of the complete decay configuration substantially reduces the unweighting time. The gain is particularly large for $W^+W^-W^+W^-$, for which the joint unweighting efficiency is only about $2\%$.
\end{enumerate}

Comparing the breakdown for the different processes and run modes (\cref{fig:timing_unweight_breakdown}) during the unweighting step, we observe that the ME calculation dominates the unweighting time by far for all but the simplest processes, where the parallel process orchestration is dominant. This is explained by the different scaling of the two computation stages with the final-state complexity. Parallel-process orchestration introduces an approximately fixed cost, which is largely independent of the process complexity. This leads to an apparently large effect for processes that are inherently fast to generate.

Comparing the \texttt{onshell} with the \texttt{madspin} modes gives an estimate of the impact of the additional computation time incurred by the modelling of the off-shell effects. For \texttt{joint} rejection, the unweighting time is increased by up to a factor of 3.

%% file: bibliography.bib
@article{Alwall:2014hca,
    author = "Alwall, J. and Frederix, R. and Frixione, S. and Hirschi, V. and Maltoni, F. and Mattelaer, O. and Shao, H. -S. and Stelzer, T. and Torrielli, P. and Zaro, M.",
    title = "{The automated computation of tree-level and next-to-leading order differential cross sections, and their matching to parton shower simulations}",
    eprint = "1405.0301",
    archivePrefix = "arXiv",
    primaryClass = "hep-ph",
    reportNumber = "CERN-PH-TH-2014-064, CP3-14-18, LPN14-066, MCNET-14-09, ZU-TH-14-14",
    doi = "10.1007/JHEP07(2014)079",
    journal = "JHEP",
    volume = "07",
    pages = "079",
    year = "2014"
}

@article{Hirschi:2011pa,
    author = "Hirschi, Valentin and Frederix, Rikkert and Frixione, Stefano and Garzelli, Maria Vittoria and Maltoni, Fabio and Pittau, Roberto",
    title = "{Automation of one-loop QCD corrections}",
    eprint = "1103.0621",
    archivePrefix = "arXiv",
    primaryClass = "hep-ph",
    reportNumber = "CERN-PH-TH-2011-031, CP3-11-07, ZU-TH-01-11",
    doi = "10.1007/JHEP05(2011)044",
    journal = "JHEP",
    volume = "05",
    pages = "044",
    year = "2011"
}

@article{Frederix:2009yq,
    author = "Frederix, Rikkert and Frixione, Stefano and Maltoni, Fabio and Stelzer, Tim",
    title = "{Automation of next-to-leading order computations in QCD: The FKS subtraction}",
    eprint = "0908.4272",
    archivePrefix = "arXiv",
    primaryClass = "hep-ph",
    reportNumber = "CERN-TH-2009-157, CP3-09-39",
    doi = "10.1088/1126-6708/2009/10/003",
    journal = "JHEP",
    volume = "10",
    pages = "003",
    year = "2009"
}

@article{Re:2014uua,
    author = "Re, Emanuele",
    title = "{Reaching NNLOPS accuracy with POWHEG and MiNLO}",
    eprint = "1401.2944",
    archivePrefix = "arXiv",
    primaryClass = "hep-ph",
    doi = "10.1393/ncc/i2014-11747-9",
    journal = "Nuovo Cim. C",
    volume = "037",
    number = "02",
    pages = "137--142",
    year = "2014"
}

@article{Campbell:2021svd,
    author = {Campbell, John M. and H{\"o}che, Stefan and Li, Hai Tao and Preuss, Christian T. and Skands, Peter},
    title = "{Towards NNLO+PS matching with sector showers}",
    eprint = "2108.07133",
    archivePrefix = "arXiv",
    primaryClass = "hep-ph",
    reportNumber = "FERMILAB-PUB-21-365-T, MCNET-21-15",
    doi = "10.1016/j.physletb.2022.137614",
    journal = "Phys. Lett. B",
    volume = "836",
    pages = "137614",
    year = "2023"
}

@article{Monni:2019whf,
    author = "Monni, Pier Francesco and Nason, Paolo and Re, Emanuele and Wiesemann, Marius and Zanderighi, Giulia",
    title = "{MiNNLO$_{PS}$: a new method to match NNLO QCD to parton showers}",
    eprint = "1908.06987",
    archivePrefix = "arXiv",
    primaryClass = "hep-ph",
    reportNumber = "CERN-TH-2019-117, LAPTH-042/19, MPP-2019-177",
    doi = "10.1007/JHEP05(2020)143",
    journal = "JHEP",
    volume = "05",
    pages = "143",
    year = "2020",
    note = "[Erratum: JHEP 02, 031 (2022)]"
}

@article{Bellm:2015jjp,
    author = "Bellm, Johannes and others",
    title = "{Herwig 7.0/Herwig++ 3.0 release note}",
    eprint = "1512.01178",
    archivePrefix = "arXiv",
    primaryClass = "hep-ph",
    reportNumber = "CERN-PH-TH-2015-289, MAN-HEP-2015-15, IFJPAN-IV-2015-13, KA-TP-18-2015, DCPT-15-142, MCNET-15-28, IPPP-15-71, HERWIG-2015-01",
    doi = "10.1140/epjc/s10052-016-4018-8",
    journal = "Eur. Phys. J. C",
    volume = "76",
    number = "4",
    pages = "196",
    year = "2016"
}

@article{Sherpa:2019gpd,
    author = "Bothmann, Enrico and others",
    collaboration = "Sherpa",
    title = "{Event Generation with Sherpa 2.2}",
    eprint = "1905.09127",
    archivePrefix = "arXiv",
    primaryClass = "hep-ph",
    reportNumber = "FERMILAB-PUB-19-218-T, SLAC-PUB-17433, IPPP/19/42, MCNET-19-11",
    doi = "10.21468/SciPostPhys.7.3.034",
    journal = "SciPost Phys.",
    volume = "7",
    number = "3",
    pages = "034",
    year = "2019"
}

@article{Richardson:2001df,
    author = "Richardson, Peter",
    title = "{Spin correlations in Monte Carlo simulations}",
    eprint = "hep-ph/0110108",
    archivePrefix = "arXiv",
    reportNumber = "CAVENDISH-HEP-2001-13, DAMTP-2001-83",
    doi = "10.1088/1126-6708/2001/11/029",
    journal = "JHEP",
    volume = "11",
    pages = "029",
    year = "2001"
}

@article{Frixione:2007zp,
    author = "Frixione, Stefano and Laenen, Eric and Motylinski, Patrick and Webber, Bryan R.",
    title = "{Angular correlations of lepton pairs from vector boson and top quark decays in Monte Carlo simulations}",
    eprint = "hep-ph/0702198",
    archivePrefix = "arXiv",
    reportNumber = "CAVENDISH-HEP-07-01, GEF-TH-09-2007, ITP-UU-07-10, NIKHEF-2007-004",
    doi = "10.1088/1126-6708/2007/04/081",
    journal = "JHEP",
    volume = "04",
    pages = "081",
    year = "2007"
}

@article{Artoisenet:2012st,
    author = "Artoisenet, Pierre and Frederix, Rikkert and Mattelaer, Olivier and Rietkerk, Robbert",
    title = "{Automatic spin-entangled decays of heavy resonances in Monte Carlo simulations}",
    eprint = "1212.3460",
    archivePrefix = "arXiv",
    primaryClass = "hep-ph",
    reportNumber = "NIKHEF-2012-021, CERN-PH-TH-2012-329",
    doi = "10.1007/JHEP03(2013)015",
    journal = "JHEP",
    volume = "03",
    pages = "015",
    year = "2013"
}

@article{Hirschi:2015iia,
    author = "Hirschi, Valentin and Mattelaer, Olivier",
    title = "{Automated event generation for loop-induced processes}",
    eprint = "1507.00020",
    archivePrefix = "arXiv",
    primaryClass = "hep-ph",
    reportNumber = "IPPP-15-35, DCPT-15-70, MCNET-15-14",
    doi = "10.1007/JHEP10(2015)146",
    journal = "JHEP",
    volume = "10",
    pages = "146",
    year = "2015"
}

@article{Brivio:2020onw,
    author = "Brivio, Ilaria",
    title = "{SMEFTsim 3.0 {\textemdash} a practical guide}",
    eprint = "2012.11343",
    archivePrefix = "arXiv",
    primaryClass = "hep-ph",
    doi = "10.1007/JHEP04(2021)073",
    journal = "JHEP",
    volume = "04",
    pages = "073",
    year = "2021"
}

@article{Brivio:2017btx,
    author = "Brivio, Ilaria and Jiang, Yun and Trott, Michael",
    title = "{The SMEFTsim package, theory and tools}",
    eprint = "1709.06492",
    archivePrefix = "arXiv",
    primaryClass = "hep-ph",
    doi = "10.1007/JHEP12(2017)070",
    journal = "JHEP",
    volume = "12",
    pages = "070",
    year = "2017"
}

@article{Brivio:2017vri,
    author = "Brivio, Ilaria and Trott, Michael",
    title = "{The Standard Model as an Effective Field Theory}",
    eprint = "1706.08945",
    archivePrefix = "arXiv",
    primaryClass = "hep-ph",
    doi = "10.1016/j.physrep.2018.11.002",
    journal = "Phys. Rept.",
    volume = "793",
    pages = "1--98",
    year = "2019"
}

@article{Aguilar-Saavedra:2008nuh,
    author = "Aguilar-Saavedra, J. A.",
    title = "{A Minimal set of top anomalous couplings}",
    eprint = "0811.3842",
    archivePrefix = "arXiv",
    primaryClass = "hep-ph",
    doi = "10.1016/j.nuclphysb.2008.12.012",
    journal = "Nucl. Phys. B",
    volume = "812",
    pages = "181--204",
    year = "2009"
}

@article{CMS:2019nrx,
    author = "Sirunyan, Albert M and others",
    collaboration = "CMS",
    title = "{Measurement of the top quark polarization and $\mathrm{t\bar{t}}$ spin correlations using dilepton final states in proton-proton collisions at $\sqrt{s} =$ 13 TeV}",
    eprint = "1907.03729",
    archivePrefix = "arXiv",
    primaryClass = "hep-ex",
    reportNumber = "CMS-TOP-18-006, CERN-EP-2019-073, CMS-PAS-TOP-18-006",
    doi = "10.1103/PhysRevD.100.072002",
    journal = "Phys. Rev. D",
    volume = "100",
    number = "7",
    pages = "072002",
    year = "2019"
}

@article{CMS:2019kzp,
    author = "Sirunyan, Albert M and others",
    collaboration = "CMS",
    title = "{Measurement of the top quark forward-backward production asymmetry and the anomalous chromoelectric and chromomagnetic moments in pp collisions at $ \sqrt{s} $ = 13 TeV}",
    eprint = "1912.09540",
    archivePrefix = "arXiv",
    primaryClass = "hep-ex",
    reportNumber = "CMS-TOP-15-018, CERN-EP-2019-270",
    doi = "10.1007/JHEP06(2020)146",
    journal = "JHEP",
    volume = "06",
    pages = "146",
    year = "2020"
}

@article{Oleari:2010nx,
    author = "Oleari, Carlo",
    editor = {Bl{\"u}mlein, Johannes and Moch, Sven-Olaf and Riemann, Tord},
    title = "{The POWHEG-BOX}",
    eprint = "1007.3893",
    archivePrefix = "arXiv",
    primaryClass = "hep-ph",
    doi = "10.1016/j.nuclphysbps.2010.08.016",
    journal = "Nucl. Phys. B Proc. Suppl.",
    volume = "205-206",
    pages = "36--41",
    year = "2010"
}

@article{Banfi:2023mhz,
    author = {Banfi, Andrea and Ferrario Ravasio, Silvia and J{\"a}ger, Barbara and Karlberg, Alexander and Reichenbach, Felix and Zanderighi, Giulia},
    title = "{A POWHEG generator for deep inelastic scattering}",
    eprint = "2309.02127",
    archivePrefix = "arXiv",
    primaryClass = "hep-ph",
    reportNumber = "CERN-TH-2023-152",
    doi = "10.1007/JHEP02(2024)023",
    journal = "JHEP",
    volume = "02",
    pages = "023",
    year = "2024"
}

@article{Hoeche:2011fd,
    author = "Hoeche, Stefan and Krauss, Frank and Schonherr, Marek and Siegert, Frank",
    title = "{A critical appraisal of NLO+PS matching methods}",
    eprint = "1111.1220",
    archivePrefix = "arXiv",
    primaryClass = "hep-ph",
    reportNumber = "SLAC-PUB-14661, IPPP-11-67, DCPT-11-134, LPN11-58, FR-PHENO-2011-019, MCNET-11-24",
    doi = "10.1007/JHEP09(2012)049",
    journal = "JHEP",
    volume = "09",
    pages = "049",
    year = "2012"
}

@article{Sherpa:2024mfk,
    author = "Bothmann, Enrico and others",
    collaboration = "Sherpa",
    title = "{Event generation with Sherpa 3}",
    eprint = "2410.22148",
    archivePrefix = "arXiv",
    primaryClass = "hep-ph",
    reportNumber = "IPPP/24/67, LTH-1385, FERMILAB-PUB-24-0748-T, ZU-TH 51/24, MCNET-24-17, CERN-TH-2024-171",
    doi = "10.1007/JHEP12(2024)156",
    journal = "JHEP",
    volume = "12",
    pages = "156",
    year = "2024"
}

@article{Alwall:2011uj,
    author = "Alwall, Johan and Herquet, Michel and Maltoni, Fabio and Mattelaer, Olivier and Stelzer, Tim",
    title = "{MadGraph 5 : Going Beyond}",
    eprint = "1106.0522",
    archivePrefix = "arXiv",
    primaryClass = "hep-ph",
    reportNumber = "FERMILAB-PUB-11-448-T",
    doi = "10.1007/JHEP06(2011)128",
    journal = "JHEP",
    volume = "06",
    pages = "128",
    year = "2011"
}

@article{Meade:2007js,
    author = "Meade, Patrick and Reece, Matthew",
    title = "{BRIDGE: Branching ratio inquiry / decay generated events}",
    eprint = "hep-ph/0703031",
    archivePrefix = "arXiv",
    month = "3",
    year = "2007"
}

@article{Denner:2024xul,
    author = "Denner, Ansgar and Lombardi, Daniele and Schwan, Christopher",
    title = "{Double-pole approximation for leading-order semi-leptonic vector-boson scattering at the LHC}",
    eprint = "2406.12301",
    archivePrefix = "arXiv",
    primaryClass = "hep-ph",
    doi = "10.1007/JHEP08(2024)146",
    journal = "JHEP",
    volume = "08",
    pages = "146",
    year = "2024"
}

@article{Denner:2000bj,
    author = "Denner, Ansgar and Dittmaier, S. and Roth, M. and Wackeroth, D.",
    title = "{Electroweak radiative corrections to e+ e- ---{\ensuremath{>}} W W ---{\ensuremath{>}} 4 fermions in double pole approximation: The RACOONWW approach}",
    eprint = "hep-ph/0006307",
    archivePrefix = "arXiv",
    reportNumber = "BI-TP-2000-06, LU-ITP-2000-003, PSI-PR-00-11, UR-1604",
    doi = "10.1016/S0550-3213(00)00511-3",
    journal = "Nucl. Phys. B",
    volume = "587",
    pages = "67--117",
    year = "2000"
}

@article{Uhlemann:2008pm,
    author = "Uhlemann, C. F. and Kauer, N.",
    title = "{Narrow-width approximation accuracy}",
    eprint = "0807.4112",
    archivePrefix = "arXiv",
    primaryClass = "hep-ph",
    doi = "10.1016/j.nuclphysb.2009.01.022",
    journal = "Nucl. Phys. B",
    volume = "814",
    pages = "195--211",
    year = "2009"
}

@article{Berdine:2007uv,
    author = "Berdine, D. and Kauer, N. and Rainwater, D.",
    title = "{Breakdown of the Narrow Width Approximation for New Physics}",
    eprint = "hep-ph/0703058",
    archivePrefix = "arXiv",
    doi = "10.1103/PhysRevLett.99.111601",
    journal = "Phys. Rev. Lett.",
    volume = "99",
    pages = "111601",
    year = "2007"
}

@article{Carrivale:2025mjy,
    author = "Carrivale, Costanza and others",
    title = "{Precise standard-model predictions for polarised Z-boson pair production and decay at the LHC}",
    eprint = "2505.09686",
    archivePrefix = "arXiv",
    primaryClass = "hep-ph",
    reportNumber = "COMETA-2025-16, IFJPAN-IV-2025-10, IPPP/25/22, MCNET-25-07",
    doi = "10.1140/epjc/s10052-025-15069-x",
    journal = "Eur. Phys. J. C",
    volume = "85",
    number = "11",
    pages = "1342",
    year = "2025"
}

@article{Durupt:2025wuk,
    author = "Durupt, Valentin and Maltoni, Fabio and Mattelaer, Olivier",
    title = "{Automated computation of spin-density matrices and quantum observables for collider physics}",
    eprint = "2510.17730",
    archivePrefix = "arXiv",
    primaryClass = "hep-ph",
    reportNumber = "IRMP-CP3-25-34",
    doi = "10.1007/JHEP04(2026)103",
    journal = "JHEP",
    volume = "04",
    pages = "103",
    year = "2026"
}

@article{KLEISS1986359,
    author = "Kleiss, R. and Stirling, W. James and Ellis, S. D.",
    title = "{A New Monte Carlo Treatment of Multiparticle Phase Space at High-energies}",
    reportNumber = "CERN-TH-4299/85",
    doi = "10.1016/0010-4655(86)90119-0",
    journal = "Comput. Phys. Commun.",
    volume = "40",
    pages = "359",
    year = "1986"
}

@article{Heimel:2022wyj,
    author = "Heimel, Theo and Winterhalder, Ramon and Butter, Anja and Isaacson, Joshua and Krause, Claudius and Maltoni, Fabio and Mattelaer, Olivier and Plehn, Tilman",
    title = "{MadNIS - Neural multi-channel importance sampling}",
    eprint = "2212.06172",
    archivePrefix = "arXiv",
    primaryClass = "hep-ph",
    reportNumber = "IRMP-CP3-22-56, MCNET-22-22, FERMILAB-PUB-22-915-T",
    doi = "10.21468/SciPostPhys.15.4.141",
    journal = "SciPost Phys.",
    volume = "15",
    number = "4",
    pages = "141",
    year = "2023"
}

@article{Heimel:2023ngj,
    author = "Heimel, Theo and Huetsch, Nathan and Maltoni, Fabio and Mattelaer, Olivier and Plehn, Tilman and Winterhalder, Ramon",
    title = "{The MadNIS reloaded}",
    eprint = "2311.01548",
    archivePrefix = "arXiv",
    primaryClass = "hep-ph",
    reportNumber = "IRMP-CP3-23-56, MCNET-23-12",
    doi = "10.21468/SciPostPhys.17.1.023",
    journal = "SciPost Phys.",
    volume = "17",
    number = "1",
    pages = "023",
    year = "2024"
}

@article{Heimel:2024wph,
    author = "Heimel, Theo and Mattelaer, Olivier and Plehn, Tilman and Winterhalder, Ramon",
    title = "{Differentiable MadNIS-Lite}",
    eprint = "2408.01486",
    archivePrefix = "arXiv",
    primaryClass = "hep-ph",
    reportNumber = "IRMP-CP3-24-23",
    doi = "10.21468/SciPostPhys.18.1.017",
    journal = "SciPost Phys.",
    volume = "18",
    number = "1",
    pages = "017",
    year = "2025"
}

@article{DeCrescenzo:2026tsp,
    author = "De Crescenzo, Giovanni and Villadamigo, Javier Mari{\~n}o and Elmer, Nina and Heimel, Theo and Plehn, Tilman and Winterhalder, Ramon and Zaro, Marco",
    title = "{MadNIS at NLO}",
    eprint = "2603.22407",
    archivePrefix = "arXiv",
    primaryClass = "hep-ph",
    reportNumber = "TIF-UNIMI-2026-2, IRMP-CP3-26-06, MCNET-26-04",
    month = "3",
    year = "2026"
}

@inproceedings{Amdahl,
    author = "Amdahl, Gene M.",
    title = "{Validity of the single processor approach to achieving large scale computing capabilities}",
    booktitle = "{Proceedings of the April 18--20, 1967, Spring Joint Computer Conference (AFIPS '67)}",
    pages = "483--485",
    publisher = "ACM",
    address = "New York, NY, USA",
    year = "1967",
    doi = "10.1145/1465482.1465560"
}

@article{Frederix:2018nkq,
    author = "Frederix, R. and Frixione, S. and Hirschi, V. and Pagani, D. and Shao, H. -S. and Zaro, M.",
    title = "{The automation of next-to-leading order electroweak calculations}",
    eprint = "1804.10017",
    archivePrefix = "arXiv",
    primaryClass = "hep-ph",
    reportNumber = "Nikhef/2018-015, TUM-HEP-1138/18, NIKHEF-2018-015, TUM-HEP-1138-18",
    doi = "10.1007/JHEP11(2021)085",
    journal = "JHEP",
    volume = "07",
    pages = "185",
    year = "2018",
    note = "[Erratum: JHEP 11, 085 (2021)]"
}

@article{Maltoni:2002qb,
    author = "Maltoni, Fabio and Stelzer, Tim",
    title = "{MadEvent: Automatic event generation with MadGraph}",
    eprint = "hep-ph/0208156",
    archivePrefix = "arXiv",
    doi = "10.1088/1126-6708/2003/02/027",
    journal = "JHEP",
    volume = "02",
    pages = "027",
    year = "2003"
}

@article{BuarqueFranzosi:2019boy,
    author = "Buarque Franzosi, Diogo and Mattelaer, Olivier and Ruiz, Richard and Shil, Sujay",
    title = "{Automated predictions from polarized matrix elements}",
    eprint = "1912.01725",
    archivePrefix = "arXiv",
    primaryClass = "hep-ph",
    reportNumber = "CP3-19-52, VBSCan-PUB-09-19, MCnet-19-24, IP/BBSR/2019-10",
    doi = "10.1007/JHEP04(2020)082",
    journal = "JHEP",
    volume = "04",
    pages = "082",
    year = "2020"
}

@article{Mattelaer:2016ynf,
    author = "Mattelaer, Olivier and Mitra, Manimala and Ruiz, Richard",
    title = "{Automated Neutrino Jet and Top Jet Predictions at Next-to-Leading-Order with Parton Shower Matching in Effective Left-Right Symmetric Models}",
    eprint = "1610.08985",
    archivePrefix = "arXiv",
    primaryClass = "hep-ph",
    reportNumber = "IPPP-16-102, CP3-16-51",
    month = "10",
    year = "2016"
}

@article{Murayama:1992gi,
    author = "Murayama, H. and Watanabe, I. and Hagiwara, Kaoru",
    title = "{HELAS: HELicity amplitude subroutines for Feynman diagram evaluations}",
    reportNumber = "KEK-91-11",
    month = "1",
    year = "1992"
}

@article{deAquino:2011ub,
    author = "de Aquino, Priscila and Link, William and Maltoni, Fabio and Mattelaer, Olivier and Stelzer, Tim",
    title = "{ALOHA: Automatic Libraries Of Helicity Amplitudes for Feynman Diagram Computations}",
    eprint = "1108.2041",
    archivePrefix = "arXiv",
    primaryClass = "hep-ph",
    reportNumber = "CP3-11-26",
    doi = "10.1016/j.cpc.2012.05.004",
    journal = "Comput. Phys. Commun.",
    volume = "183",
    pages = "2254--2263",
    year = "2012"
}

@article{Jezo:2015aia,
    author = "Je{\v{z}}o, Tom{\'a}{\v{s}} and Nason, Paolo",
    title = "{On the Treatment of Resonances in Next-to-Leading Order Calculations Matched to a Parton Shower}",
    eprint = "1509.09071",
    archivePrefix = "arXiv",
    primaryClass = "hep-ph",
    doi = "10.1007/JHEP12(2015)065",
    journal = "JHEP",
    volume = "12",
    pages = "065",
    year = "2015"
}

@article{Frederix:2021zsh,
    author = "Frederix, Rikkert and Tsinikos, Ioannis and Vitos, Timea",
    title = "{Probing the spin correlations of $t{\bar{t}} $ production at NLO QCD+EW}",
    eprint = "2105.11478",
    archivePrefix = "arXiv",
    primaryClass = "hep-ph",
    reportNumber = "LU-TP 21-13",
    doi = "10.1140/epjc/s10052-021-09612-9",
    journal = "Eur. Phys. J. C",
    volume = "81",
    number = "9",
    pages = "817",
    year = "2021"
}

@article{Bernreuther:2015yna,
    author = "Bernreuther, Werner and Heisler, Dennis and Si, Zong-Guo",
    title = "{A set of top quark spin correlation and polarization observables for the LHC: Standard Model predictions and new physics contributions}",
    eprint = "1508.05271",
    archivePrefix = "arXiv",
    primaryClass = "hep-ph",
    reportNumber = "TTK-15-16",
    doi = "10.1007/JHEP12(2015)026",
    journal = "JHEP",
    volume = "12",
    pages = "026",
    year = "2015"
}

@article{Argyropoulos:2024yxo,
    author = "Argyropoulos, Spyros and Haisch, Ulrich and Kalaitzidou, Ilia",
    title = "{Novel collider signatures in the type-I 2HDM+a model}",
    eprint = "2404.05704",
    archivePrefix = "arXiv",
    primaryClass = "hep-ph",
    reportNumber = "MPP-2024-66",
    doi = "10.1007/JHEP07(2024)263",
    journal = "JHEP",
    volume = "07",
    pages = "263",
    year = "2024"
}

@article{Campbell:2010ff,
    author = "Campbell, John M. and Ellis, R. K.",
    editor = {Bl{\"u}mlein, Johannes and Moch, Sven-Olaf and Riemann, Tord},
    title = "{MCFM for the Tevatron and the LHC}",
    eprint = "1007.3492",
    archivePrefix = "arXiv",
    primaryClass = "hep-ph",
    reportNumber = "FERMILAB-CONF-10-244-T",
    doi = "10.1016/j.nuclphysbps.2010.08.011",
    journal = "Nucl. Phys. B Proc. Suppl.",
    volume = "205-206",
    pages = "10--15",
    year = "2010"
}

@mastersthesis{Heirebaudt:2023msc,
    author = "Heirebaudt, Quentin",
    title = "{MadSpin Density: a new approach to the computation of matrix elements}",
    school = "Universit\'e catholique de Louvain",
    year = "2023",
    url = "https://hdl.handle.net/2078.2/35143"
}

@article{Buckley:2011ms,
    author = "Buckley, Andy and others",
    title = "{General-purpose event generators for LHC physics}",
    eprint = "1101.2599",
    archivePrefix = "arXiv",
    primaryClass = "hep-ph",
    reportNumber = "CAVENDISH-HEP-10-21, CERN-PH-TH-2010-298, DCPT-10-202, IPPP-10-101, KA-TP-40-2010, LU-TP-10-28, MAN-HEP-2010-23, SLAC-PUB-14333, HD-THEP-10-24, MCNET-11-01",
    doi = "10.1016/j.physrep.2011.03.005",
    journal = "Phys. Rept.",
    volume = "504",
    pages = "145--233",
    year = "2011"
}

@article{Nason:2004rx,
    author = "Nason, Paolo",
    title = "{A New method for combining NLO QCD with shower Monte Carlo algorithms}",
    eprint = "hep-ph/0409146",
    archivePrefix = "arXiv",
    reportNumber = "BICOCCA-FT-04-11",
    doi = "10.1088/1126-6708/2004/11/040",
    journal = "JHEP",
    volume = "11",
    pages = "040",
    year = "2004"
}

@article{Frixione:2007vw,
    author = "Frixione, Stefano and Nason, Paolo and Oleari, Carlo",
    title = "{Matching NLO QCD computations with Parton Shower simulations: the POWHEG method}",
    eprint = "0709.2092",
    archivePrefix = "arXiv",
    primaryClass = "hep-ph",
    reportNumber = "BICOCCA-FT-07-9, GEF-TH-21-2007",
    doi = "10.1088/1126-6708/2007/11/070",
    journal = "JHEP",
    volume = "11",
    pages = "070",
    year = "2007"
}

@article{Alioli:2010xd,
    author = "Alioli, Simone and Nason, Paolo and Oleari, Carlo and Re, Emanuele",
    title = "{A general framework for implementing NLO calculations in shower Monte Carlo programs: the POWHEG BOX}",
    eprint = "1002.2581",
    archivePrefix = "arXiv",
    primaryClass = "hep-ph",
    reportNumber = "DESY-10-018, SFB-CPP-10-22, IPPP-10-11, DCPT-10-22",
    doi = "10.1007/JHEP06(2010)043",
    journal = "JHEP",
    volume = "06",
    pages = "043",
    year = "2010"
}

@article{Stuart:1991xk,
    author = "Stuart, Robin G.",
    title = "{Gauge invariance, analyticity and physical observables at the Z0 resonance}",
    reportNumber = "CINVESTAV-FIS-1-91",
    doi = "10.1016/0370-2693(91)90653-8",
    journal = "Phys. Lett. B",
    volume = "262",
    pages = "113--119",
    year = "1991"
}

@article{Aeppli:1993rs,
    author = "Aeppli, Andre and van Oldenborgh, Geert Jan and Wyler, Daniel",
    title = "{Unstable particles in one loop calculations}",
    eprint = "hep-ph/9312212",
    archivePrefix = "arXiv",
    reportNumber = "PSI-PR-93-22, ZU-TH-30-93, TTP-93-33",
    doi = "10.1016/0550-3213(94)90195-3",
    journal = "Nucl. Phys. B",
    volume = "428",
    pages = "126--146",
    year = "1994"
}

@article{Frixione:2002ik,
    author = "Frixione, Stefano and Webber, Bryan R.",
    title = "{Matching NLO QCD computations and parton shower simulations}",
    eprint = "hep-ph/0204244",
    archivePrefix = "arXiv",
    reportNumber = "CAVENDISH-HEP-02-01, LAPTH-905-02, GEF-TH-2-2002",
    doi = "10.1088/1126-6708/2002/06/029",
    journal = "JHEP",
    volume = "06",
    pages = "029",
    year = "2002"
}

@article{Alwall:2006yp,
    author = "Alwall, J. and others",
    title = "{A Standard format for Les Houches event files}",
    eprint = "hep-ph/0609017",
    archivePrefix = "arXiv",
    reportNumber = "FERMILAB-PUB-06-337-T, CERN-LCGAPP-2006-03",
    doi = "10.1016/j.cpc.2006.11.010",
    journal = "Comput. Phys. Commun.",
    volume = "176",
    pages = "300--304",
    year = "2007"
}

@article{Denner:1999gp,
    author = "Denner, Ansgar and Dittmaier, S. and Roth, M. and Wackeroth, D.",
    title = "{Predictions for all processes e+ e- ---{\ensuremath{>}} 4 fermions + gamma}",
    eprint = "hep-ph/9904472",
    archivePrefix = "arXiv",
    reportNumber = "BI-TP-99-10, PSI-PR-99-12",
    doi = "10.1016/S0550-3213(99)00437-X",
    journal = "Nucl. Phys. B",
    volume = "560",
    pages = "33--65",
    year = "1999"
}

@article{Denner:2005fg,
    author = "Denner, Ansgar and Dittmaier, S. and Roth, M. and Wieders, L. H.",
    title = "{Electroweak corrections to charged-current e+ e- ---{\ensuremath{>}} 4 fermion processes: Technical details and further results}",
    eprint = "hep-ph/0505042",
    archivePrefix = "arXiv",
    reportNumber = "MPP-2005-23, PSI-PR-05-05",
    doi = "10.1016/j.nuclphysb.2011.09.001",
    journal = "Nucl. Phys. B",
    volume = "724",
    pages = "247--294",
    year = "2005",
    note = "[Erratum: Nucl.Phys.B 854, 504--507 (2012)]"
}

@article{Glover:1988fe,
    author = "Glover, E. W. Nigel and van der Bij, J. J.",
    title = "{VECTOR BOSON PAIR PRODUCTION VIA GLUON FUSION}",
    reportNumber = "CERN-TH-5247/88",
    doi = "10.1016/0370-2693(89)91099-X",
    journal = "Phys. Lett. B",
    volume = "219",
    pages = "488--492",
    year = "1989"
}

@article{Grzadkowski:2010es,
    author = "Grzadkowski, B. and Iskrzynski, M. and Misiak, M. and Rosiek, J.",
    title = "{Dimension-Six Terms in the Standard Model Lagrangian}",
    eprint = "1008.4884",
    archivePrefix = "arXiv",
    primaryClass = "hep-ph",
    reportNumber = "IFT-9-2010, TTP10-35",
    doi = "10.1007/JHEP10(2010)085",
    journal = "JHEP",
    volume = "10",
    pages = "085",
    year = "2010"
}

@article{ATLAS:2019zrq,
    author = "Aaboud, Morad and others",
    collaboration = "ATLAS",
    title = "{Measurements of top-quark pair spin correlations in the $e\mu$ channel at $\sqrt{s} = 13$ TeV using $pp$ collisions in the ATLAS detector}",
    eprint = "1903.07570",
    archivePrefix = "arXiv",
    primaryClass = "hep-ex",
    reportNumber = "CERN-EP-2019-034",
    doi = "10.1140/epjc/s10052-020-8181-6",
    journal = "Eur. Phys. J. C",
    volume = "80",
    number = "8",
    pages = "754",
    year = "2020"
}

@article{Caola:2015psa,
    author = {Caola, Fabrizio and Melnikov, Kirill and R{\"o}ntsch, Raoul and Tancredi, Lorenzo},
    title = "{QCD corrections to ZZ production in gluon fusion at the LHC}",
    eprint = "1509.06734",
    archivePrefix = "arXiv",
    primaryClass = "hep-ph",
    reportNumber = "TTP15-035, CERN-PH-TH-2015-226, FERMILAB-PUB-15-398-T",
    doi = "10.1103/PhysRevD.92.094028",
    journal = "Phys. Rev. D",
    volume = "92",
    number = "9",
    pages = "094028",
    year = "2015"
}

@article{Denner:2010jp,
    author = "Denner, A. and Dittmaier, S. and Kallweit, S. and Pozzorini, S.",
    title = "{NLO QCD corrections to WWbb production at hadron colliders}",
    eprint = "1012.3975",
    archivePrefix = "arXiv",
    primaryClass = "hep-ph",
    reportNumber = "FR-PHENO-2010-041, PSI-PR-10-14, ZH-TH-19-10",
    doi = "10.1103/PhysRevLett.106.052001",
    journal = "Phys. Rev. Lett.",
    volume = "106",
    pages = "052001",
    year = "2011"
}

@article{Bevilacqua:2010qb,
    author = "Bevilacqua, Giuseppe and Czakon, Michal and van Hameren, Andreas and Papadopoulos, Costas G. and Worek, Malgorzata",
    title = "{Complete off-shell effects in top quark pair hadroproduction with leptonic decay at next-to-leading order}",
    eprint = "1012.4230",
    archivePrefix = "arXiv",
    primaryClass = "hep-ph",
    reportNumber = "TTK-10-56, IFJPAN-IV-2010-9, WUB-10-26",
    doi = "10.1007/JHEP02(2011)083",
    journal = "JHEP",
    volume = "02",
    pages = "083",
    year = "2011"
}

@article{Ballestrero:2017bxn,
    author = "Ballestrero, Alessandro and Maina, Ezio and Pelliccioli, Giovanni",
    title = "{$W$ boson polarization in vector boson scattering at the LHC}",
    eprint = "1710.09339",
    archivePrefix = "arXiv",
    primaryClass = "hep-ph",
    reportNumber = "VBSCAN-PUB-02-17",
    doi = "10.1007/JHEP03(2018)170",
    journal = "JHEP",
    volume = "03",
    pages = "170",
    year = "2018"
}

@article{Denner:2021csi,
    author = "Denner, Ansgar and Pelliccioli, Giovanni",
    title = "{NLO EW and QCD corrections to polarized ZZ production in the four-charged-lepton channel at the LHC}",
    eprint = "2107.06579",
    archivePrefix = "arXiv",
    primaryClass = "hep-ph",
    doi = "10.1007/JHEP10(2021)097",
    journal = "JHEP",
    volume = "10",
    pages = "097",
    year = "2021"
}

@article{Frixione:1995ms,
    author = "Frixione, S. and Kunszt, Z. and Signer, A.",
    title = "{Three jet cross-sections to next-to-leading order}",
    eprint = "hep-ph/9512328",
    archivePrefix = "arXiv",
    reportNumber = "SLAC-PUB-7073, SLAC-PUB-95-7073, ETH-TH-95-42",
    doi = "10.1016/0550-3213(96)00110-1",
    journal = "Nucl. Phys. B",
    volume = "467",
    pages = "399--442",
    year = "1996"
}

@article{Kilian:2007gr,
    author = "Kilian, Wolfgang and Ohl, Thorsten and Reuter, Jurgen",
    title = "{WHIZARD: Simulating Multi-Particle Processes at LHC and ILC}",
    eprint = "0708.4233",
    archivePrefix = "arXiv",
    primaryClass = "hep-ph",
    reportNumber = "DESY-11-126, EDINBURGH-2010-36, FR-PHENO-2010-037, SI-HEP-2010-18",
    doi = "10.1140/epjc/s10052-011-1742-y",
    journal = "Eur. Phys. J. C",
    volume = "71",
    pages = "1742",
    year = "2011"
}

@article{Moretti:2001zz,
    author = "Moretti, Mauro and Ohl, Thorsten and Reuter, Jurgen",
    editor = "Behnke, Ties and Bertolucci, S. and Heuer, Rolf D. and Miller, David and Richard, Francois and Settles, Ron and Telnov, V. and Zerwas, Peter",
    title = "{O'Mega: An Optimizing matrix element generator}",
    eprint = "hep-ph/0102195",
    archivePrefix = "arXiv",
    reportNumber = "IKDA-2001-06, LC-TOOL-2001-040",
    pages = "1981--2009",
    month = "2",
    year = "2001"
}

@article{Lange:2001uf,
    author = "Lange, D. J.",
    editor = "Erhan, S. and Schlein, P. and Rozen, Y.",
    title = "{The EvtGen particle decay simulation package}",
    doi = "10.1016/S0168-9002(01)00089-4",
    journal = "Nucl. Instrum. Meth. A",
    volume = "462",
    pages = "152--155",
    year = "2001"
}

@article{Jadach:1990mz,
    author = "Jadach, Stanislaw and Kuhn, Johann H. and Was, Zbigniew",
    title = "{TAUOLA: A Library of Monte Carlo programs to simulate decays of polarized tau leptons}",
    reportNumber = "CERN-TH-5856-90",
    doi = "10.1016/0010-4655(91)90038-M",
    journal = "Comput. Phys. Commun.",
    volume = "64",
    pages = "275--299",
    year = "1990"
}

@article{Jadach:1993hs,
    author = "Jadach, S. and Was, Z. and Decker, R. and Kuhn, Johann H.",
    title = "{The tau decay library TAUOLA: Version 2.4}",
    reportNumber = "CERN-TH-6793-93",
    doi = "10.1016/0010-4655(93)90061-G",
    journal = "Comput. Phys. Commun.",
    volume = "76",
    pages = "361--380",
    year = "1993"
}

@article{Davidson:2010rw,
    author = "Davidson, N. and Nanava, G. and Przedzinski, T. and Richter-Was, E. and Was, Z.",
    title = "{Universal Interface of TAUOLA Technical and Physics Documentation}",
    eprint = "1002.0543",
    archivePrefix = "arXiv",
    primaryClass = "hep-ph",
    reportNumber = "IFJPAN-IV-2009-10",
    doi = "10.1016/j.cpc.2011.12.009",
    journal = "Comput. Phys. Commun.",
    volume = "183",
    pages = "821--843",
    year = "2012"
}

@article{Czyczula:2012ny,
    author = "Czyczula, Z. and Przedzinski, T. and Was, Z.",
    title = "{TauSpinner Program for Studies on Spin Effect in tau Production at the LHC}",
    eprint = "1201.0117",
    archivePrefix = "arXiv",
    primaryClass = "hep-ph",
    reportNumber = "CERN-PH-TH-20011-307",
    doi = "10.1140/epjc/s10052-012-1988-z",
    journal = "Eur. Phys. J. C",
    volume = "72",
    pages = "1988",
    year = "2012"
}

@article{Denner:2019vbn,
    author = "Denner, Ansgar and Dittmaier, Stefan",
    title = "{Electroweak Radiative Corrections for Collider Physics}",
    eprint = "1912.06823",
    archivePrefix = "arXiv",
    primaryClass = "hep-ph",
    reportNumber = "FR-PHENO-019",
    doi = "10.1016/j.physrep.2020.04.001",
    journal = "Phys. Rept.",
    volume = "864",
    pages = "1--163",
    year = "2020"
}

@article{Denner:2020bcz,
    author = "Denner, Ansgar and Pelliccioli, Giovanni",
    title = "{Polarized electroweak bosons in ${\bf \text{W}^+\text{W}^-}$ production at the LHC including NLO QCD effects}",
    eprint = "2006.14867",
    archivePrefix = "arXiv",
    primaryClass = "hep-ph",
    doi = "10.1007/JHEP09(2020)164",
    journal = "JHEP",
    volume = "09",
    pages = "164",
    year = "2020"
}

@article{Denner:2020eck,
    author = "Denner, Ansgar and Pelliccioli, Giovanni",
    title = "{NLO QCD predictions for doubly-polarized WZ production at the LHC}",
    eprint = "2010.07149",
    archivePrefix = "arXiv",
    primaryClass = "hep-ph",
    doi = "10.1016/j.physletb.2021.136107",
    journal = "Phys. Lett. B",
    volume = "814",
    pages = "136107",
    year = "2021"
}

@article{Czakon:2010td,
    author = "Czakon, M.",
    title = "{A novel subtraction scheme for double-real radiation at NNLO}",
    eprint = "1005.0274",
    archivePrefix = "arXiv",
    primaryClass = "hep-ph",
    doi = "10.1016/j.physletb.2010.08.036",
    journal = "Phys. Lett. B",
    volume = "693",
    pages = "259--268",
    year = "2010"
}

@article{Czakon:2014oma,
    author = "Czakon, M. and Heymes, D.",
    title = "{Four-dimensional formulation of the sector-improved residue subtraction scheme}",
    eprint = "1408.2500",
    archivePrefix = "arXiv",
    primaryClass = "hep-ph",
    reportNumber = "TTK-14-16",
    doi = "10.1016/j.nuclphysb.2014.11.006",
    journal = "Nucl. Phys. B",
    volume = "890",
    pages = "152--227",
    year = "2014"
}

@article{Czakon:2019tmo,
    author = "Czakon, Micha{\l} and van Hameren, Andreas and Mitov, Alexander and Poncelet, Rene",
    title = "{Single-jet inclusive rates with exact color at $ \mathcal{O} $ ($ {\alpha}_s^4 $)}",
    eprint = "1907.12911",
    archivePrefix = "arXiv",
    primaryClass = "hep-ph",
    doi = "10.1007/JHEP10(2019)262",
    journal = "JHEP",
    volume = "10",
    pages = "262",
    year = "2019"
}

@article{Poncelet:2021jmj,
    author = "Poncelet, Rene and Popescu, Andrei",
    title = "{NNLO QCD study of polarised W$^{+}$W$^{-}$ production at the LHC}",
    eprint = "2102.13583",
    archivePrefix = "arXiv",
    primaryClass = "hep-ph",
    reportNumber = "CAVENDISH--HEP--21/03, CAVENDISH-HEP-21/03",
    doi = "10.1007/JHEP07(2021)023",
    journal = "JHEP",
    volume = "07",
    pages = "023",
    year = "2021"
}

@article{Pellen:2021vpi,
    author = "Pellen, Mathieu and Poncelet, Rene and Popescu, Andrei",
    title = "{Polarised W+j production at the LHC: a study at NNLO QCD accuracy}",
    eprint = "2109.14336",
    archivePrefix = "arXiv",
    primaryClass = "hep-ph",
    reportNumber = "CAVENDISH-HEP-21/13, FR-PHENO-2021-11",
    doi = "10.1007/JHEP02(2022)160",
    journal = "JHEP",
    volume = "02",
    pages = "160",
    year = "2022"
}

@article{Pellen:2022fom,
    author = "Pellen, Mathieu and Poncelet, Rene and Popescu, Andrei and Vitos, Timea",
    title = "{Angular coefficients in $\hbox {W}+\hbox {j}$ production at the LHC with high precision}",
    eprint = "2204.12394",
    archivePrefix = "arXiv",
    primaryClass = "hep-ph",
    reportNumber = "CAVENDISH--HEP--22/04, FR-PHENO-2022-04, Lund-22-24",
    doi = "10.1140/epjc/s10052-022-10641-1",
    journal = "Eur. Phys. J. C",
    volume = "82",
    number = "8",
    pages = "693",
    year = "2022"
}

@article{Le:2022lrp,
    author = "Le, Duc Ninh and Baglio, Julien",
    title = "{Doubly-polarized WZ hadronic cross sections at NLO QCD + EW accuracy}",
    eprint = "2203.01470",
    archivePrefix = "arXiv",
    primaryClass = "hep-ph",
    reportNumber = "CERN-TH-2022-027",
    doi = "10.1140/epjc/s10052-022-10887-9",
    journal = "Eur. Phys. J. C",
    volume = "82",
    number = "10",
    pages = "917",
    year = "2022"
}

@article{Le:2022ppa,
    author = "Le, Duc Ninh and Baglio, Julien and Dao, Thi Nhung",
    title = "{Doubly-polarized WZ hadronic production at NLO QCD+EW: calculation method and further results}",
    eprint = "2208.09232",
    archivePrefix = "arXiv",
    primaryClass = "hep-ph",
    reportNumber = "CERN-TH-2022-110",
    doi = "10.1140/epjc/s10052-022-11032-2",
    journal = "Eur. Phys. J. C",
    volume = "82",
    number = "12",
    pages = "1103",
    year = "2022"
}

@article{Dao:2023kwc,
    author = "Dao, Thi Nhung and Le, Duc Ninh",
    title = "{NLO electroweak corrections to doubly-polarized $W^+W^-$ production at the LHC}",
    eprint = "2311.17027",
    archivePrefix = "arXiv",
    primaryClass = "hep-ph",
    doi = "10.1140/epjc/s10052-024-12579-y",
    journal = "Eur. Phys. J. C",
    volume = "84",
    number = "3",
    pages = "244",
    year = "2024"
}

@article{Dao:2024ffg,
    author = "Dao, Thi Nhung and Le, Duc Ninh",
    title = "{Polarized $W^+W^-$ pairs at the LHC: Effects from bottom-quark induced processes at NLO QCD + EW}",
    eprint = "2409.06396",
    archivePrefix = "arXiv",
    primaryClass = "hep-ph",
    doi = "10.1140/epjc/s10052-025-13835-5",
    journal = "Eur. Phys. J. C",
    volume = "85",
    number = "1",
    pages = "108",
    year = "2025"
}

@article{ATLAS:2023zrv,
    author = "Aad, Georges and others",
    collaboration = "ATLAS",
    title = "{Evidence of pair production of longitudinally polarised vector bosons and study of CP properties in ZZ {\textrightarrow} 4{\ensuremath{\ell}} events with the ATLAS detector at $ \sqrt{s} $ = 13 TeV}",
    eprint = "2310.04350",
    archivePrefix = "arXiv",
    primaryClass = "hep-ex",
    reportNumber = "CERN-EP-2023-199",
    doi = "10.1007/JHEP12(2023)107",
    journal = "JHEP",
    volume = "12",
    pages = "107",
    year = "2023"
}

@article{Denner:2022riz,
    author = "Denner, Ansgar and Haitz, Christoph and Pelliccioli, Giovanni",
    title = "{NLO QCD corrections to polarized diboson production in semileptonic final states}",
    eprint = "2211.09040",
    archivePrefix = "arXiv",
    primaryClass = "hep-ph",
    doi = "10.1103/PhysRevD.107.053004",
    journal = "Phys. Rev. D",
    volume = "107",
    number = "5",
    pages = "053004",
    year = "2023"
}

@article{Denner:2023ehn,
    author = "Denner, Ansgar and Haitz, Christoph and Pelliccioli, Giovanni",
    title = "{NLO EW corrections to polarised W+W{\ensuremath{-}} production and decay at the LHC}",
    eprint = "2311.16031",
    archivePrefix = "arXiv",
    primaryClass = "hep-ph",
    reportNumber = "COMETA-2023-01, MPP-2023-268",
    doi = "10.1016/j.physletb.2024.138539",
    journal = "Phys. Lett. B",
    volume = "850",
    pages = "138539",
    year = "2024"
}

@article{Denner:2024tlu,
    author = "Denner, Ansgar and Haitz, Christoph and Pelliccioli, Giovanni",
    title = "{NLO EW and QCD corrections to polarised same-sign WW scattering at the LHC}",
    eprint = "2409.03620",
    archivePrefix = "arXiv",
    primaryClass = "hep-ph",
    reportNumber = "COMETA-2024-22, MPP-2024-178",
    doi = "10.1007/JHEP11(2024)115",
    journal = "JHEP",
    volume = "11",
    pages = "115",
    year = "2024"
}

@article{Ball:2012cx,
    author = "Ball, Richard D. and others",
    title = "{Parton distributions with LHC data}",
    eprint = "1207.1303",
    archivePrefix = "arXiv",
    primaryClass = "hep-ph",
    reportNumber = "EDINBURGH-2012-08, IFUM-FT-997, FR-PHENO-2012-014, RWTH-TTK-12-25, CERN-PH-TH-2012-037, SFB-CPP-12-47",
    doi = "10.1016/j.nuclphysb.2012.10.003",
    journal = "Nucl. Phys. B",
    volume = "867",
    pages = "244--289",
    year = "2013"
}

@article{Ball:2013hta,
    author = "Ball, Richard D. and Bertone, Valerio and Carrazza, Stefano and Del Debbio, Luigi and Forte, Stefano and Guffanti, Alberto and Hartland, Nathan P. and Rojo, Juan",
    collaboration = "NNPDF",
    title = "{Parton distributions with QED corrections}",
    eprint = "1308.0598",
    archivePrefix = "arXiv",
    primaryClass = "hep-ph",
    reportNumber = "EDINBURGH-2013-20, FR-PHENO-2013-008, CERN-PH-TH-2013-075, Edinburgh 2013/20, IFUM-1014-FT, FR-PHENO-2013-008,
  CERN-PH-TH/2013-075",
    doi = "10.1016/j.nuclphysb.2013.10.010",
    journal = "Nucl. Phys. B",
    volume = "877",
    pages = "290--320",
    year = "2013"
}

@article{Catani:2001cc,
    author = "Catani, S. and Krauss, F. and Kuhn, R. and Webber, B. R.",
    title = "{QCD matrix elements + parton showers}",
    eprint = "hep-ph/0109231",
    archivePrefix = "arXiv",
    reportNumber = "CERN-TH-2000-367, CAVENDISH-HEP-00-03",
    doi = "10.1088/1126-6708/2001/11/063",
    journal = "JHEP",
    volume = "11",
    pages = "063",
    year = "2001"
}

@article{Basu:2025zds,
    author = "Basu, Trina and Ruiz, Richard",
    title = "{The Four Polarizations of the $W$ at High Energies}",
    eprint = "2512.10015",
    archivePrefix = "arXiv",
    primaryClass = "hep-ph",
    reportNumber = "IFJPAN-IV-2025-23, COMETA-2025-45",
    month = "12",
    year = "2025"
}

@article{Basu:2026zmz,
    author = "Basu, Trina and Ruiz, Richard",
    title = "{Polarization interference in exclusive $V+$jets at all orders in $\alpha_s$}",
    eprint = "2606.28496",
    archivePrefix = "arXiv",
    primaryClass = "hep-ph",
    reportNumber = "IFJPAN-IV-2026-13, COMETA-2026-09",
    month = "6",
    year = "2026"
}

@article{Javurkova:2024bwa,
    author = "Javurkova, Martina and Ruiz, Richard and de S{\'a}, Rafael Coelho Lopes and Sandesara, Jay",
    title = "{Polarized ZZ pairs in gluon fusion and vector boson fusion at the LHC}",
    eprint = "2401.17365",
    archivePrefix = "arXiv",
    primaryClass = "hep-ph",
    reportNumber = "IFJPAN-IV-2024-2, COMETA-2024-01",
    doi = "10.1016/j.physletb.2024.138787",
    journal = "Phys. Lett. B",
    volume = "855",
    pages = "138787",
    year = "2024"
}

@article{Frixione:2023hwz,
    author = "Frixione, Stefano and Amoroso, Simone and Mrenna, Stephen",
    title = "{Matrix element corrections in the Pythia8 parton shower in the context of matched simulations at next-to-leading order}",
    eprint = "2308.06389",
    archivePrefix = "arXiv",
    primaryClass = "hep-ph",
    reportNumber = "FERMILAB-PUB-23-413-CSAID",
    doi = "10.1140/epjc/s10052-023-12154-x",
    journal = "Eur. Phys. J. C",
    volume = "83",
    number = "10",
    pages = "970",
    year = "2023"
}
